\documentclass{aa}
\usepackage{lscape}
\usepackage[varg]{txfonts}
\usepackage{graphicx}
\usepackage{booktabs}
\usepackage{amsmath}
\usepackage{colortbl}
\usepackage{gensymb}
\usepackage{subfig}
\usepackage{rotating}
\usepackage{amssymb}
\usepackage{latexsym}
\usepackage{ifthen}
\usepackage{enumitem}

\usepackage{mathrsfs}
\usepackage{amsfonts}
\def\cgs{ ${\rm erg~cm}^{-2}~{\rm s}^{-1}$ } 

\newcommand{\asec} {\mbox{$^{\prime \prime}$} }
\newcommand{\amin} {\mbox{$^{\prime}$}}

\def\gsimeq{\hbox{\raise0.5ex\hbox{$>\lower1.06ex\hbox{$\kern-1.07em{\sim}$}$}}} 
\def\lsimeq{\hbox{\raise0.5ex\hbox{$<\lower1.06ex\hbox{$\kern-1.07em{\sim}$}$}}} 

\begin{document}

\title{Triggering nuclear and galaxy activity in the Bullet cluster\thanks{Based on archival {\it Chandra} observations (PI: Murray, identification number: 554, 3184; PI: Markevitch 4984, 4985, 4986, 5355, 5356, 5357,5358, 5361) and archival data from the database of the Infrared Array Camera (IRAC) on the Spitzer Space Telescope (PI: Gonzales, Program ID: 40593). Based on observations made with the Wide Field Imager (WFI) on the European Southern Observatory/MPG 2.2 m telescope at La Silla, Chile (PI: Fiore, Program ID: 70.A-0594(A)) and the VIsible MultiObject Spectrograph (VIMOS) mounted at the 8.2 m Very Large Telescope (Cerro Paranal, Chile) (PI: Bongiorno, Program ID: 090.A-0267).}}

\author{S. Puccetti\inst{1}, F. Fiore\inst{2,3}, A. Bongiorno\inst{3}, K. Boutsia\inst{4}, R. Fassbender\inst{5,6}, M. Verdugo\inst{7}}

\offprints{S. Puccetti, \email{simonetta.puccetti@asi.it} }

\authorrunning{S. Puccetti et al. }
\titlerunning{\textsc{AGN in the Bullet cluster environment}}

\institute{$^1$Agenzia Spaziale Italiana-Unita' di Ricerca Scientifica, Via del Politecnico, 00133 Roma, Italy;\\$^2$INAF Osservatorio Astronomico di Trieste, Via G. Tiepolo 11, Trieste, Italy;\\ $^3$INAF Osservatorio Astronomico di Roma, via Frascati 33, 00078 Monte Porzio Catone, Italy;\\ $^4$Las Campanas Observatory, Carnegie Observatories, Casilla 601, La Serena, Chile;\\ $^5$Max-Planck-Institut f\"ur extraterrestrische Physik (MPE), Postfach 1312, Giessenbachstr.,  85741 Garching, Germany; $^6$OmegaLambdaTec GmbH, Lichtenbergstr. 8, 85748 Garching, Germany;\\ $^7$Department for Astrophysics, University of Vienna, Türkenschanzstr. 17, 1180 Vienna, Austria.}
\date{29 November 2019}

\abstract{The analysis of the cluster environment is a valuable instrument to investigate the origin of AGN and star-forming galaxies gas fuelling and trigger mechanisms. To this purpose, we present a detailed analysis of the point--like X--ray sources in the Bullet cluster field. Thanks to $\sim600$~ks {\it Chandra} observations, we produced a catalogue of 381 X--ray point sources up to a distance of $\sim$1.5 virial radius and with flux limits $\sim 1\times 10^{-16}$ and $ \sim 8\times10^{-16}$ erg cm$^{-2}$ s$^{-1}$ in the 0.5--2 keV and 2--10 keV bands, respectively. We found a strong (up to a factor 1.5-2) and significant ($\ge$4$\sigma$) over--density in the full region studied $0.3R_{200}<R<1.5R_{200}$. We identified optical (R band) and infrared (Spitzer/IRAC) counterparts for $\sim$84\% and $\sim$48\% of the X--ray sources, respectively. We obtained new spectroscopic redshifts for 106 X-ray sources and collected from the literature additional 13 spectroscopic and 8 photometric redshifts of X-ray sources. 29 X-ray sources turned out to be cluster members. Spectroscopic and photometric redshifts of optical and infrared sources have been also collected, and these sources were used as ancillary samples. We used the multi--wavelength data to characterise the nature of the Bullet cluster X--ray point sources. We find that the over-density in the region $0.3R_{200}<R<R_{200}$ is likely due to X-ray AGN (mostly obscured) and star-forming galaxies both associated to the cluster, while in the more external region it is likely mostly due to background AGN.  
The fraction of cluster galaxies hosting an X-ray detected AGN is 1.0$\pm$0.4$\%$, nearly constant with the radius, a fraction similar to that reported in other clusters of galaxies at similar redshift. The fraction of X-ray bright AGN (L$_{2-10keV}$$>$10$^{43}$ ergs s$^{-1}$) in the region $0.3R_{200}<R<R_{200}$ is $0.5^{+0.6}_{-0.2}$$\%$, higher than that in other clusters at similar redshift and more similar to the AGN fraction in the field.
Finally, the spatial distributions of AGN and star--forming galaxies, selected also thanks to their infrared emission, appear similar, thus suggesting that both are triggered by the same mechanism. }


\maketitle


\section{Introduction}

It is generally recognised that active galactic nuclei (AGN) are powered by gas accretion onto super-massive black holes (SMBHs). However, both the origin of this gas and the detailed mechanism(s) triggering its accretion toward the SMBH are not clear yet. It is quite likely that galaxy interactions, and in particular major mergers of gas--rich galaxies, are the main triggering mechanisms of luminous AGN (quasars). The high luminosity of quasars must be produced by relatively high accretion rates (of the order of $\approx M_{\odot}/yr$), that must be sustained for typical AGN timescales (of the order of $10^7$yr), requiring that a relatively large fraction of the galactic gas is channelled toward the nucleus. In mergers, the fraction of galactic gas which is able to flow to the centre is related to the variation of the gas specific angular momentum induced by the interactions, which is in turn connected to the mass ratio of the merging partners. If the merging partners have similar masses the variation of the specific angular momentum reaches a maximum, as the fraction of gas accreting to the nucleus (Cavaliere \& Vittorini 2000, Menci et al. 2014 and references therein). Lower luminosity AGN such as Seyfert galaxies have fuelling requirements that can be $10^3-10^4$ lower than luminous quasars, and in these cases a number of additional, less efficient triggering mechanisms can be at work, including internal galaxy dynamics like disk instabilities,
large--scale bars, turbulence in the interstellar medium and weakly non-axisymmetric variations in the gravitational potential (see e.g., Simkin et al. 1980; Elmegreen et al. 1998; Genzel et al. 2008, Hopkins \& Quataert 2011, Menci et al. 2014). A further mode of gas accretion onto SMBH in the central galaxies of groups and cluster of galaxies, as well as isolated elliptical galaxies, is the so called chaotic cold gas accretion (Gaspari et al. 2013, Gaspari \& Sadowski 2017 and references therein), or precipitation (Voit et al. 2015). In these cases, low level but continuous nuclear activity, mainly in the form of mechanical power, is produced by a rain of cold gas clouds, condensing out from hot atmospheres and funnelled toward the nucleus via inelastic collisions. Finally, also ram pressure is a possible mechanism for feeding AGN, as suggested by Poggianti et al. (2017). They found a high incidence of AGN among a sample of "jellyfish" galaxies in clusters, and suggest that this may be due to ram pressure causing gas to flow towards the centre and triggering the nuclear activity.

Disentangling between galaxy interactions, internal galaxy dynamics, and chaotic cold accretion as the main trigger mechanism of AGN that are not very luminous is extremely difficult. A promising possibility is to study AGN demography in dense environments, such as groups and clusters of galaxies. Indeed, these are natural sites of frequent galaxy interactions, whose rate depends on the inverse of mass of the group/cluster $M$, since the merger rate scales with $1/\sigma^3 \sim 1/M$, where $\sigma$ is the velocity dispersion of a cluster of mass $M$ (Mamon et al. 1992). Furthermore, the rate of galaxy interaction likely depends also on the distance to the cluster core\footnote{The cluster core is the region inside the radius at which the surface number density equals one-half its peak value (average value of 0.17$\pm$0.02 Mpc for H$_0=70$ km s$^{-1}$ Mpc$^{-1}$, Bahcall, N. A. 1973)}, because the probability of galaxy interactions is correlated with the density and anti--correlated with the velocity, and therefore it should be higher in the cluster outskirts, where there is low galaxy velocity dispersion ($<300$km/s, Binney \& Tremaine 1987) and high in--falling galaxy density, than in the core.

X--ray surveys are extremely efficient tools to select AGN. Indeed, 5--10 keV X--rays are capable to penetrate column densities up to $\sim 10^{24}$ cm$^{-2}$, allowing the selection of moderately obscured AGN. In addition, X-ray surveys allow the detection of low luminosity AGN, which are difficult to select in optical surveys, because their optical light is diluted by the host galaxy stellar emission. Furthermore, AGN are the dominant population in the X-rays, because they represent the majority ($\sim$80\%) of the X--ray sources making more efficient their spectroscopic observation.
  
  After the first pioneering studies on single or small samples of clusters the first systematic study of a relatively large sample of clusters (51 clusters at z=0.3-0.7) was performed by Ruderman \& Ebeling (2005). They detected a total of about 500 sources at fluxes corresponding to moderately luminous AGN and found a significant over--density of X-ray sources with respect to the field in the central 0.5 Mpc. Gilmour et al. (2009) found again a significant over--density of X-ray sources in a sample of 148 clusters at z=0.1-0.9, with a flat distribution up to $\sim1$~Mpc.  Ehlert et al. in a series of three papers (2013, 2014, 2015) analysed Chandra data of an increasing number of clusters (43 to 135) at z=0.2-0.9,  finding up to ten thousands X-ray sources, more that an order of magnitude improvement with respect to previous works. They found an highly significant over--density of X-ray sources per unit solid angle with respect to the field, concentrated at radii $<0.5$~Mpc (or $0.5~R/R_{500}$\footnote{$R_{500}$, $R_{200}$, $R_{100}$ are the radii within which the density contrast with the critical density of the universe is 500, 200 and 100 respectively. $R_{100}$ ($\sim$ the virial radius) is $\simeq 1.3 \times R_{200}$.}), and following a power law profile $R^{-0.6}$. Ehlert et al. (2014) also studied the fraction of X-ray active galaxies to the total number of galaxies brighter than R=23 (cluster members $+$foreground \& background sources), finding that in  cluster cores this is $\approx3$ times lower than in the field (when excluding the central dominant galaxies from the calculation). The fraction of active galaxies increases with the projected distance from the cluster centre and become consistent with the field fraction of active galaxies at $R\sim2.5~R_{500}$. Finally, Ehlert et al. (2015)  found that the number density of X-ray sources in cluster fields scales with the mass of the cluster as $M^{-1.2}$, consistent with the expectation based on the merging rates.\\
\indent The over--densities of X-ray sources with respect to the field reported in the papers cited above are not surprising, because clusters are over--dense regions of the Universe, showing galaxy densities much higher than in the field. A different question is whether the fraction of galaxies hosting an AGN is different in clusters with respect to the field, and how this fraction changes with the projected radius and the mass of the cluster. Haines et al. (2012) analysed a sample of 26 clusters at z=0.15--0.3.  Using Chandra data and optical spectroscopy they found a total of 48 X-ray AGN among cluster members, with 0.3--7 keV luminosities in the range $2\times 10^{41}-10^{44}$. Their luminosity function is a factor 7.5 higher than the field luminosity function, again confirming an over--density of X-ray sources in cluster fields. However, this over-density is completely due to the much higher density of galaxies in clusters, and in fact the same authors found an average fraction of active galaxies in their cluster sample
of $0.73 \pm 0.14\%$, which is not higher than in the field.  Similar results were obtained by Martini et al. (2007) and Haggard et al. (2010), who found AGN fractions of 1\% and 1.19$^{+0.11}_{-0.08}\%$ in clusters and in the field, respectively, at redshift similar to that of the Bullet cluster. Limiting the analysis to AGN with L$_{2-10keV}> 10^{43}$ erg s$^{-1}$ Martini et al. (2013) find a smaller AGN fraction, 0.107$^{+0.057}_{-0.039}\%$, about six times smaller than in the field (Haggard 2012, private communication in Martini et al. 2013, reports a field AGN fraction of  0.64$^{+0.04}_{-0.05}\%$). Finally, Ehlert et al. (2015) noted that HST images of 23 spectroscopically confirmed cluster AGN members, suggest that they are hosted in merger systems, and conclude that the triggering of AGN is predominantly due to mergers in clusters, while in the fields more or different mechanisms can be at work.\\
\indent
In this paper we re-examine in a single peculiar field covered with unprecedented X-ray sensitivity over a wide area all above issues: i) over--density of X-ray sources with respect to the field and their spatial distribution; and ii) fraction of (cluster members$+$foreground \& background sources) X-ray active galaxies; iii) fraction of cluster X-ray active galaxies. We target the Bullet cluster, a massive galaxy cluster at z$=$0.296 undergoing a major merger event. The $R_{200}$ of the main cluster component is $\sim2.14$ Mpc (Springel \& Farrar 2007, Clowe et al. 2004), which is $\sim$8.07 $\amin$ at the cluster distance according to a cosmology H$_0=70$ km s$^{-1}$ Mpc$^{-1}$, $\Omega_M=0.27$, $\Omega_\Lambda=0.73$. In this cluster the star formation rate (SFR) is higher than in other clusters of similar mass (Chung et al. 2010; Rawle et al. 2010), suggesting more galaxy interactions, or, more in general, more gas--rich galaxies and more efficient SFR triggering mechanisms, than usual. All these peculiar factors may also trigger enhanced AGN activity.  To probe nuclear activity down to $10^{41}$ erg s$^{-1}$ we exploit ten precious archival {\it Chandra} observations, covering a wide region around the cluster (up to $\sim$1.6--2~$\times$~R$_{200}$), with a sensitivity (flux limits $\sim 1\times 10^{-16}$ \cgs and $8\times10^{-16}$\cgs in the 0.5–2 keV and 2–10 keV bands, respectively) hardly ever reached in other clusters. Chandra archival data of the Bullet cluster were used in previous publications, but never in their entirety or full sensitivity. Gilmour et al. (2009) used only the first two Chandra observations for a total exposure time of $\sim100$~ks, about 1/6 of the total available today. Ehlert et al. (2013, 2014, 2015) restrict their analyses to the central 6\amin~radius region, corresponding to projected distances from the centre $<$~0.75$\times$~$R_{200}$, and to fluxes greater than $3\times10^{-15}$\cgs and $1.5\times10^{-14}$\cgs in the 0.5–2 keV and 2–10 keV bands, respectively.

The paper is organised as follows: in Section 2, we describe the data used in this work, which are from: X--ray ({\it Chandra}), optical (ESO2.2m/WFI), infrared (Spitzer/IRAC archive, Fazio et al. 2004) and spectra from VIMOS. In Section 3 we present the X--ray catalogue.  In Section 4 we show the X--ray number counts and the source over--density. In Section 5 we combine the Chandra data with archival optical/NIR photometry. The archival spectroscopy as well as new optical spectroscopy obtained with VLT/VIMOS are discussed in Section 6. Finally the results and conclusions are discussed in Section 7.

\section{The data}

\subsection{{\it Chandra} observations}

We analysed ten archival {\it Chandra} observations of the Bullet
field, performed from October 2000 trough August 2004 with the ACIS--I
camera, for a total exposure time of $\sim$584.2 ksec (observation
identification numbers: 554, 3184, 4984, 4985, 4986, 5355, 5356, 5357,
5358, 5361). We reprocessed all the observations using the ${\it
  chandra\_repro}$ script included in the {\it Chandra} Interactive
Analysis Observations (CIAO) software (v4.5; Fruscione et
al. 2006). For the observations taken in {\it very faint} (VF)
data--mode we applied the appropriate ACIS particle background
cleaning as suggested in the {\it Chandra} analysis guide, enabling VF
data mode background processing in the {\it chandra\_repro}
script. For each observation, we produced calibrated, cleaned and
energy filtered events files and exposure maps.

\begin{figure*}
\begin{center}
\includegraphics[angle=0,height=12.5truecm,width=13.0truecm]{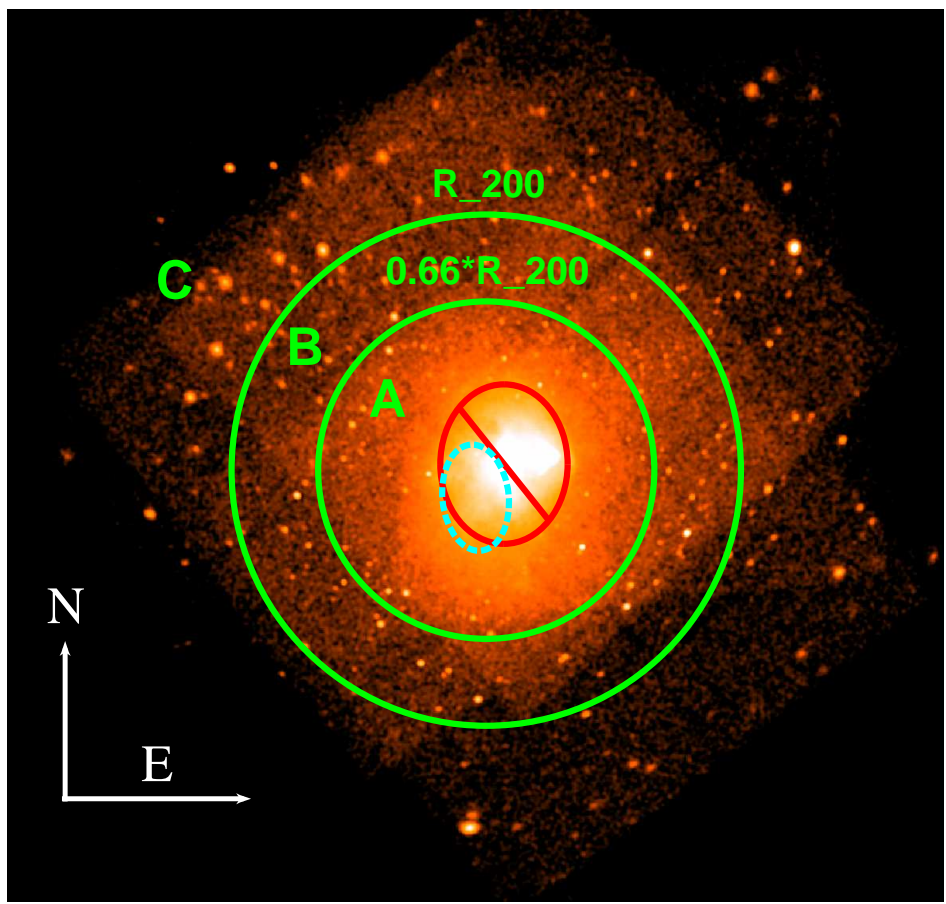}
\caption{Image of the mosaic of the ten {\it Chandra} observations (see Sect. 2.1) of the Bullet cluster field in the 0.5--7 keV energy range. The green circles are centred on the main cluster and have radii of $\sim5.3\amin$ and $\sim 8.07\amin$, corresponding to 0.66 and 1$\times R_{200}$. The observations cover up to $\sim 2\times R_{200}$ (i.e. $\sim 1.5 \times$ virial radii). The red elliptical region marks the area with the strongest ICM emission  at distance less than $< 0.3\times R_{200}$ (i.e. $\sim 0.2\times$  virial radii), which has been excluded from our analysis. The cyan dashed elliptical region is centred in the peak of the dark matter distribution, which is significantly shifted eastward from the peak of the X-ray emission (Clowe et al. 2004, 2006).
\label{image}}
\end{center}
\end{figure*}

Before generating the mosaic used for sources detection we brought
each observation to the same astrometric system. To this aim, we first
produced a sample of X--ray bright sources, following the procedure
described in Sect. 3.1 and using a probability threshold of
$3\times10^{-7}$. Then, we identified the optical R$_{c}$ band (see
Sect. 2.2) counterparts of the X--ray bright sources by matching the
X--ray and the R$_{c}$ positions within a maximum distance of
2$\asec$. We excluded sources which are not point--like and/or with
R$_{c}$ magnitudes outside the range 16.5--22.5. Using this restricted
magnitude range only minimises systematic effects introduced by bright
stars (saturation) and faint background objects
(misidentification). Finally we evaluated the Right Ascension and
Declination astrometric corrections as the median of the X--ray
position$-$Optical position values. The Right Ascension and
Declination astrometric corrections were all smaller than $1\asec$.
The astrometric corrected event files are finally used to generate the
mosaics of the event files (see Fig. \ref{image}) and exposure maps by
using the ${\it merge\_{obs}}$ CIAO tool.

\subsection{Wide Field Imager (WFI) data}

Optical imaging in the R$_{c}$ (BB\#RC/162\_ESO844) and B
(BB\#B/123\_ESO878) bands were obtained on February and March 2003
with the Wide Field Imager (WFI) on the ESO/MPG 2.2 m telescope at La
Silla, Chile. The detector is a 4$\times$2 mosaic with 2k$\times$4k
CCDs, having a pixel scale of 0.238$\asec$/pixel and a large field of
view (34$\amin \times$33$\amin$). The data were collected in service
mode and splitted in six individual exposures of 700 sec in the
R$_{c}$ band and 500 sec in the B band, following a dither pattern in
order to cover the gaps between the CCD chips.

We reduced each image separately, using standard procedures (bias
subtraction and flat fielding using dome flats) and aligned it to a
common coordinates system using the USNO catalogue. We then stacked the
astrometric corrected images using SWarp (Bertin et al. 2002) to
produce one mosaic in each band. The total exposure time for each
mosaic is 4200 sec in the R$_{c}$ band and 3000 sec in the B band. The
photometric zero points (i.e. zp), determined using Landolt standard
star fields obtained during the same night, are 24 mag and 24.55 mag
in the R$_{c}$ and B band, respectively.

We produced photometric catalogues on each band using SExtractor (Bertin
\& Arnouts 1996). We used aperture photometry with a diameter of
2~$\times$~FWHM in each band, and corrected it for flux losses due to
aperture size. The correction has been calculated adding the average
value of ($<$ mag$_{\rm AUTO}$$-$mag$_{\rm APER}$$>$)$_{\rm stellar}$
for bright stellar sources, to the total magnitude of each source. The
10~$\sigma$ magnitude limit of our imaging data is 23.51 mag and 23.98
mag for the R$_{c}$ and B band, respectively.
For the X--sources that do not have a detected optical counterpart we
calculated upper limits of their R$_{c}$ and B band magnitudes. To
this aim we generated simulated optical images with artificial sources
at the positions of these X--ray sources, and run Sextractor in dual
mode, using the simulated image for detection and the real image to
calculate the aperture magnitudes at the same positions. In positions
where the flux in the aperture is lower than the error, the magnitude
is calculated using the error value as input flux in the equation:
mag$=-2.5\times$ log(flux)$+$zp.

\subsection{IRAC data}

\begin{figure*}
\begin{center}
\includegraphics[angle=0,height=12.5truecm,width=12truecm]{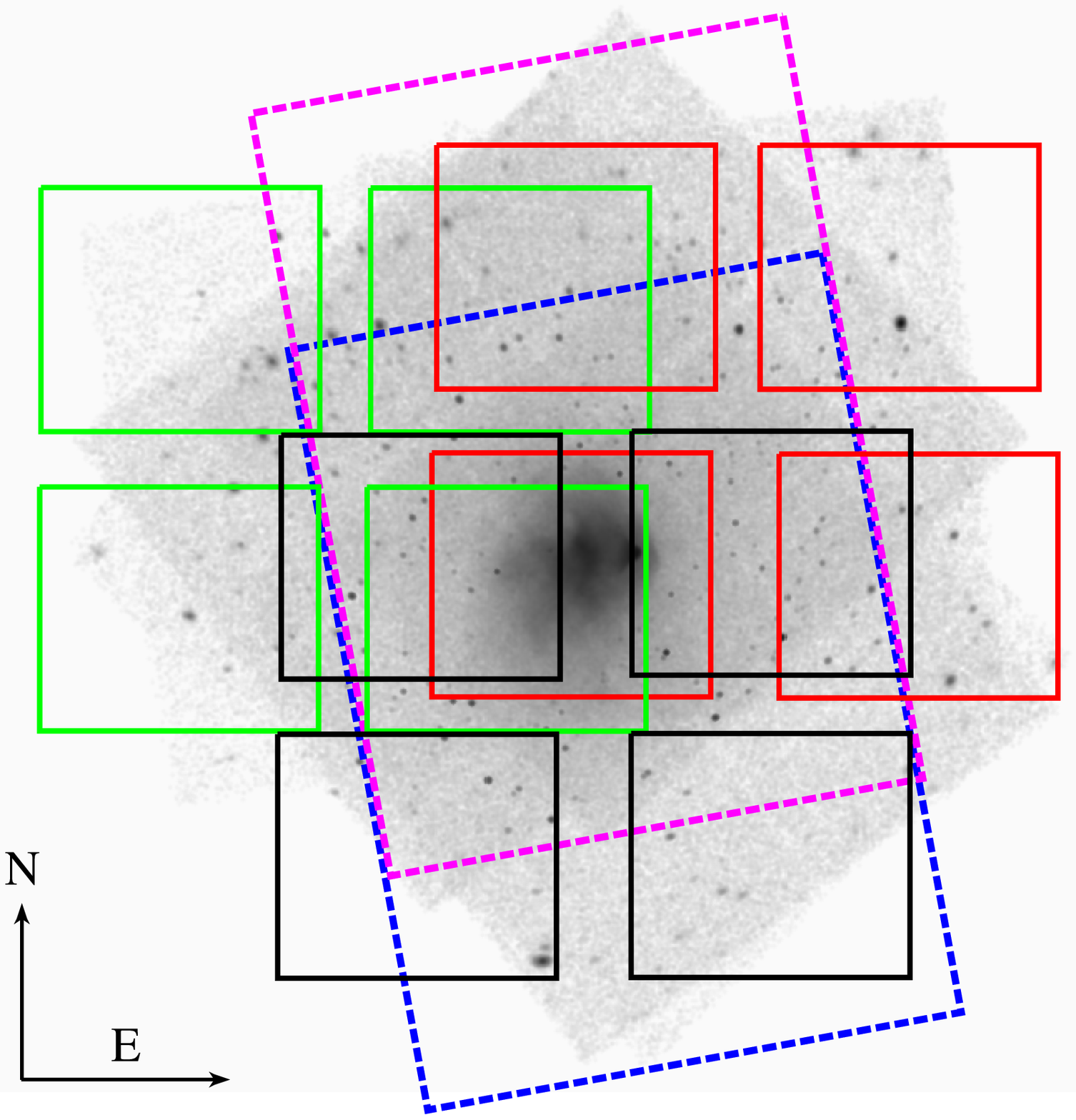}
\caption{Gray scale image of the mosaic of the ten {\it Chandra} observations of the Bullet cluster field in the 0.5--7 keV energy range (see  Fig. \ref{image}). The green, red and black squares mark three VIMOS fields. The magenta and blue dashed boxes indicate the portion of the field covered by IRAC ch1--ch3 and IRAC ch2--ch4 photometry respectively.
\label{coverage}}
\end{center}
\end{figure*}

We retrieved SPITZER/IRAC images of the Bullet field from the IRSA
archive. Images in the four IRAC bands (ch1$=$3.6 $\mu m$, ch3$=$5.8
$\mu m$, ch2$=$4.5 $\mu m$ and ch4$=$8.0 $\mu m$) were obtained on
2007 November 14 (PI: Gonzales, Program ID: 40593) with a cycling,
small scale dither-pattern and a full array read--out mode. We used
the reduced and astrometrically calibrated BCD mosaics along
with their respective uncertainty and covariance maps available on the
IRAC database. The total area covered by the images in all four
channels is 15$\amin\times$15$\amin$ and the mosaics have a pixel
scale of 0.61$\asec$/pixel (see Fig. \ref{coverage}).

We produced photometric catalogues in each channel using SExtractor (Bertin \& Arnouts 1996). We have used a detection threshold of $3\sigma$ and a minimum detection area of 6 pixels (ch1 and ch2) or of 9 pixels (ch3 and ch4). For each channel we used the Vega zero--points found in the IRAC manual, which are 18.8 mag  (ch1), 18.32 mag (ch2), 17.83 mag (ch3) and 17.20 mag (ch4) and correction apertures of 0.127 mag, 0.130 mag, 0.145 mag and 0.126 mag respectively, based on the manual values. The $5\sigma$ depth was found to be: 16.9 mag, 16.4 mag, 15.6 mag and 15.0 mag in ch1, ch2, ch3 and ch4 respectively.

\subsection{VIMOS data}

Multi--object spectroscopic observations was performed in February and March 2013 (PI: Bongiorno, Program ID: 090.A-0267) using the VIsible MultiObject Spectrograph (VIMOS) mounted at the 8.2 m Very Large Telescope (Cerro Paranal, Chile). Optical spectra were obtained using the medium resolution grism and the filter GG475 which covers the spectral range from 5000~\AA~ to ~8000~\AA, with a resolution R$=$580.

The central area of the Bullet cluster was covered with three overlapping VIMOS masks as shown in Fig. \ref{coverage}, targeting 253 sources brighter than 24 mag in the R$_{c}$ band. 145 of them were extracted from the {\it Chandra} catalogue (see Sect.  3.2), 83 from the IRAC catalogue (see Sect. 2.3) and 25 from the WFI catalogue (see Sect. 2.2). Spectra are the result of exposures of about 1800 s each. 1 hour net exposure time was requested in service mode for each mask. However, for mask 2 and 3, observations were repeated more than once to meet the requested observational conditions. Combining all the good quality data, the final net exposure times obtained are 1 hour for mask 1, 3 hours for mask 2 and 2 hours for mask 3.\\
\indent Data reduction was performed using the ESO ESOREX pipeline. Spectra were then flux calibrated using standard stars observed during the same nights at comparable airmass and 1D spectra were extracted from the 2D images of the masks using the IRAF task \textit{apall}. The 1D spectrum were extracted for 218 out of the full sample of 253 sources for which a slit was placed in the masks.  Finally, redshifts were measured using the IRAF task \textit{rvidlines}. Following Le Fèvre et al. (2013) we also assigned to each redshift a quality flag, ranging from 1 for very insecure redshifts to 4 corresponding to secure ones, with 0 (24 sources) corresponding to no redshift assigned.

\section{X--ray point sources}

\subsection{Source detection and characterisation}

The ten {\it Chandra} pointings are strongly overlapping but not all coaxial, to ensure a deep and uniform sensitivity along the full cluster. Each position in the mosaic was observed with up to seven different Point Spread Functions (PSFs), requiring the development of a specific procedure to detect and characterise serendipitous sources. To detect sources we used the {\it pwdetect} code (Damiani et al.  1997), which was originally developed for the analysis of ROSAT data, and was then adapted for the analysis of {\it Chandra} and XMM--Newton data.  This method is particularly well suited for cases in which the PSF is varying across the image, as for {\it Chandra} images, since {\it pwdetect} is based on the wavelet transform of the X--ray image, i.e., a convolution of the image with a ``generating wavelet'' kernel, which depends on position and length scale, that is a free parameter. For the {\it Chandra} data, the length scale is varied from 0.35$\asec$ to 16$\asec$ in steps of $\sqrt{2}$.\\
\indent The preliminary step was to run {\it pwdetect} on the {\it Chandra}
mosaic to produce a first source position list, using a
non-conservative probability threshold corresponding to hundreds of
spurious sources. The final X--ray catalogue will be generated by
screening this list according to the following procedure.

\begin{enumerate}
\item
For each source candidate we determined the counts extraction radius
$R_{best}$ that provides the best signal to noise ratio (SNR) on the
mosaic event file. We used circular regions centred on the {\it
  pwdetect} position, starting from a radius of 1$\asec$ and
increasing the radius in steps of 0.25$\asec$ up to 16$\asec$.

\item
We then extracted source counts $T_i$ at the position of the source
candidate on each individual {\it Chandra} field {\it i} using
circular regions with radius $R_{best}$.

\item
We evaluated background counts $Bs_i$ at the position of each source candidate from the background maps
computed by {\it pwdetect} (see Damiani et al.  1997) using single
event files. 

\item
The net source counts (i.e. $C_i=T_i-Bs_i$) were
corrected for the encircled count fraction of the PSF ($f_{PSF}$) at the off--axis angle
$\theta_i$ (the distance of the source position from the aim
point of each  {\it Chandra} pointing), as calibrated by the
CXC\footnote{http://cxc.harvard.edu/caldb/}:
$Ccor_i=C_i/f_{PSF}$ .

\item
The
multiple field aperture photometry values were combined to
produce a single set of values as in the following:

\begin{description}
\item[$\bullet$] total source counts $T=\sum_{i=1}^{10 }T_i$;
\item[$\bullet$] total background counts  $Bs=\sum_{i=1}^{10 }Bs_i$; 
\item[$\bullet$] total net PSF corrected source counts  $Ccor=\sum_{i=1}^{10 }Ccor_i$;
\item[$\bullet$] source  count rate $CR={{Ccor}\over{\sum_{i=1}^{10 } Expo_i}}$, where
  Expo$_i$ are the vignetting corrected exposure times from the
  exposure maps of each single observation. 
\end{description}

\item
At each candidate source is then assigned a Poisson probability of
being background fluctuations, according to the formula:
\begin{equation}
P_{Poisson}=e^{-(Bs)}\sum_{i=T}^{\infty} { (Bs)^i \over i!}
\end{equation}

\item
Finally, the {\it pwdetect} source list is cut at a threshold of
P$_{Poisson}=2\times10^{-5}$, which allowed us to have a
reliable catalogue (see e.g., C--COSMOS survey, Elvis et al. 2009,
Puccetti et al. 2009). The final list was also visually
inspected to identify obviously spurious detections (e.g., on the wings
of the {\it Chandra} PSF around bright sources).

\item
As additional test, we applied this very same procedure to the C--COSMOS data, 
finding results consistent with the official catalogue (Elvis et al. 2009).
\end{enumerate}

\subsection{X--ray catalogues}

We produced X--ray source catalogues in three energy bands: 0.5--2 keV (soft band, S), 2--7 keV (hard band, H) and 4--7 keV (very hard band HH).  Exposure maps were computed at energies of 1.17, 2.6 and 4.5 keV, respectively, which correspond to the energy mean values over the energy S, H and HH energy ranges, weighted with the spectral model used to convert count rates in fluxes (see below).

We performed the source detection in the whole {\it Chandra} field, but we excluded from our analysis the central region of the Bullet cluster, where the ICM emission is strongest and makes difficult if not impossible to detect point like sources.  The excluded region, marked by the red ellipse in Fig. \ref{image}, includes the cluster core ($\lsimeq$ 1$\amin$), and corresponds to an ICM background level $\gtrsim 6 \times10^{-7}$ct/s/arcs$^2$, $\gtrsim 5\times10^{-7}$ct/s/arcs$^2$ and $\gtrsim 2.5\times10^{-7}$ct/s/arcs$^2$ for the S, H and HH band, respectively.

The count rates (CR) were converted to fluxes ($F_x$) using the formula: $F_x=CR/(CF\cdot10^{11}$), where $CF$ is the energy conversion factor, that is evaluated by PIMMS\footnote{http://cxc.harvard.edu/toolkit/pimms.jsp} using an absorbed power$-$law spectrum with $\Gamma=1.4$\footnote{$f_E \propto E^{-\Gamma}$ with $\Gamma$$=$$\alpha_E$+1 and $\alpha_E$ energy index.} and Galactic column density toward the position of the Bullet cluster $N_H=4.9\times10^{20}$ cm$^{-2}$, and suitable for the observing cycle 5, during which most of the observations are taken (8 out of 10). In order to compare our results with the literature, count rates estimated in the S, H and HH band were extrapolated to fluxes in the 0.5--2~keV, 2--10~keV and 5--10~keV bands respectively. We used $CF=1.82$, 0.34 and 0.22 for the S, H and HH band, respectively.

We detected 317, 254 and 148 point--like sources in the S, H and HH bands (see Tab. \ref{cat}). The three source lists are merged in one catalogue, using a matching radius of 2.5$\asec$. The final catalogue is made up by 381 unique sources.

\begin{table}
\footnotesize
\caption{Number of sources detected in each X--ray band.}
\begin{center}
\begin{tabular}{lccc}
\hline
Band & N$^a$ & flux$_{min}$$^b$ & flux$_{max}$$^c$\\
\hline
S&  317 & 0.08 & 188.8 \\
H&  254 & 0.76 & 248.5 \\
HH&  148 & 0.92 & 112.3 \\
\hline
S+H+HH  & 119 \\
S+H & 75\\
H+HH & 25 \\
S only & 123 \\
H only & 35\\
HH only & 4\\
\hline 
\end{tabular}
\end{center}

$^{a}$ number of detected sources with detection significance level $\leq$ 2$\times10^{-5}$; $^{b}$ flux of the faintest source in unit of 10$^{-15}$ erg cm$^{-2}$ s$^{-1}$ ; $^{c}$ flux of the brightest source in unit of 10$^{-15}$ erg cm$^{-2}$ s$^{-1}$.
\label{cat}
\end{table} 

\section{X--ray number counts and excess density}

\subsection{Survey sensitivity}

As mentioned in the previous sections, the ten {\it Chandra}  pointings are mostly overlapping but not--coaxial, then at each sky position we have a mixture of different PSFs and vignetting factors, a situation similar to the C--COSMOS case (Elvis et al. 2009, Puccetti et al. 2009). For this reason we computed sensitivity maps and sky-coverages (the total area in the sensitivity maps covered down to a given flux limit) in the S, H and HH band following the analytical method presented in Puccetti et al. (2009).

\begin{figure}
\includegraphics[angle=0,height=8truecm]{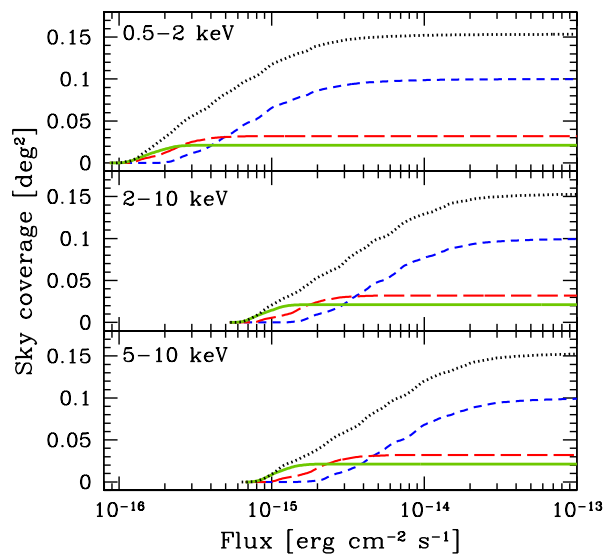}
\caption{From {\it top} to {\it bottom panel}: sky--coverage calculated as in Sect. 4.1 in the S, H and HH energy bands excluding the area with the strongest ICM emission (black dotted lines, see also Fig. \ref{image}) and the A (green solid lines), B (red long dashed lines) and C (blue short dashed lines) regions (see Fig. \ref{image} and Sect. 4.1).
\label{skycov}}
\end{figure}

We evaluated the sky-coverages in the three energy bands S, H and HH for the whole {\it Chandra} field around the Bullet cluster, and also for three regions A, B and C in Fig. \ref{image} at increased distance from the cluster centre.  The three regions A, B and C are defined as following:

\begin{description}

\item[A]: circular region centred on the main cluster with radius$=0.66 \times R_{200}$, excluding the area with the strongest ICM emission at $R< 0.3\times R_{200}$ (red elliptical region in Fig. \ref{image});

\item[B]: annular region centred on the main cluster with inner radius $R=0.66 \times R_{200}$ and outer radius $R=R_{200}$;

\item[C]: area outside the regions A and ($R_{200}<R\leq2\times R_{200}$).
\end{description} 
Fig.\ref{skycov} shows the sky--coverages in the S, H and HH band for the whole {\it Chandra} Bullet field and the three regions A, B and C.

\subsection{X-ray number counts and excess density in the {\it Chandra} field}

The sky--coverages are used to evaluate the integral X--ray number counts using the following equation:

\begin{equation}
N(>{\it S})=\sum_{i=1}^{N_{\it S}}{ 1 \over {\Omega_i}} deg^{-2}
\end{equation}

\noindent
where N$_{\it S}$ is the total number of detected sources with fluxes larger than {\it S}, and $\Omega_i$ is the sky-coverage at the flux of the i--th source.

\begin{figure*}
\centering
\includegraphics[width=6.cm]{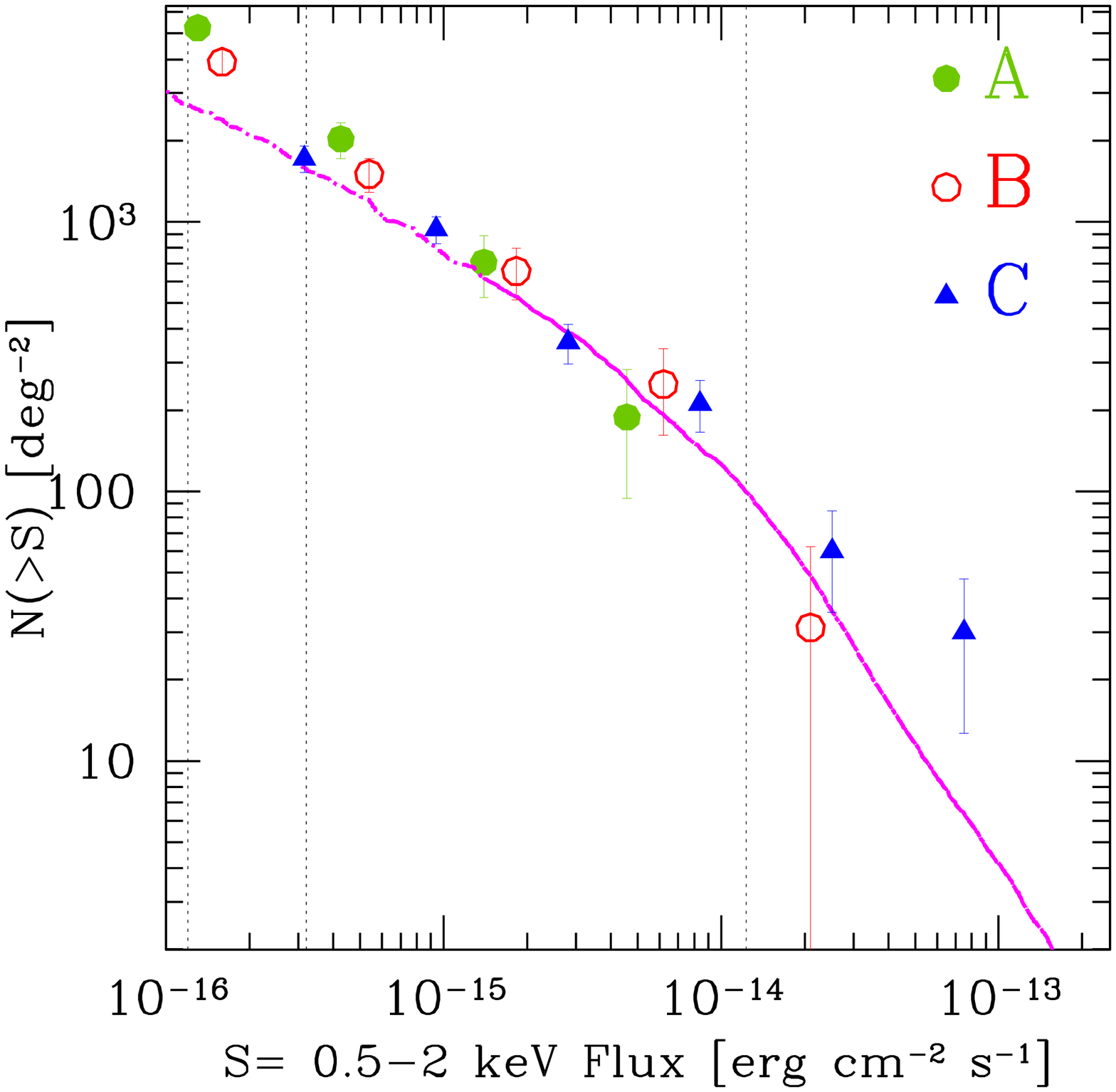}\includegraphics[width=6.cm]{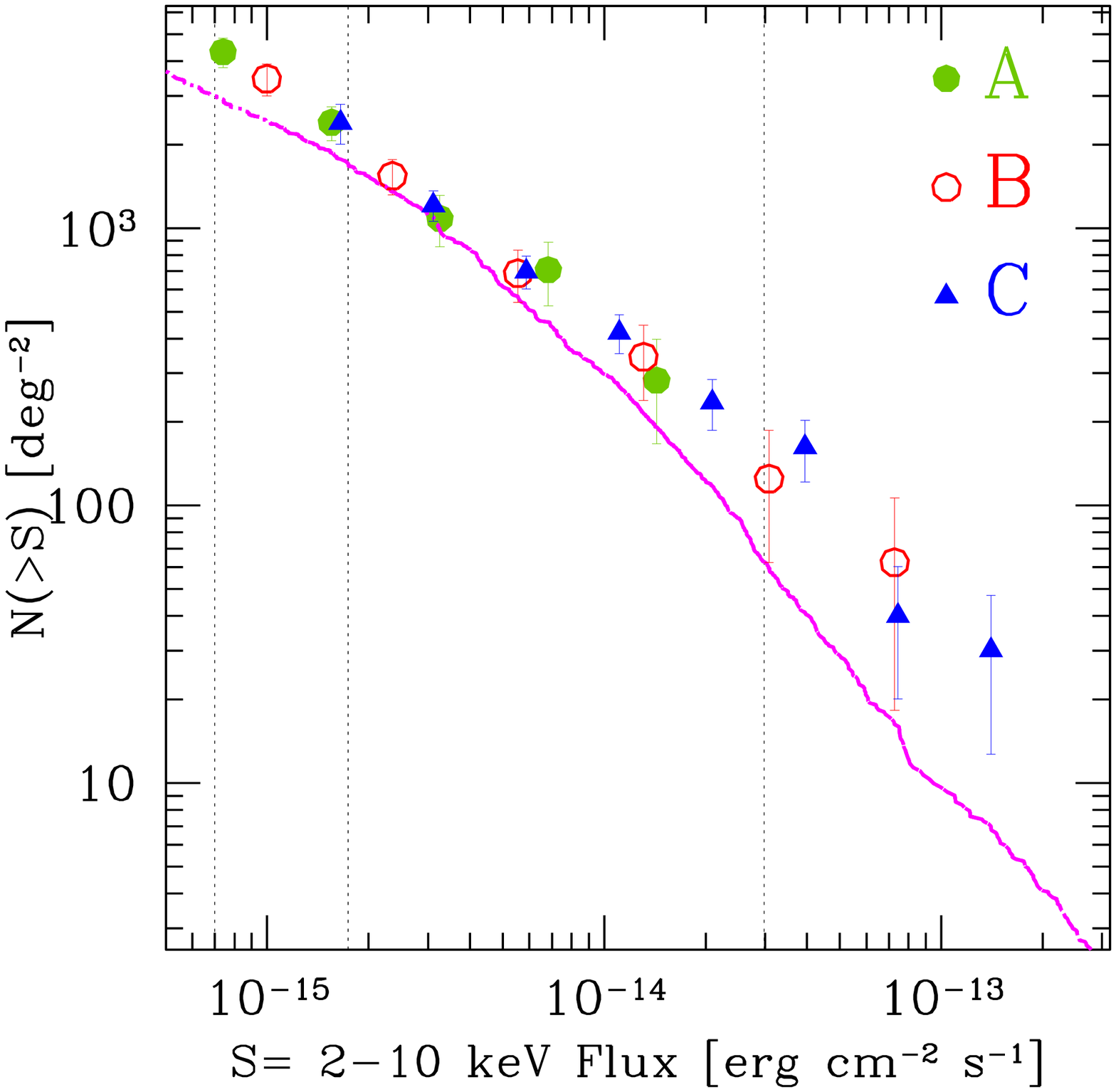}\includegraphics[width=6.cm]{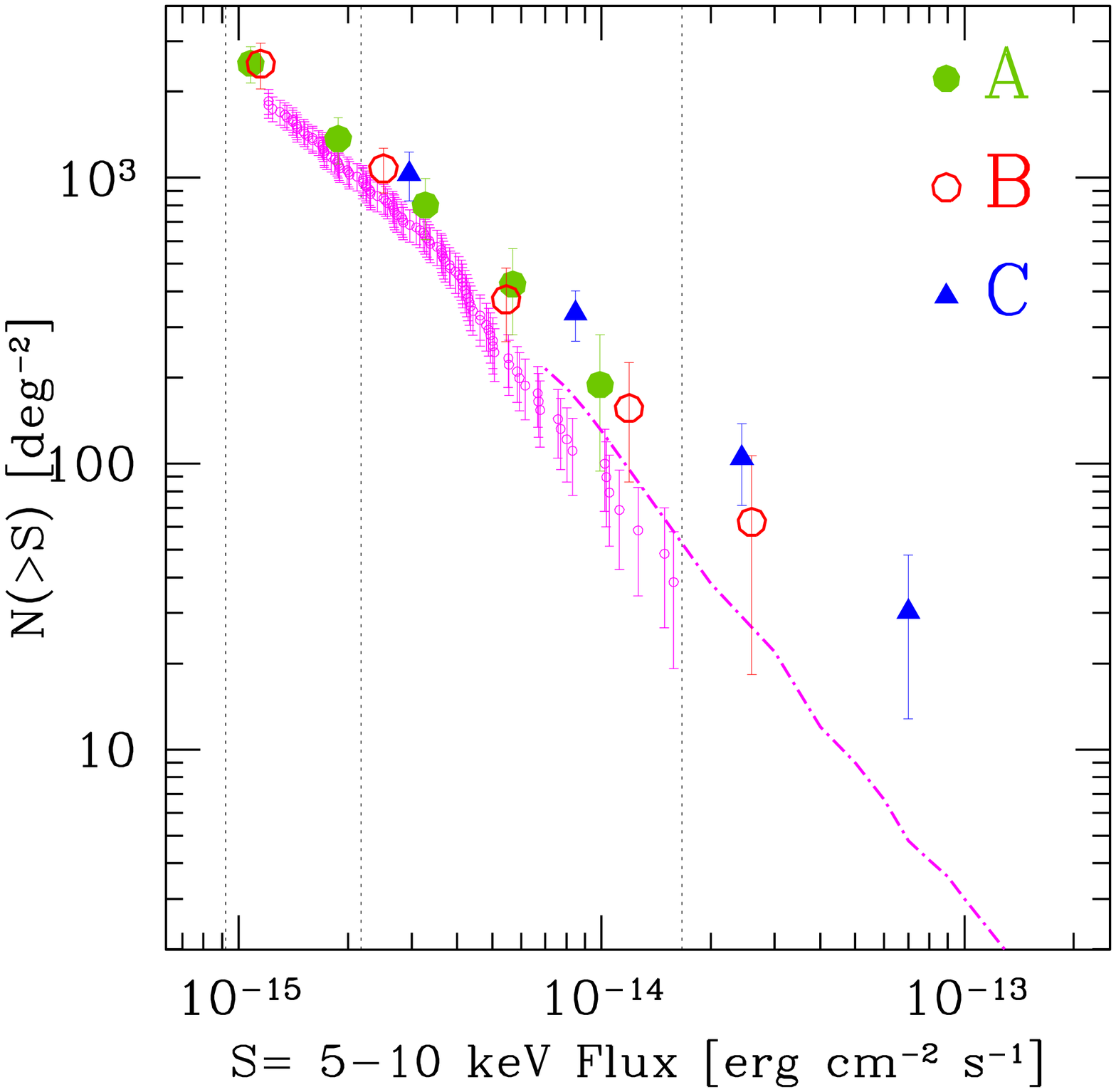}
\includegraphics[width=6.cm]{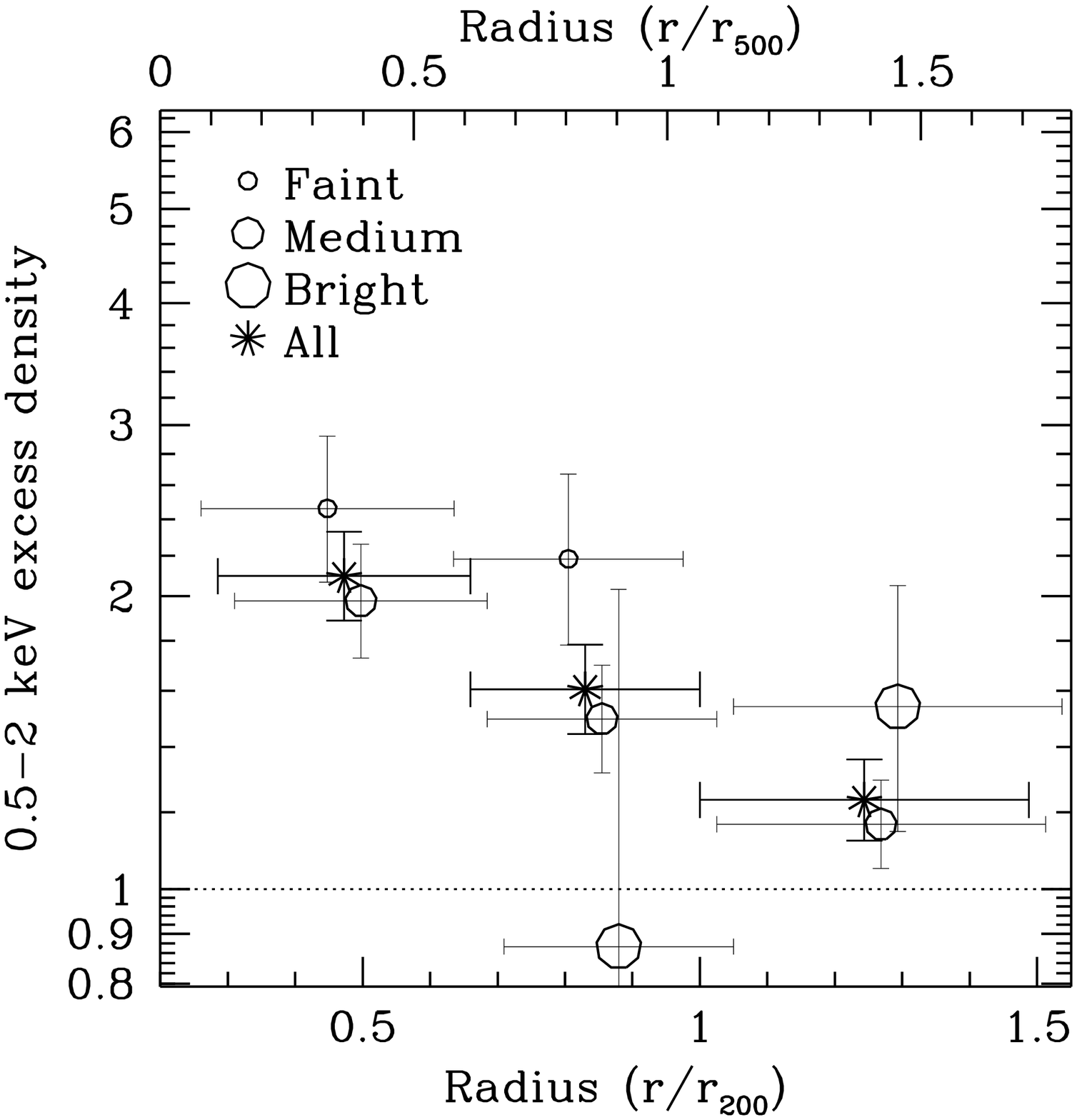}\includegraphics[width=6.cm]{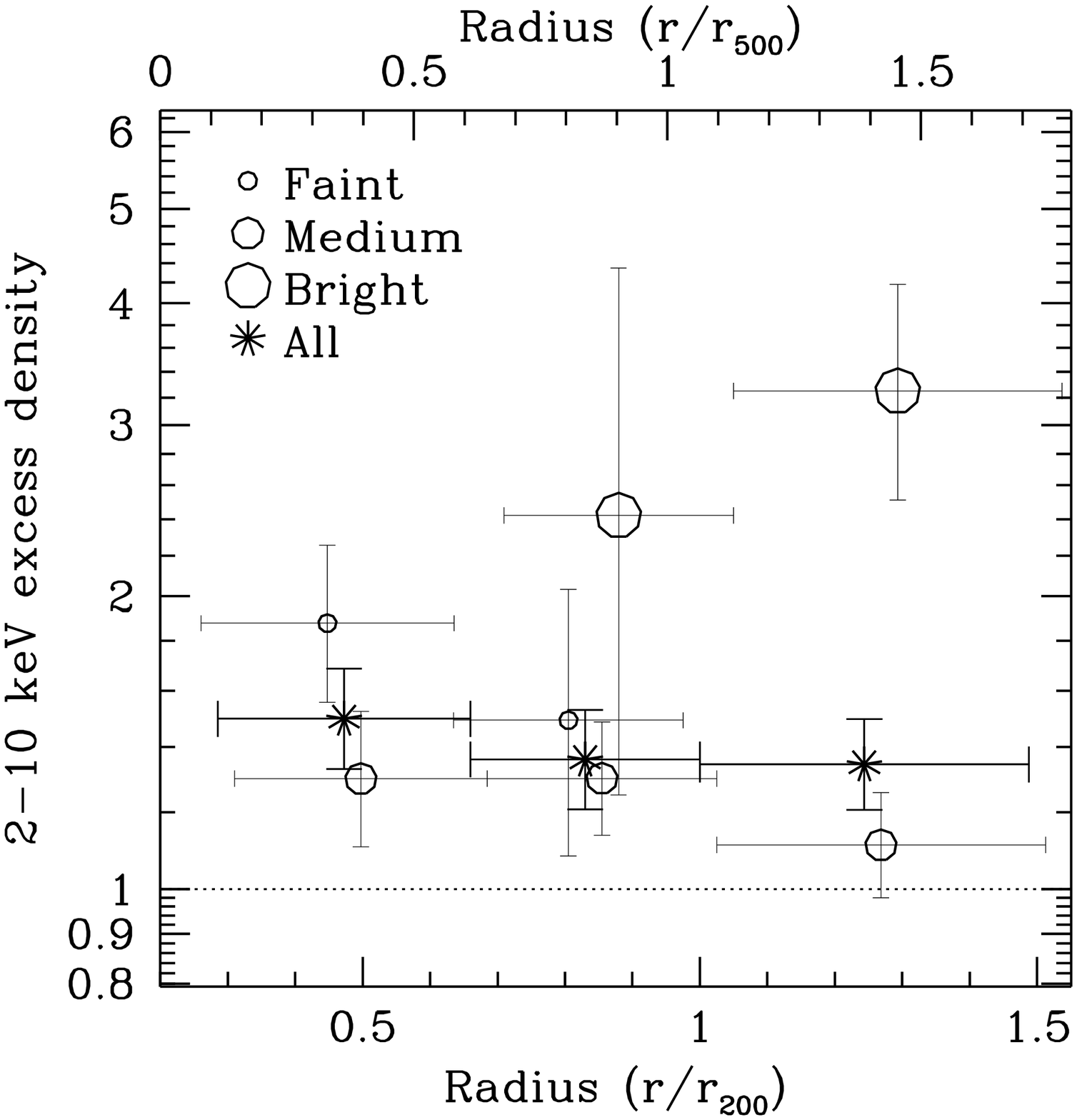}\includegraphics[width=6.cm]{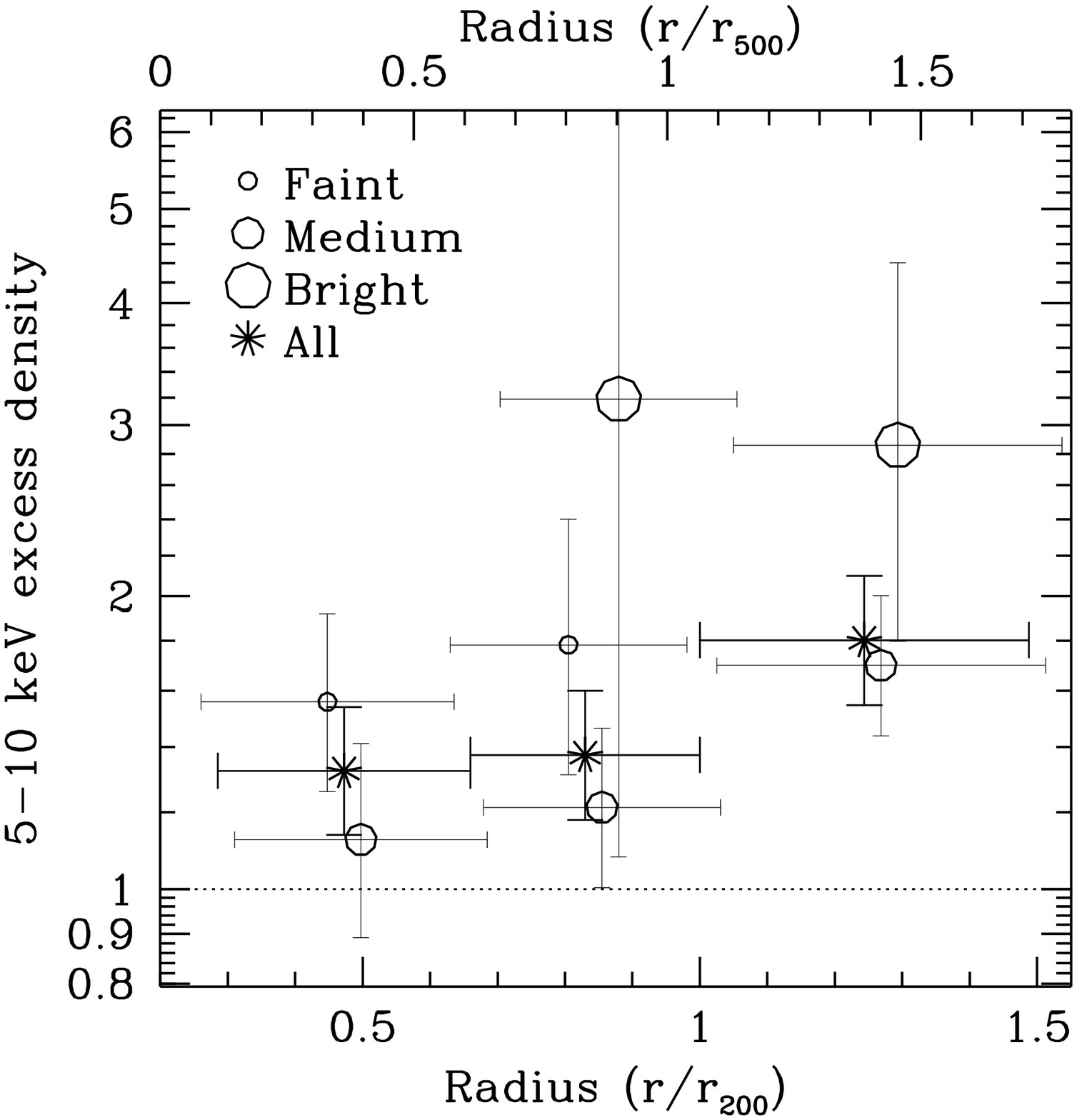}
\caption{{\it Upper panels}: the X--ray number counts of the X-ray point sources in the Bullet cluster regions A (green solid circles), B (red open circles) and C (blue solid triangles). {\it Left panel}: 0.5--2 keV, magenta solid line = field number counts from  Moretti et al. (2003). {\it Central panel}: 2-10 keV, magenta solid line = field number counts from  Moretti et al. (2003). {\it right panel} 5-10 keV, magenta  dashed-dot line = HELLAS2XMM 5-10 keV number counts from Baldi et al. (2002); magenta open circles =  CDF-S 5--10 keV number counts from Luo et al. (2008). The black dotted vertical lines identify the three flux ranges: faint, medium and bright. {\it Bottom panels}: excess densities with respect to field number counts as a function of the distance from the cluster centre in the three energy bands 0.5--2 keV ({\it left panel}), 2--10 keV ({\it central panel}) and 5--10 keV  ({\it right panel}).  Faint flux sources = small open circles; medium flux sources = medium open circles; bright flux sources = large open circles; all sources = stars. Note that the excess densities could include some contribution from cosmic variance ($<1.25, <1.20$ and $<1.10$ in the A, B and C regions, respectively, see Sect. 4.3).
\label{lnlsrad}}
\end{figure*}

Fig. \ref{lnlsrad} (upper panels) compares the X-ray number counts in the three regions A, B and C to those in {\it field} surveys. The X--ray number counts in all studied X-ray bands S, H and HH are systematically higher than the ones from {\it field} surveys. We estimated the excess density and uncertainties as follows:

\begin{equation}
Excess~density = {N_{obs} \over N_{exp}}
\end{equation}

\noindent
where N$_{obs}$ are the observed number of sources and N$_{exp}$ are the number of the expected sources evaluated from the compilation of Moretti et al. (2003) for the S and H band, and from the Luo et al. (2008) and Baldi et al. (2002) determinations for the HH band. The upper ($+\Delta$ Excess) and lower ($-\Delta$ Excess) uncertainties are evaluated using the formulae of Gehrels (1986) for Poisson statistics for the number of sources (i.e.  $+N_{obs}=1+ \sqrt{N_{obs}+0.75}$, $-N_{obs}=\sqrt{N_{obs}-0.25}$).\\
\indent The excess densities for the whole {\it Chandra} Bullet field and for the three regions A, B, and C are reported in Tab. \ref{tabexc}.\\
\begin{table*}
  \caption{Excess density}
  \begin{center}
\begin{small}
    \begin{tabular}{lcccc|ccc|cc|cc}
      \hline
&& Total && &&Region A && Region B && Region C \\
\hline
Band & Flux$^a$ & Excess~den$^b$ & $\sigma^c$ & Excess~den$\_N^d$ &  Flux$^a$ & Excess~den$^b$ & $\sigma^c$ &  Excess~den$^b$ & $\sigma^c$ &  Excess~den$^b$ & $\sigma^c$ \\
\hline
S &  $> 0.1$ &  1.50$\pm$0.08 & 6.2 &   105$^{+19}_{-18}$  &  $> 0.1$ & 2.10$\pm$0.05 &24 & 1.61$\pm$0.05 &13  &  1.24$\pm$0.04 &5.6 \\
H &  $> 0.8$ &   1.39$\pm$0.09 & 4.3 &   71$^{+17}_{-16}$  & $> 0.8$  &1.50$\pm$0.05  &9.6 & 1.36$\pm$0.05 & 7.0&   1.34$\pm$0.05 &7.3 \\   
HH & $> 0.9$ &    1.47$\pm$0.12 & 3.9 &  47$^{+13}_{-12}$  &  $> 0.9$ & 1.32$\pm$0.07  & 4.9& 1.37$\pm$0.07 &5.6   &  1.80$\pm$0.07 &12 \\
\hline
S & $ 0.1-0.8 $  &   1.91$^{+0.16}_{-0.15}$ & 6.1  &  75$^{+14 }_{-13}$   & 0.1-0.3 & 2.5$^{+0.5}_{-0.4}$ & 3.7&  2.2$^{+0.5}_{-0.4}$ & 2.9 & &\\                   
S & $ 0.8-1.6 $   &   1.47$^{+0.21}_{-0.18}$ &  2.4 &  20$^{+9}_{-8}$     & 0.3-12 & 2.0$^{+0.3}_{-0.2}$ & 3.9&  1.5$^{+0.2}_{-0.2}$ & 2.8& 1.2$^{+0.1}_{-0.1}$ &1.4  \\
S & $ 1.6-9.6 $    &   1.11$^{+0.15}_{-0.13}$ &  0.8 &  7$^{+10}_{-8}$    & $>$12	&	   &&	0.9$^{+1.2}_{-0.6}$ & 0.1 & 1.5$^{+0.5}_{-0.4}$ &1.4 \\ 
S & $ 9.6-56 $   &   1.02$^{+0.29}_{-0.23}$ &  0.1 &  0$^{+5}_{-4}$  & & &  & & &  &\\
S & $ 56-189 $ &   2.38$^{+2.33}_{-1.31}$ &  1.1 &  2$^{+3}_{-2}$  & & &  & & &  &\\
\hline
H & $ 0.8-1.6 $      &    1.64$^{+0.31}_{-0.27}$ & 2.4  & 15$^{+7}_{-6}$   & 0.8-1.7 & 1.9$^{+0.4}_{-0.3}$ & 2.7 & 1.5$^{+0.5}_{-0.4}$ & 1.2 & & \\ 
H & $ 1.6-9.6 $      &    1.30$^{+0.11}_{-0.11}$ & 2.7  & 35$^{+13}_{-12}$   & 1.7-29.8 &1.3$^{+0.2}_{-0.2}$ &1.5& 1.3$^{+0.2}_{-0.2}$ &1.8 & 1.1$^{+0.1}_{-0.1}$ & 0.8\\
H & $ 9.6-56 $     &    1.40$^{+0.21}_{-0.18}$ & 2.2  & 16$^{+9}_{-8}$   & $>$29.8		 & &   & 2.4$^{+1.9}_{-1.2}$  & 1.2 & 3.3$^{+0.9}_{-0.7}$ & 3.0\\
H & $ 56-259 $   &    2.49$^{+1.23}_{-0.87}$ & 1.7  &  5$^{+4}_{-3}$   & & &  & & & & \\  
\hline
HH &  $ 0.9-1.6 $  &      1.60$^{+0.45}_{-0.36}$ & 1.7   & 8$^{+6}_{-4}$     & 0.9-2.2 & 1.5$^{+0.4}_{-0.3}$ & 1.9& 1.8$^{+0.6}_{-0.5}$ & 1.7 & &\\                  
HH &  $ 1.6-9.6 $  &      1.29$^{+0.14}_{-0.13}$ & 2.2   & 22$^{+11}_{-10}$  & 2.2-16.6 &1.1$^{+0.3}_{-0.2}$ & 0.5 & 1.2$^{+0.3}_{-0.2}$ & 1.0 & 1.7$^{+0.3}_{-0.3}$ & 2.7 \\
HH &  $ 9.6-56 $ &      2.26$^{+0.51}_{-0.42}$ & 3     & 16$^{+6}_{-5}$      & $>$16.6 &	&	  & 3.2$^{+4.2}_{-2.1}$ & 1.0 & 2.9$^{+1.5}_{-1.1}$ & 1.8\\  
HH &  $ 56-189 $ &    3.46$^{+3.39}_{-1.91}$ & 1.3   & 2$^{+3}_{-2}$   & & &  & & & &\\
\hline
    \end{tabular}
    \end{small}
\end{center}
$^a$Flux interval where the excess density is evaluated. The flux is   in unit of $1\times10^{-15}$ erg cm$^{-2}$ s$^{-1}$ ; $^b$Excess
  density with 1$\sigma$ uncertainties; $^c \sigma$ confidence level of the excess, which is evaluated as $\sigma = {Excess~density -1 \over \sigma Excess~density}$; $^d$ excess density in source
  number (see also Fig. \ref{lnlsrad}). Note that the excess densities could include some contribution from cosmic variance ($<1.25, 1.20$ and 1.10 in the A, B and C regions, respectively, see Sect. 4.3).
\label{tabexc}
\end{table*} 
\indent We estimate excess densities with respect to field number counts of a factor 1.5--2 at $\sim6 \sigma$ in the S band and $\sim 4 \sigma$ in the H and HH bands (see Tab. \ref{tabexc}).  The analysis of the excess density in flux bins highlights a different behaviour of the soft and hard sources. 
At the faintest fluxes, we found significant excess density in all the three energy bands. At medium and high fluxes, the excess density is significant in the H and HH band only.\\
\indent Fig. \ref{lnlsrad} (bottom panels) shows the excess densities with respect to {\it field} number counts in three energy bands as a function of R$_{200}$. The excess density of the soft sources in the full flux range (stars) decreases with the distance from the cluster centre, while the excess density of the hard sources stays about constant and the excess density of the very hard sources increases with the distance from the cluster centre.\\
\indent The excess densities are evaluated also in three flux ranges, faint, medium and bright, identified in Tab. \ref{tabexc} and by the black dotted lines in the upper panels of Fig. \ref{lnlsrad}.  Faint sources are detected only up to a distance of $\sim$R$_{200}$ (regions A and B), and show an excess density up to a factor $\sim$2.  Medium and bright sources are detected in all three regions. The excess density of the soft/medium flux sources decreases with the distance from the cluster centre. The excess density of the hard/medium flux sources shows a similar behaviour, but it becomes consistent with no excess at distances $>R_{200}$.  On the contrary, the excess densities of the hard/bright sources and the medium and bright/very hard sources is up to a factor of $\sim$3--4 at distances from the centre of the cluster $>R_{200}$.  In general, bright X-ray point sources are missing near the cluster centre and show an excess density at distance $>$R$_{200}$, in particular in the H and HH energy bands. The X--ray catalogue contains 23 sources brighter than $3\times10^{-14}$ erg cm$^{-2}$ s$^{-1}$ in the 2-10 keV band, which are located at distance from the main cluster greater than $\sim$6.1\amin, and 19 out of 23 are located at distance greater than $\sim$8.5\amin (i.e. greater than $R_{200}$).

\subsection{Cosmic variance?}

Systematic uncertainty due to cosmic variance can be significant, especially for surveys covering small areas.  For example, Cappelluti et al. (2005) found a cosmic variance of X-ray sources of the order of 15-25\% on scales similar to our region A. To better evaluate how the X--ray cosmic variance ($\sigma_x$) can affect our analysis, we estimated the excess density in the Chandra--COSMOS and CDF-S fields in regions of sizes similar to those of the regions A, B and C defined in the Bullet cluster field, and in the three flux bins adopted in Tab. \ref{tabexc}. For each region we evaluated the excess density of X-ray sources using exactly the same procedure followed for the Bullet cluster field. We repeated the exercise at ten not-overlapping, different positions in the Chandra-COSMOS and CDF-S fields. We found that the excess density in these regions is always $<1.25$ for regions of the size of region A, $<1.20$ for regions of the size of region B and $<1.10$ for regions of the size of region C. We conclude that cosmic variance is not likely the cause of the large excess densities found in the Bullet cluster field.

\subsection{X-ray properties}

We calculated the hardness ratio (H-S)/(H+S) of the hard X-ray sources in the regions A, B and C as a function of the flux. The hardness ratio distribution of the sources in the whole Bullet cluster field is consistent with that of the  C--COSMOS sources (using the Kolmogorov--Smirnov test).

We investigated in more detail the hardness ratio distribution in the A, B and C regions, by selecting only X--ray sources with 2--10 keV flux $>2\times10^{-15}$ erg cm$^{-2}$ s$^{-1}$, for which the average statistical error on (H-S)/(H+S) is $0.2\pm 0.1$.  In each region we evaluated the fraction of the hard sources N$_{hard}$, defined as the fraction of source with $(H-S)/(H+S)>0.5$ to the total. This hardness ratio corresponds to a column density $N_H>5\times10^{21}$ cm$^{-2}$, using an absorbed power-law model with photon index 1.9. We found that the fraction of hard sources N$_{hard}$ is different in the three regions, being $73\pm2\%,~60\pm2\%$ and $68\pm$2\% in the A, B and C regions, respectively, suggesting less obscured X-ray sources in region B than in region A and C.

\begin{figure*}
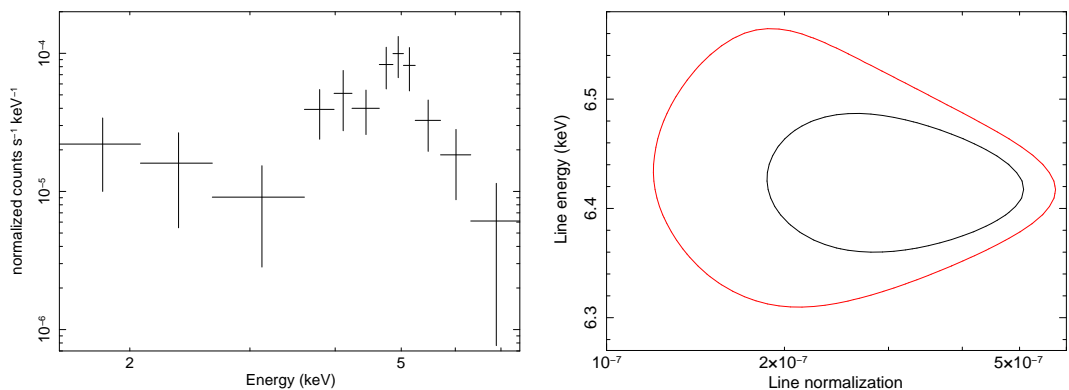

\centering 
\includegraphics[angle=270,width=7.1cm]{spe22.ps}
\includegraphics[angle=270,width=7.1cm]{fe22.ps}
\caption{{\it Left panel}: the {\it Chandra} spectrum of source \#13 (see Tab. \ref{souzbullet}) grouped to have at least 10 counts per energy bin. {\it Right Panel}: 68\% and 90\% confidence contours of the Fe$K_{\alpha}$ line energy at redshift $\sim$0.3 as a function of the line normalisation.}
\label{spettri2b}
\end{figure*}

We also analysed the {\it Chandra} spectra of bright individual sources. We found that a source with extreme hardness ratios, (H-S)/(H+S)$=$0.93, (HH-H)/(HH+H)$=$0.15, has a spectrum dominated by a strong emission line, corresponding to the Fe$K_{\alpha}$ transition at a redshift consistent with that of the Bullet cluster. The {\it Chandra} spectrum is well fitted with a pure reflection component (\textsc{xspec} \textsc{pexrav} model (Magdziarz \& Zdziarski 1995) plus an emission line (\textsc{xspec} \textsc{zgauss} model) with equivalent width $\sim$2~keV (see Fig. \ref{spettri2b} and Tab. \ref{tabfit}), making the source a good candidate Compton Thick (CT) AGN (e.g., Puccetti et al. 2014). The source lies in region A, at distance $\sim 5 \amin$ from the cluster centre.

\begin{table}
\caption{Best--fitting parameters of the candidate CT source}
\begin{tabular}{lcc}
\hline
Parameter& Value & Units  \\ [1ex]
\hline
Redshift & 0.31$_{-0.02}^{+0.01}$ \\
Fe$K_{\alpha}$ EW$^a$ & 2.3$_{-1.4}^{+2.7}$ & keV \\
$\chi^2$/d.o.f.  &  0.26/2  &  \\ [1ex]                               
$\chi^{2}_{\nu}$/$\chi^2$  prob. & 0.13/88\%\\ [1ex]    
2-10 keV Luminosity$^b$ & 1.8 & 10$^{42}$ erg s$^{-1}$\\ [1ex]
\hline
\end{tabular}

Best--fit values with uncertainties at the 90\% confidence level for one parameter of interest ($\Delta\chi^2=$2.706), obtained by fitting the {\it Chandra} spectrum grouped to have at least 30 total counts per bin.  $^a$ line equivalent width; $^b$ observed luminosity.

\label{tabfit}
\end{table}

\section{Multi--wavelength properties}

\subsection{Optical--Infrared counterparts of the X--ray sources}

We identified the counterparts of the 381 X--ray sources in the optical and infrared bands, using the R$_{c}$, B band and near infrared IRAC catalogues presented in Sect. 2.2 and 2.3. We searched for counterparts by matching the X--ray and optical/infrared positions within a maximum radius of 3$\asec$, which is well suited for Chandra mosaics as demonstrated for the Chandra--COSMOS survey (Civano et al. 2012).

\begin{figure}
\includegraphics[angle=0,height=8truecm]{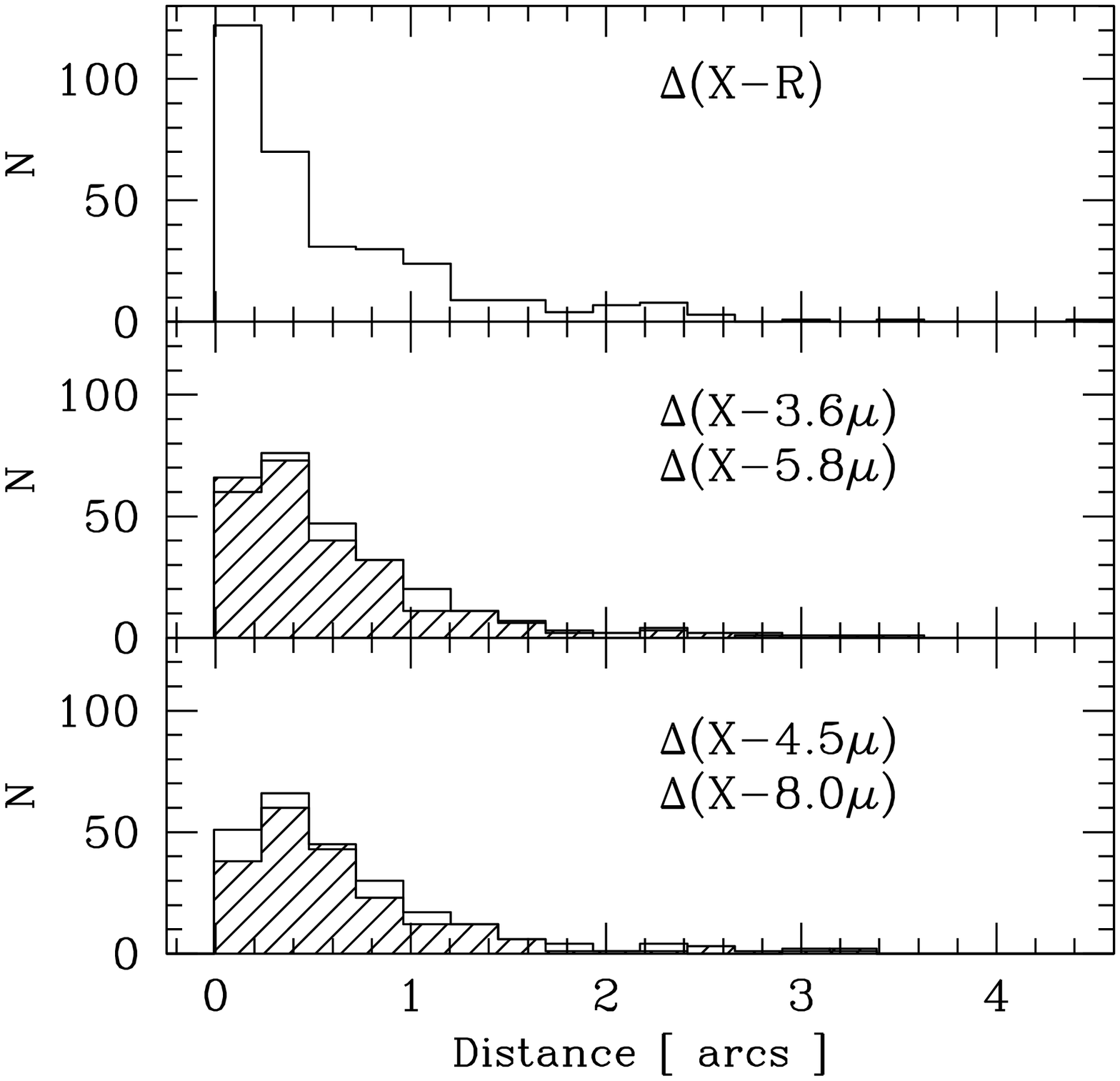}
\caption{{\it Top panel}: Histogram of the distances between the X--ray sources and their optical counterparts. {\it Middle panel}: Histogram of the distances between the X--ray sources and their counterparts in IRAC ch1 (empty histogram), and IRAC ch2 (dashed histogram). {\it Bottom panel}: Histogram of the distances between the X--ray sources and their counterparts in IRAC ch3 (empty histogram), and IRAC ch4 (dashed histogram).
\label{istoxo}}
\end{figure}

We matched first the X--ray and the optical catalogues. We excluded 24 sources after visual inspection, mostly lying in background/saturated regions in the optical images. We identified the R$_{c}$--band counterparts for a total of 320 X--ray sources (see Fig. \ref{istoxo}). 258/320 (302/320) of the identified sources lie within 1$\asec$ (2$\asec$) from the X-ray centroids. Two X-ray sources are associated to optical galaxies even if they are off--nuclear. In additional nine cases the association between the X-ray and the optical source is considered tentative, because the relative distance is greater than 2$\asec$. Fig. \ref{istoxo} shows the distribution of the distance between the X-ray sources and their optical counterparts. This distribution is consistent with the analogous one produced for the Chandra--COSMOS multi--wavelength catalogue (Civano et al. 2012). For the X--sources without an optical counterpart, we calculated the magnitude upper limits as described in Sect. 2.2.

We cross--correlated the catalogue of the optical counterparts, or the catalogue of the X--ray sources when an optical counterparts is not available, with the IRAC catalogues presented in Sect. 2.3. We identified 264, 251, 237 and 206, 3.6 $\mu m$, 4.5 $\mu m$, 5.8 $\mu m$ and 8.0 $\mu m$ counterparts, respectively. 181 X--ray sources have counterparts in all four IRAC channel. Fig. \ref{istoxo} shows the distribution of the distance between the X-ray sources and their near infrared counterparts.

\subsection{Infrared properties}

\begin{figure}
\begin{center}
\includegraphics[width=8cm]{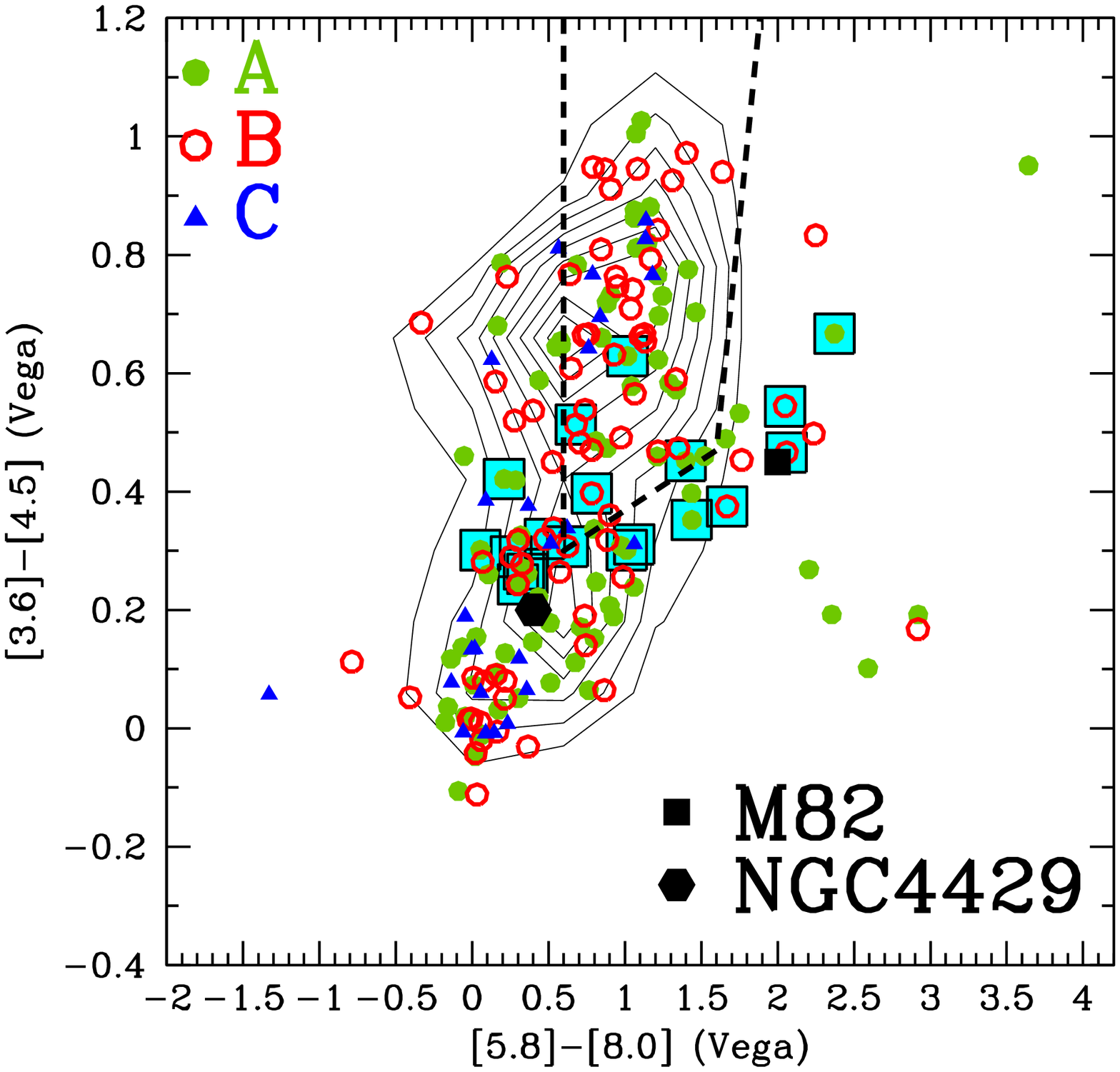}
\caption{IRAC colour--colour diagram. Green solid circles = sources in region A; red open dots = sources in region B; blue solid triangles = sources in region C. The cyan full squares mark sources at the Bullet's redshift (see Sect. 6). The black contour levels represent C--COSMOS sources. The black solid square and diamond represent two galaxy templates, \textsc{M82} and \textsc{NGC 4429}, respectively, which were moved at the redshift of the Bullet. The black dashed line marks the empirically determined region, largely populated by luminous unobscured AGN (Stern et al. 2005).
\label{irac}}
\end{center}
\end{figure}

Fig. \ref{irac} shows a IRAC colour--colour diagram for the sources in the regions A, B and C of the Bullet cluster field (note that the IRAC coverage is smaller than the {\it Chandra} field, see Fig. \ref{coverage}).  Using the selection criterion by Stern et al. (2005) to identify luminous unobscured AGN, the fraction of candidates unobscured AGN in the Bullet field is (40$\pm$2)\%, which is lower than that in the C--COSMOS field (i.e. $\sim 48\%$). We note however a concentration of Bullet cluster sources with IRAC colours close to zero, much higher than C--COSMOS sources (16$\pm$2\% of the Bullet cluster sources versus $\sim 5\%$ of the C--COSMOS sources). If we exclude these sources we obtain a fraction of luminous unobscured AGN in the Bullet cluster field consistent with the C--COSMOS field. 22 out of the 24 brightest hard source (2--10 keV flux $>$ $10^{-14}$ erg cm$^{-2}$ s$^{-1}$) with IRAC counterparts are selected as luminous unobscured AGN according to the Stern et al. (2005) criteria.

\begin{figure}
\begin{center}
\includegraphics[width=6.5cm]{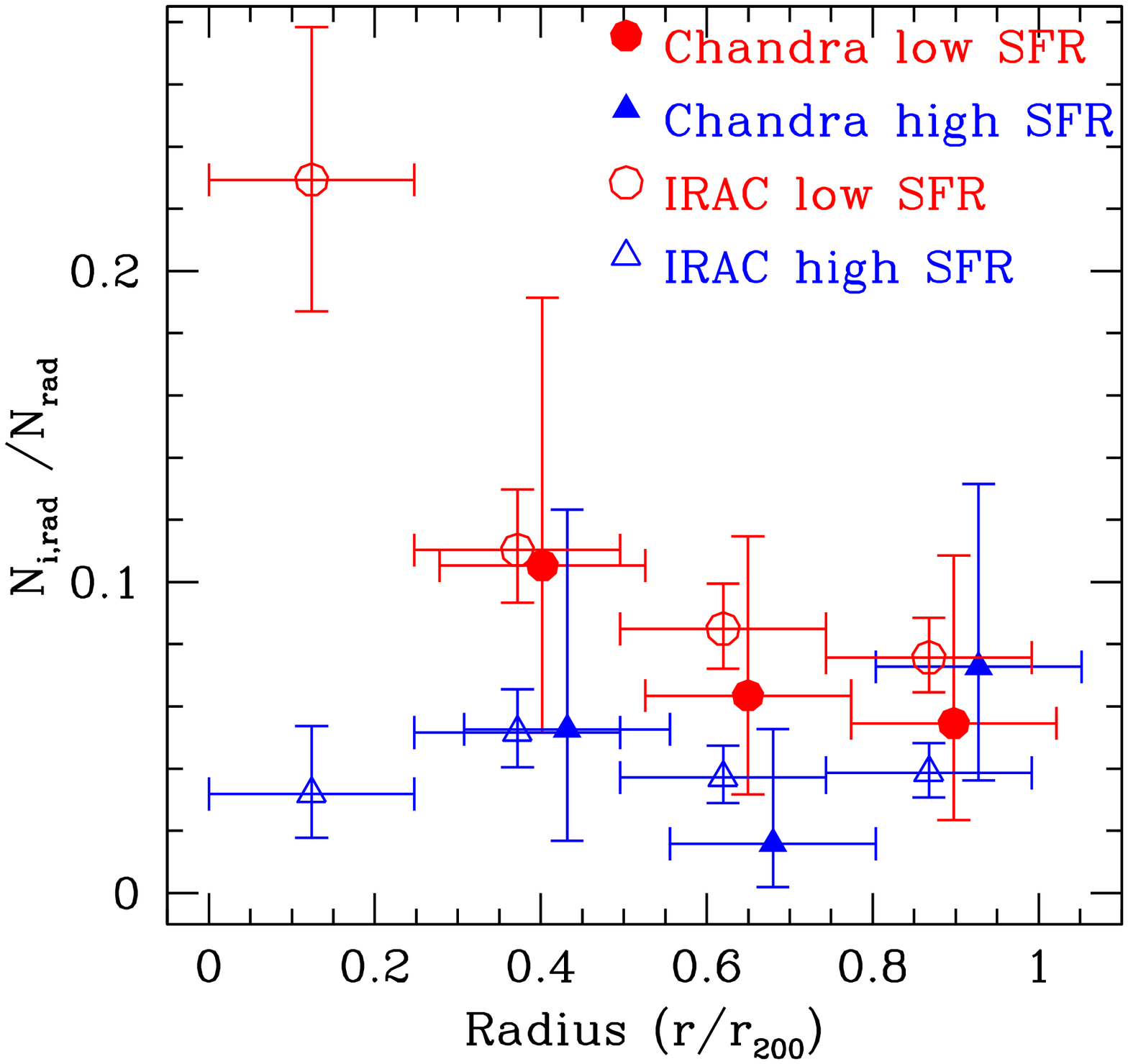}
\caption{Ratio between the number of IRAC and {\it Chandra} {\it LSFRG} and {\it HSFRG} to the total samples as a function of the distance from the cluster centre.  Red solid dots = {\it Chandra LSFRG}; blue solid triangles = {\it Chandra HSFR}; red open dots = IRAC {\it LSFRG}; blue open triangles = IRAC {\it HSFG}.
\label{iracfra}}
\end{center}
\end{figure}

We tried to quantify the fraction of high and low
star--formation galaxies in the Bullet cluster IRAC field using the colour--colour locus at the Bullet's redshift of two non-evolving galaxy templates from Devriendt et al. (1999):
\textsc{M82} a starburst galaxy with $[5.8]-[8.0]=2$ and $[3.6]-[4.5]=0.45$, and \textsc{NGC 4429} an S0/Sa galaxy with a star--formation about 4000 times lower (with $[5.8]-[8.0]=0.4$ and $[3.6]-[4.5]=0.2$). We classified the sources according to the following criteria:

\begin{itemize}
\item  {\it high SFR galaxies, HSFRG}: \begin{equation}
1.5 < [5.8]-[8.0]< 2.2~  \& ~ 0.35 < [3.6]-[4.5] < 0.55 
\end{equation}
 \item {\it low SFR galaxies, LSFRG}:
 \begin{equation}0.2 < [5.8]-[8.0]< 0.6~ \& ~ 0.1 < [3.6]-[4.5] < 0.3 
\end{equation}

\end{itemize}
We divided the area, covered by the four IRAC channels, in four radial bins, from zero to R$_{200}$. For each bin we evaluated the ratio between the number of {\it HSFRG} and {\it LSFRG} to the total (see Fig. \ref{iracfra}). We also calculated similar ratios for {\it Chandra} sources with IRAC counterparts.  We found that the fraction of IRAC {\it LSFRG} are dominant near the cluster centre, and decreases at increasing radii. Conversely, the fraction of {\it HSFRG} is constant with the radius, and remains at all radii smaller than the fraction of {\it LSFRG}.  We found a similar behaviour for {\it Chandra} sources with IRAC counterparts, although with poorer statistics.

\subsection{Fraction of X-ray active galaxies (cluster members + foreground \& background sources)}

\begin{figure*}
\centering
\includegraphics[angle=0,height=7.5truecm]{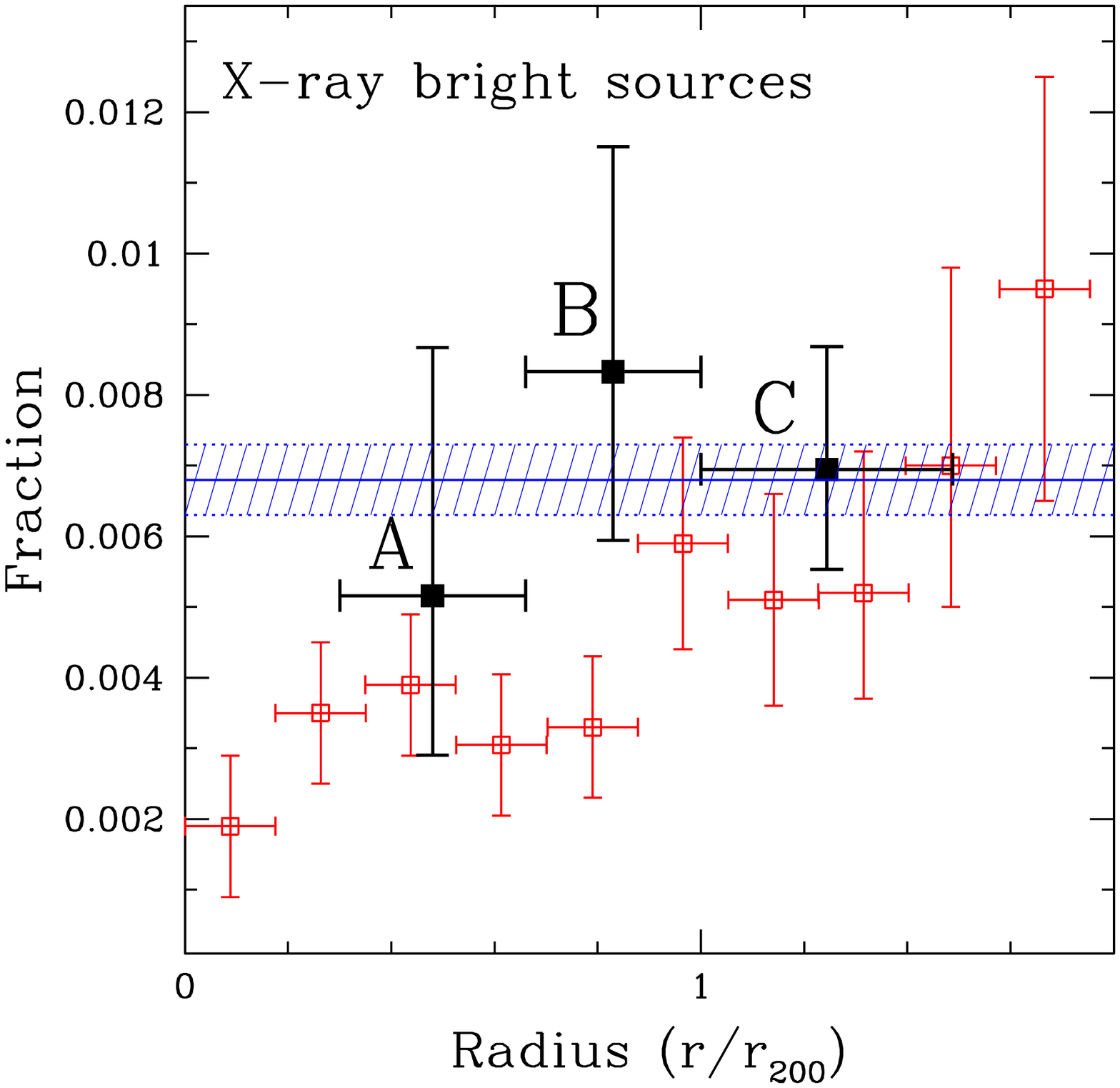} \includegraphics[angle=0,height=7.5truecm]{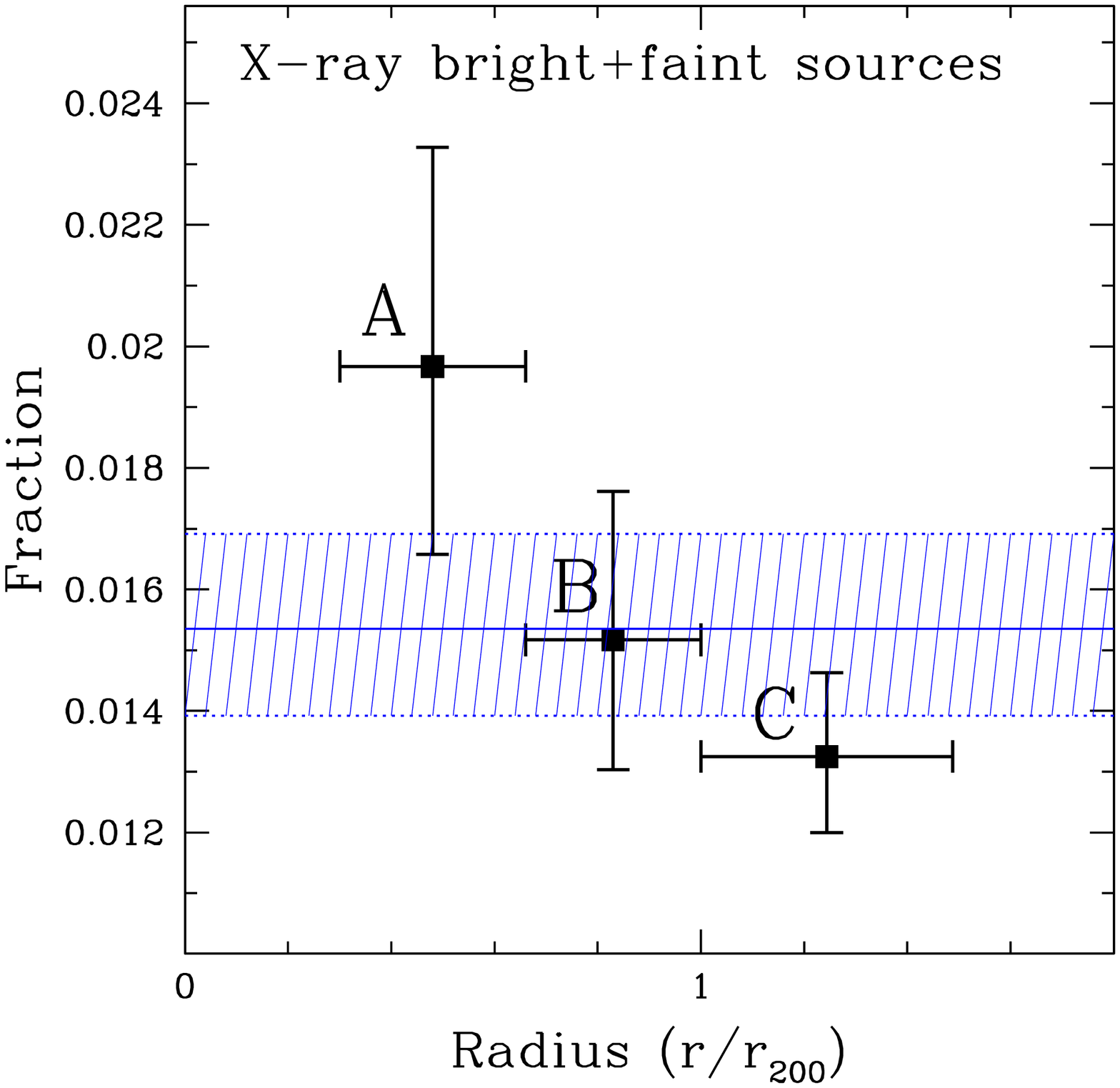}
\caption{The fraction of X-ray bright galaxies as a function of the distance from the cluster centre in units of R$_{200}$. {\it Left  panel}: black solid squares = Bullet cluster X-ray sources with 0.5--8 keV flux $> 1\times10^{-14}$ \cgs; red open squares = the sample of 42 cluster by Ehlert et al. (2014). The average field AGN fraction in the COSMOS--Legacy survey is marked by the blue solid line, with 1$\sigma$ uncertainties identified by the dashed region. {\it Right panel}: black squares = Bullet cluster sources with 0.5--2 keV keV flux $ \ge 3\times10^{-16}$ \cgs. The average field AGN fraction in the CDF-S 4 Msec survey is marked by the blue solid line, with 1$\sigma$ uncertainties identified by the dashed region.}
\label{frax}
\end{figure*}

Clusters are over--dense regions of the Universe, and therefore it is not surprising to find excess of X-ray sources with respect to the field.  A different question is how the fraction of active galaxies in the Bullet cluster field compares with the fraction of active galaxies in other cluster fields and in random fields, not targeting dense region of the Universe. To start investigating these issues we first evaluated the fraction of optical galaxies (cluster members + foreground \& background galaxies) hosting an X--ray source, and then compared it to the fractions of active galaxies in other cluster fields (Ehlert et al. (2014) and in the COSMOS--Legacy field (Civano et al. 2016). Following Ehlert et al. (2014) we:
\begin{enumerate}
\item 
selected X--ray sources with 0.5--8 keV flux larger than $1\times10^{-14}$ \cgs (X-ray bright galaxies);
\item
selected galaxies with aperture magnitude R in the range 18.8--23 mag (see Sections 2.2 and 5.1). We used R aperture magnitudes with 3$\asec$ diameter, coherently with the magnitudes in the Legacy--COSMOS catalogue and in Ehlert et al. (2013) work.
\item
excluded stars from the analysis using the SEXTRACTOR CLASS$\_$STAR parameter larger than 0.98 and B$-R$ colour larger than 2. For the COSMOS--Legacy field we marked and excluded stars using the flag in the Ilbert et al. (2009) and Marchesi et al. (2016) catalogues.
\end{enumerate}
The fraction of galaxies hosting an X--ray source is then given by

\begin{equation}
Fraction = {N_{XR} \over N_{R}}
\end{equation}

\noindent
where N$_{XR}$ is the number of galaxies hosting an X--ray source and N$_{R}$ is the total number of R band selected galaxies. The fraction of X-ray bright galaxies is plotted in Fig. \ref{frax} (left panel).  The upper ($+\Delta$ Fraction) and lower ($-\Delta$ Fraction) uncertainties are evaluated using the Gehrels (1986) formulae for Poisson statistics for N$_{XR}$ and N$_{R}$.

The average fraction of X-ray bright galaxies in the Bullet field is 0.0070$\pm^{+0.0013}_{-0.0011}$, consistent with the value of 0.0068$\pm$0.0005, evaluated for the COSMOS-Legacy field (see left panel of Fig. \ref{frax}). The fraction in the three spatial regions A, B and C is also consistent within the error with the COSMOS-Legacy value. Considering the punctual values, these fractions are somewhat higher than those of the Ehlert et al. (2014). However, the rather large uncertainties prevent us from drawing any strong conclusion.

We also evaluated the fraction of galaxies hosting X--ray sources at X--ray fluxes lower than $1\times10^{-14}$ \cgs. At these fluxes the survey incompleteness cannot be neglected, and therefore the number of X-ray sources has been corrected for the sky--coverage in Fig. \ref{skycov}. We then evaluated the fraction of galaxies hosting X--ray sources at S flux $\ge 3\times10^{-16}$ \cgs and R in the range 18.--24.5 mag, to avoid optical incompleteness. As a comparison sample, in this case we used the CDF-S 4 Msec survey (Xue et al. 2011, Lehmer et al. 2012), which is more suitable for a comparison with the Bullet cluster field at the above deep flux limit. The fraction of galaxies hosting X--ray sources in the three Bullet cluster regions B and C is consistent with the average field X--ray sources fraction in the CDF-S within the uncertainties (see right panel of Fig. \ref{frax}). In region A we found a fraction of X-ray galaxies (clusters members + foreground \& background galaxies) higher than in the CDF-S field, although the difference is only a $\sim$ 1$\sigma$ (see right panel of Fig. \ref{frax}).

\section{Analysis of the source populations based on their redshifts}

\subsection{VIMOS spectroscopic redshifts}

\begin{figure}
\includegraphics[width=9.7cm]{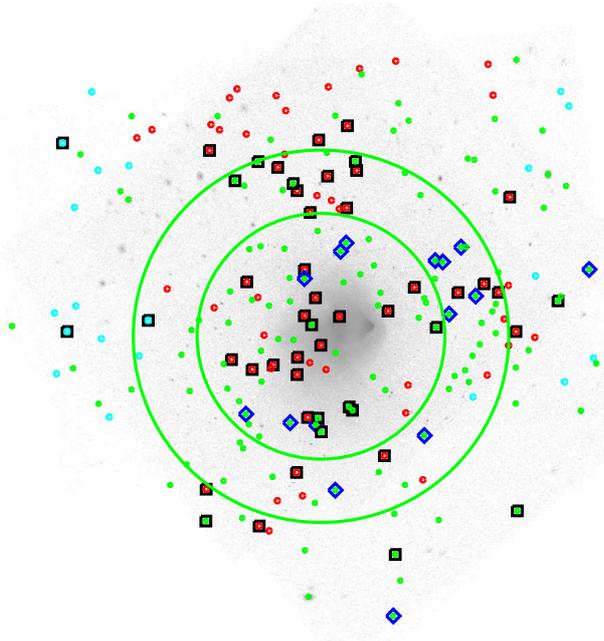}
\caption{Image mosaic of the ten {\it Chandra} observations of the Bullet field in the 0.5--7 keV energy range (see Fig. \ref{image}). The green circles are centred on the main cluster and have increasing radius of 0.66$\times R_{200}$ and $R_{200}$ (see Fig. \ref{image}). Points of different colours correspond to sources selected from different catalogues with a measured VIMOS redshift: green circles = Chandra sources; cyan circles = WFI sources; red circles = IRAC sources. Black squares identify the sources with redshifts z$\pm \Delta$z=(0.296$\pm$0.014), corresponding to that of the Bullet cluster. Blue diamonds identify {\it Chandra} sources with redshifts z$\pm \Delta$z=(0.296$\pm$0.014), but redshift taken from the literature.}
\label{redshift}
\end{figure}

Spectroscopic redshifts have been measured on VIMOS spectra of  194 sources, 106 of which are X-ray selected sources, 69 are near infrared selected sources and 19 are optically selected sources. Given the velocity dispersion of the main cluster components (i.e. 1400 km/s, Bradač et al. 2006), we estimated the 3$\sigma$ redshift extension of the cluster to be z$\pm \Delta$z=0.296$\pm$0.014.  We found that 49 out of the 194 VIMOS sources have redshift consistent with this range (see Fig. \ref{redshift}). Tab. \ref{souzbullet} includes the redshifts of these 49 cluster members derived from the VIMOS spectra, which are also shown in Fig. \ref{spettri} of the Appendix.  Tab. \ref{sounozbullet} of the Appendix, lists the redshifts of the remaining 145 field point--like sources in the field.

\subsection{Redshifts: total sample}

\begin{figure}
\includegraphics[width=9cm]{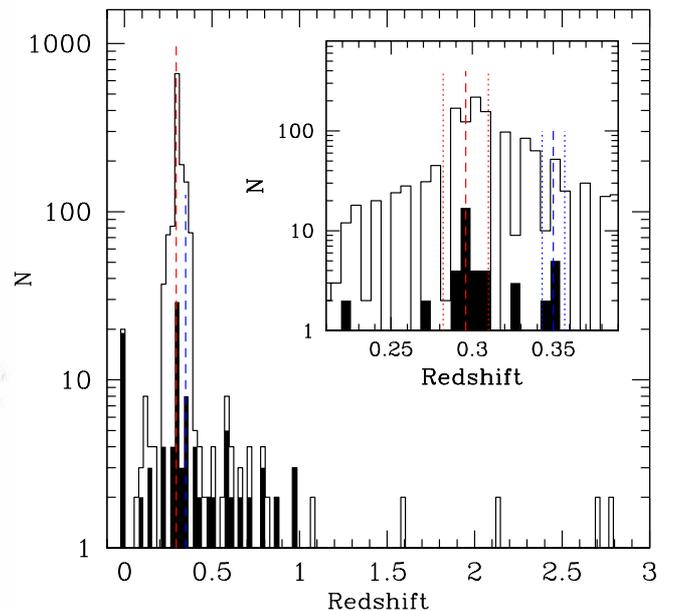}
\caption{Redshift distribution of the entire analysed sample (open histogram) and of the X-ray sources (filled histogram). The red dashed line marks the Bullet cluster's redshift, the blue dashed line marks the redshift of a background cluster (Rawle et al. 2010). {\it Inset panel}: zoom of the redshift range corresponding to the Bullet cluster, i.e. z$\pm \Delta$z=(0.296$\pm$0.014).}
\label{isredshift}
\end{figure}

We added to our VIMOS sample, 1201 redshifts from the literature, 231 spectroscopic and 970 photometric (Foex et al. 2017, Chung et al. 2010, Barrena et al. 2002, Guzzo et al. 2009, Rex et al. 2010, Rawle et al. 2012, Bradač et al. 2009, Jones et al. 2009, Zhang et al. 2016, Souchay et al 2015, Liang et al. 2000, Johansson et al. 2010, Sereno 2015, Stott et al. 2011). We thus consider a total of 1395 sources with spectroscopic or photometric redshift. 127 out of the 1395 sources are X-ray selected sources, for which we collected 119 spectroscopic redshifts and 8 photometric redshifts. Fig. \ref{isredshift} shows the redshift
distribution of the entire redshift sample (open histogram) and of the X-ray sources (filled one). The redshift distribution shows a statistically significant peak at the Bullet cluster's redshift. 29 X--ray sources have redshift consistent with the Bullet cluster one and 25 out of the 29 sources have spectroscopic redshift.

In the Bullet cluster field there is a background cluster discovered by {\it Herschel} at z$\pm\Delta$z=0.350$\pm$0.007 (Rawle et al. 2010). The full source sample redshift distribution has a statistically significant peak at z$\sim$0.35, while no statistically significant excess at z$\sim$0.35 is seen in the X--ray selected
sources.

\subsection{Bullet cluster galaxies with X-ray emission}

Our sample includes 29 X--ray sources at the Bullet cluster redshift. 14 redshifts have been determined from our VIMOS spectra, 15 are from the literature and in only 4 cases the redshift has been evaluated through photometric techniques (see Tab. \ref{souzbullet}). \\
\indent Optical spectra were used to provide a first rough classification, based on the presence of emission lines (emission line galaxies, ELGs), or absorption lines (absorption line galaxies, ALGs).

X-ray luminosities (2-10 keV) have been evaluated from H band count rates, assuming a power$-$law spectrum with $\Gamma=$1.9, reduced at low energies by photoelectric absorption. We considered both Galactic absorption along the line of sight and absorption at the redshift of the source. The luminosities include the k-correction. The rest frame column density $N_H$ is estimated using the hardness ratio (H-S)/(H+S) and the same model spectrum. This column density is likely a lower limit to the true one.  In fact, non uniform absorbers, or ionised absorbers of a given column density $N_H$ may allow significant leaking of low energy photons, producing hardness ratios much smaller than that corresponding to a uniform, cold gas cloud of the same column density $N_H$. Furthermore, low energy photons may also be produced by scattering or reflection, as often observed in Compton Thick sources, where a power-law component with normalisation 0.01-0.1 that of the direct power-law often emerges (Comastri et al. 2007, Ueda et al. 2007, Kawamuro et al. 2016; Marchesi et al. 2017, Tanimoto et al. 2018). These limitations of the hardness redshift analysis are clearly highlighted by the case of source \#13, the candidate CT AGN discussed in Sect. 4.4. Note that the optical spectroscopy redshift is fully consistent with that evaluated from the X-ray spectrum.  The iron line equivalent width suggests a column density $>10^{24}$ cm$^{-2}$, that is, about 10 times higher than the limit estimated from the hardness ratio analysis. In summary, the luminosities reported in Tab. \ref{souzbullet} may under-estimate the intrinsic luminosities, if the real spectra of the sources are more complex than our simplistic assumption.

\begin{figure*}
\centering
\includegraphics[width=6.5cm]{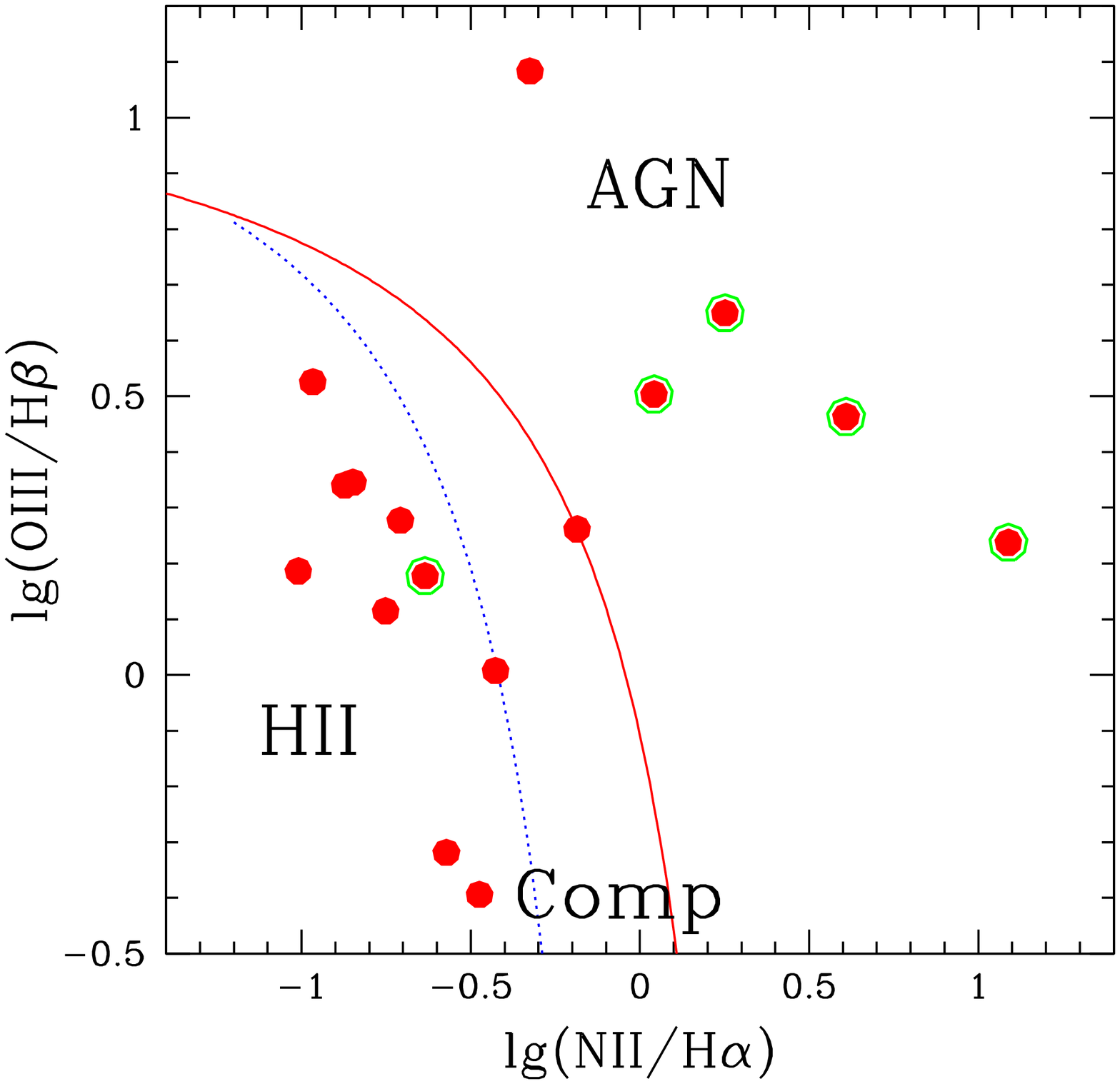} 
\includegraphics[width=6.5cm]{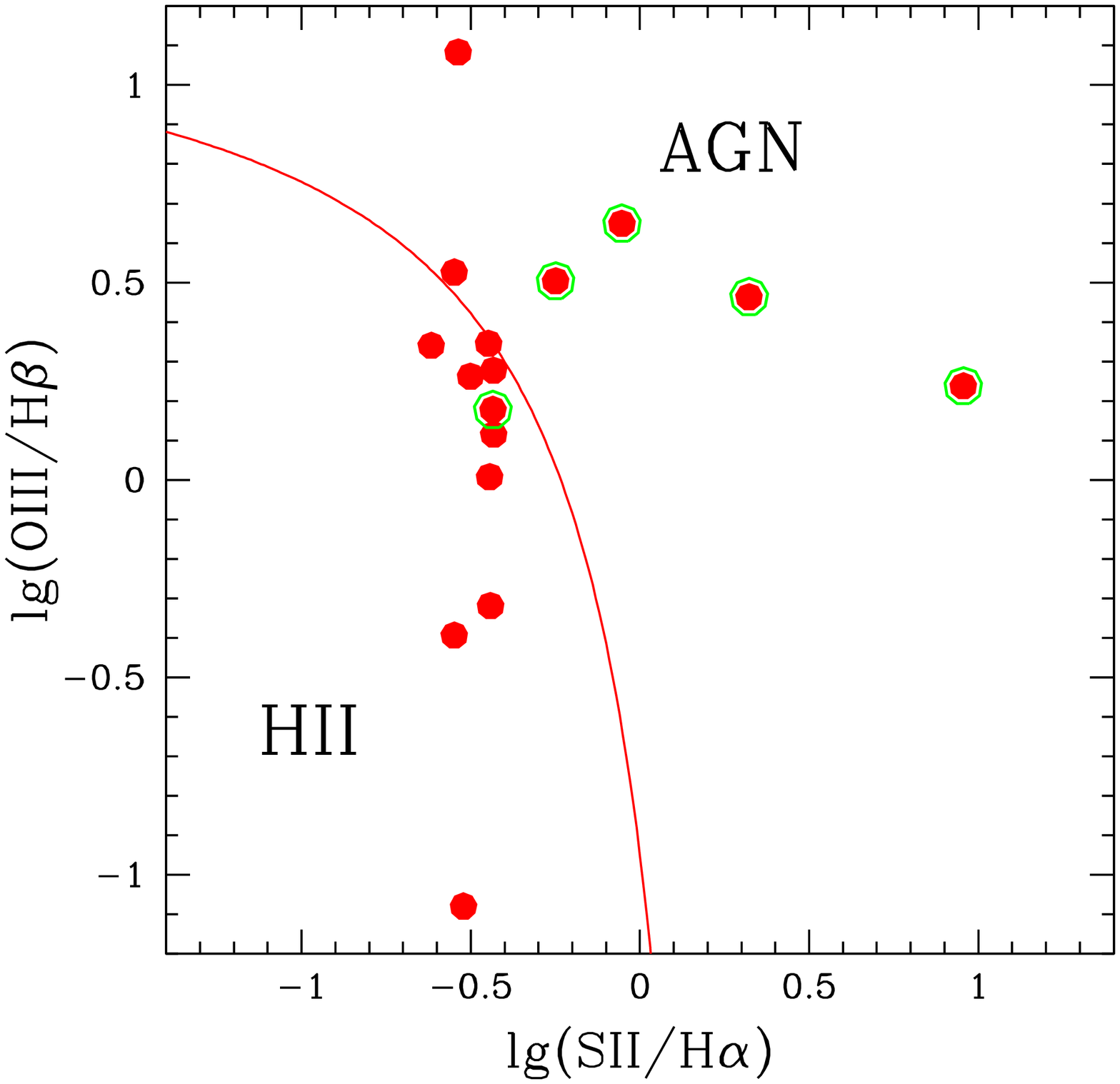}
\caption{{\it Left panel}: [OIII]/H$_\beta$ vs. [NII]/H$_\alpha$ diagnostic diagram for the Bullet's sources. The Kewley et al. (2001) extreme starburst line and the Kauffmann et al. (2003) classification line are shown as red solid and blue dashed lines, respectively. The green open dots mark the X--ray sources at the Bullet redshift classified as AGN according to their 2--10 keV luminosity. {\it Right panel}: [OIII]/H$_\beta$ vs. [SII]/H$_\alpha$ diagnostic diagram. Symbols are as in {\it left panel}.}
\label{BPT}
\end{figure*}

Keeping in mind the caveats briefly mentioned above, we assess the AGN nature of a source based on its X-ray luminosity and hardness ratio. We classify as {\it bona fide} AGN the seven objects with L$_{2-10~keV}\gtrsim 10^{42}$ erg s$^{-1}$, because the X-ray luminosity produced by star-formation, estimated from the narrow H${\alpha}$ luminosity (Kennicutt et al. 1994), and using the Lehmer et al. (2016) or Ranalli et al. (2003) SFR -- X-ray luminosity correlations to convert the SFR into an X-ray luminosity, is always much smaller than the above threshold (note that six of the seven sources have optical spectra classified as ELG, while for the seventh we do not have information on the optical spectrum).  Emission line optical spectra were classified also using diagnostic diagrams (BPT: Baldwin, Phillips \& Terlevich, 1981, Veilleux \& Osterbrock, 1987). We found that four out of the five X--ray sources classified as AGN using their X-ray emission are recognised as AGN also using diagnostic diagrams
(see Fig. \ref{BPT}).

We classify as {\it candidate} AGN the six objects with 41.44$\le$log(L$_{2-10~keV}$)$\le$41.84 and (S-H)/(S+H)$\ge$0.5 because the high hardness ratio suggests significant obscuration, and, as explained above, the $N_H$ evaluated from a single hardness ratio may under-estimate the real one.  Also in these six cases the X-ray luminosity produced by star-formation is smaller than the observed X-ray luminosity (note that two of the six sources have optical spectra classified as ALG and for the remaining four we do not have information on the optical spectrum).

Finally, we classify as normal galaxies the remaining 16 objects with 40.95$\le$log(L$_{2-10~keV}$)$\le$41.64. Out of these 16 sources, 14 are not detected in the H band and therefore, only have an upper limits on their hardness ratio. Stacking together the counts of the 14 sources produces a significant detection in the H band (more than 3$\sigma$), and the resulting stacked hardness ratio is $\sim$-0.12, suggesting no or very little absorption and a steep spectrum. In these sources an active nucleus is either not present, or accreting at very low regimes (e.g., Low Luminosity AGN, LLAGN). The remaining two low luminosity sources detected in the H band have hardness ratios = 0.25--0.37, consistent with that of a an unobscured AGN spectrum. Nevertheless, we conservatively classify these two sources as normal galaxies.

The integrated X--ray emission from normal galaxies (not hosting an active nucleus) is due to three contributions: 1) high-mass X-ray binaries (HMXBs), 2) low-mass X-ray binaries (LMXBs), and 3) thermal emission from a hot gas. We quantified the possible contribution to the X--ray luminosity of the 29 {\it Chandra} sources from these three components.  Mineo et al. (2012) show that HMXBs are a good tracer of recent star formation, and their 0.5-8 keV luminosity scale linearly with the SFR. We then evaluated the HMXB X-ray luminosities from the $\sim$24$\mu m$ SFR (calculated from Spitzer MIPS and WISE archived data) and found that HMXBs can contribute $<1\%$, $<5\%$, $<25\%$ of the observed X-ray luminosity for the seven {\it bona fide} AGN, the six {\it candidate} AGN, and the 16 normal galaxies, respectively. We evaluated the possible contribution from LMXBs or the hot gaseous coronae using the predicted relations between X-ray and K luminosities from Kim \& Fabbiano (2004) and Sun et al. (2007), respectively, using  2MASS archive data. Also in this case the contribution of LMXBs and hot halos to the seven {\it bona fide} AGN and the six {\it candidate} resulted negligible ($<0.5-4\%$ and  $<5-10\%$ respectively), confirming that the X-ray emission from this sources likely originate from accreting SMBHs. In the normal galaxies the contribution of LMXBs and hot aloes to the X-ray emission could reach $\sim30-35\%$.

Tab. \ref{souzbullet} also reports the VIMOS results for 32 IRAC and 3 WFI selected sources at the Bullet cluster redshift.

In summary, we identify 29 {\it Chandra} sources at the redshift of the Bullet cluster: 7 {\it bona fide} AGN, 6 {\it candidate AGN} and 16 normal galaxies/LLAGN (6 ELGs, 5 ALGs and 5 with unknown optical classification).

\begin{landscape}
  \begin{table}
      \centering
  \caption{Cluster members properties.}
    \begin{small}
\begin{tabular}{lccccccccccccccc}
      \hline
      Id$^{a}$ & R.A. & Dec. & z  &     fl$^{b}$ & Instr.$^{c}$  &   Type$^{d}$ & Type$_X$$^e$ & Type$_{BPT}$$^f$& Reg.$^g$ & Dis.$^h$ & Log(Lx$_H$)$^i$ & N$_H$$^l$ & X/O$^m$ & SFR$^n$  & SFR$_{H{\alpha}}$$^o$\\
 \hline
1$^{p}$ &   104.30458  & -55.90027 &  0.31    &  & Chandra  & unc &  Galaxy                          &                            & C  &   11.88  &     41.32  &    0      & $<$0.02  & 35.2  &   \\
2  &  104.34416 &  -55.92304 &  0.297   &  3  &  Chandra &  ELG  &  {\it Bona fide} AGN              &  {\it Candidate} AGN       &  C    &  10.31   &  42.05  &  $>$3 &  0.48    &  183 & 3\\                                  
3 &  104.39524 &  -56.07483 &  0.303   &  2  &  Chandra &  ELG &  Galaxy                             &  Galaxy                    &  C    &  11.32   &  41.61  &  0          &  $<$5.3  &  69.2 & 10 \\                                  
4 &  104.39787 &  -55.94532 &  0.298   &  2  &  IRAC    &  ALG  &                                    &                            &  C    &  8.382   &       &           &        \\                                          
5 &  104.40642 &  -55.8482  &  0.2957  &  2  &  IRAC    &  ALG  &                                    &                            &  C    &  10.12   &       &           &        \\                                          
6 &  104.42037 &  -55.91705 &  0.291   &  2  &  IRAC    &  ALG  &                                    &                            &  B    &  7.865   &       &           &        \\                                          
7 &  104.43885 &  -55.91093 &  0.293   &  2  &  IRAC    &  ELG &                                     &  {\it Bona fide} AGN       &  B    &  7.37    &       &           &      & & 1  \\                                          
8$^{p}$  &  104.44949 &  -55.91978 &  0.296   &     &  Chandra &  ELG  &  Galaxy                     &                             &  B    &  6.875   &  41.48  &  $<$2.5  &  $<$0.04  &  51.7\\                                  
9$^{p}$ &   104.46833  & -55.88452 &  0.296   &  & Chandra  & unc & Galaxy                           &                            & B  &   7.147  &     41.45  &   $<$2.0  & $<$0.02  & 48.2  &   \\
10  &  104.47203 &  -55.91725 &  0.2952  &  2  &  IRAC    &  ALG  &                                  &                            &  B    &  6.19    &       &           &        \\                                          
11$^{p}$  &  104.48333 &  -55.93286 &  0.2941 &     &  Chandra &  unc                                &                             Galaxy  & &  B    &  5.592   &  40.95  &  $<$0.01  &  $<$0.08 & 15.3 \\                                  
12$^{p}$ &   104.49166  & -55.89522 &  0.2951  &  & Chandra  & unc &  {\it Candidate} AGN            &                             & B  &   6.140  &     41.45  &   $>$4.0  & 0.13  & 48.3  &   \\
13 &  104.49973 &  -55.94242 &  0.289   &  3  &  Chandra &  ELG &  {\it Bona fide} AGN               &  {\it Bona fide} AGN        &  A    &  4.97    &  42.25(43.41$^{q}$)   &  6.0(100$^{q}$)   &  0.14   &  306 & 1 \\
14$^{p}$ &  104.50123 &  -55.89424 &  0.2879  &     &  Chandra &  unc            & {\it Candidate} AGN &                         &  B    &  5.907   &  41.72  &  2.0     &  0.08  &    90.5   \\                                  
15$^{p}$ &  104.51473 &  -56.02061 &  0.296   &     &  Chandra &  ELG  &  {\it Bona fide} AGN        &                              &  B    &  6.181   &  43.3  &  3.0     &  1.7   &   3185   \\                                  
16 &  104.52784 &  -55.91356 &  0.297   &  4  &  IRAC    &  ELG &                                    &  Galaxy                     &  A    &  4.54    &       &           &     & & 9   \\                                          
17 &  104.55173 &  -56.1065  &  0.303   &  3  &  Chandra &  ALG  &  Galaxy                           &                            &  C    &  9.976   &  41.64  &  $<$0.15  &  $<$0.03   &   74.2\\                                  
18$^{p}$ &   104.55416  & -56.15088 &  0.31    &  & Chandra  & unc &   {\it Bona fide} AGN           &                            & C  &   12.51  &     42.4   &   $>$1.0  & 0.43  & 424  &   \\
19 &  104.56145 &  -55.93076 &  0.298   &  2  &  IRAC    &  ALG  &                                   &                             &  A    &  3.081   &       &           &        \\                                          
20 &  104.56586 &  -56.03525 &  0.299   &  3  &  IRAC    &  ELG &                                    &  {\it Candidate} AGN       &  B    &  5.85    &       &           &       & & 7 \\                                          
21 &  104.60170 &  -55.82924 &  0.289   &  3  &  IRAC    &  ELG &                                    &  Galaxy                   &  B    &  7.345   &       &           &       & &9 \\                                          
22 &  104.60330 &  -55.82266 &  0.288   &  2  &  Chandra &  ALG  &  Galaxy                           &                                 &  B    &  7.721   &  41.0   &  $<$0.07  &  $<$0.01  &     17.3 \\                                  
23 &  104.60683 &  -56.0024  &  0.292   &  2  &  Chandra &  ALG  &  Galaxy                           &                                 &  A    &  3.479   &  41.52  &  $<$1.5  &  $<$0.02   &  56.9\\                                  
24 &  104.61115 &  -56.00005 &  0.298   &  3  &  Chandra &  ALG  &  Galaxy                           &                                 &  A    &  3.294   &  41.53  &  0          &  $<$0.02 &  57.9\\                                  
25 &  104.61352 &  -55.79699 &  0.287   &  3  &  IRAC    &  ELG &                                    &  {\it Candidate} AGN         &  C    &  9.189   &       &           &   & & 4     \\                                          
26 &  104.61445 &  -55.85632 &  0.295   &  2  &  IRAC    &  ALG  &                                   &                               &  B    &  5.667   &       &           &        \\                                          
27$^{p}$ &   104.61500  & -55.88166 &  0.2962  &  & Chandra  & unc &   Galaxy                        &                                    & A  &   4.180  &     41.21  &   $<$1.0  & $<$0.07  & 27.8  &   \\
28$^{p}$ &  104.62214 &  -55.88763 &  0.296   &     &  Chandra &  ELG  &  Galaxy                     &                                       &  A    &  3.775   &  41.12  &  $<$0.1  &  $<$0.02  &  22.6 \\                                  
29 &  104.62348 &  -55.93472 &  0.29    &  2  &  IRAC    &  ALG  &                                   &                               &  *    &  1.167   &       &           &        \\                                          
30$^{p}$ &  104.62877 &  -56.06013 &  0.296   &     &  Chandra &  ELG  &  Galaxy                     &                                       &  B    &  6.698   &  41.38  &  $<$0.6  &  $<$0.05  &  41.0\\                                  
31 &  104.63875 &  -55.83334 &  0.292   &  2  &  IRAC    &  ALG  &                                   &                               &  B    &  6.944   &       &           &        \\                                          
32 &  104.64674 &  -56.01799 &  0.298   &  3  &  Chandra &  ELG &  {\it Bona fide} AGN               &  {\it Bona fide} AGN                            &  A    &  4.141   &  42.19  &  7.5     &  0.33  &  419 & 3 \\                                  
33 &  104.64738 &  -55.95546 &  0.302   &  2  &  IRAC    &  ALG  &                                   &                               &  *    &  0.389  &       &           &        \\                                          
34 &  104.65010 &  -55.80724 &  0.3     &  2  &  IRAC    &  ALG  &                                   &                               &  C    &  8.505   &       &           &        \\                                          
35 &  104.65097 &  -56.0081  &  0.304   &  4  &  Chandra &  ELG &  Galaxy                            &  Galaxy                        &  A    &  3.55    &  41.38  &  $<$1.0  &  $<$0.06 &  40.7 & 6\\          
\hline
\end{tabular}
  \end{small}
  {\it continue on the next page}
  \end{table} 
\end{landscape}
\begin{landscape}
  \begin{table}
    \ContinuedFloat
      \centering
  \caption{Cluster members properties.}
    \begin{small}
\begin{tabular}{lccccccccccccccc}
      \hline
      Id$^{a}$ & R.A. & Dec. & z  &     fl$^{b}$ & Instr.$^{c}$  &   Type$^{d}$ & Type$_X$$^e$ & Type$_{BPT}$$^f$& Reg.$^g$ & Dis.$^h$ & Log(Lx$_H$)$^i$ & N$_H$$^l$ & X/O$^m$ & SFR$^n$  & SFR$_{H{\alpha}}$$^o$\\
 \hline
36$^{n}$ &  104.65373 &  -56.01291 &  0.296   &     &  Chandra &  ELG  &  Galaxy                     &                                       &  A    &  3.843   &  41.29  &  0.0        &  0.04  &   33.4   \\                                  
37 &  104.65424 &  -55.9212  &  0.293   &  2  &  IRAC    &  ALG  &                                   &                               &  *    &  1.684   &       &           &        \\                                          
38 &  104.65907 &  -55.94071 &  0.297   &  3  &  Chandra &  ALG  &  {\it Candidate} AGN              &                                               &  *    &  0.639   &  41.44  &  0.6     &  0.03    &  47.1  \\                                  
39 &  104.66104 &  -55.8596  &  0.304   &  3  &  IRAC    &  ALG  &                                   &                               &  B    &  5.383   &       &           &        \\                                          
40 &  104.66428 &  -56.00758 &  0.295   &  3  &  IRAC    &  ALG  &                                   &                               &  A    &  3.564   &       &           &        \\                                          
41 &  104.66824 &  -55.90079 &  0.289   &  2  &  IRAC    &  ALG  &                                   &                               &  A    &  2.977   &       &           &        \\                                          
42 &  104.66830 &  -55.93417 &  0.293   &  2  &  IRAC    &  ALG  &                                   &                               &  *    &  1.139   &       &           &        \\                                          
43$^{p}$ &   104.66833  & -55.90733 &  0.31    &  & Chandra  & unc & {\it Candidate} AGN             &                                                & A  &   2.599  &     41.46  &   $>$0.4  & 0.19  & 48.6  &   \\
44 &  104.67715 &  -55.96416 &  0.3     &  2  &  IRAC    &  ALG  &                                   &                               &  *    &  1.361   &       &           &        \\                                          
45 &  104.67759 &  -55.84383 &  0.297   &  3  &  IRAC    &  ALG  &                                   &                               &  B    &  6.392   &       &           &        \\                                          
46 &  104.67772 &  -55.97663 &  0.3     &  3  &  IRAC     &  ALG &                                   &                               &  A    &  1.953   &       &           &        \\                                          
47 &  104.67843 &  -56.04731 &  0.292   &  3  &  IRAC    &  ELG &                                    &  {\it Candidate} AGN         &  B    &  5.993   &       &           &      & & 2  \\                                          
48 &  104.68266 &  -55.83871 &  0.301   &  2  &  Chandra &  ELG &  {\it Bona fide} AGN               &  {\it Bona fide} AGN                            &  B    &  6.724   &  42.87  &  $>$10  &  0.56   &   1262 & 0 \\                                  
49$^{p}$ &   104.68666  & -56.01138 &  0.31    &  & Chandra  & unc & {\it Candidate} AGN             &                                                  & A  &   3.973  &     41.48  &   $>$2.5  & 0.05  & 50.9  &  \\
50 &  104.70230 &  -55.82701 &  0.2903  &  2  &  IRAC    &  ALG  &                                   &                               &  B    &  7.551   &       &           &        \\                                          
51 &  104.70824 &  -55.96969 &  0.297   &  3  &  IRAC    &  ALG  &                                   &                               &  A    &  2.402   &       &           &        \\                                          
52 &  104.72647 &  -56.08598 &  0.299   &  3  &  IRAC    &  ALG  &                                   &                               &  C    &  8.64    &       &           &        \\                                          
53 &  104.72728 &  -55.82286 &  0.297   &  3  &  Chandra &  ALG  &  Galaxy                           &                                 &  B    &  8.036   &  41.47  &  $<$0.001        &  $<$0.04  &  50.4\\                                  
54 &  104.73517 &  -55.97295 &  0.303   &  2  &  IRAC    &  ALG  &                                   &                               &  A    &  3.291   &       &           &        \\                                          
55 &  104.74191 &  -55.90965 &  0.295   &  2  &  IRAC    &  ALG  &                                   &                               &  A    &  3.969   &       &           &        \\                                          
56$^{p}$ &   104.74333  & -56.00516 &  0.2956  &  & Chandra  & unc &  Galaxy                         &                                    & A  &   4.670  &     41.22  &   0.1     & 0.01  & 28.4 &   \\
57 &  104.75672 &  -55.83665 &  0.296   &  2  &  Chandra &  ALG  & {\it Candidate} AGN               &                                                &  B    &  7.687   &  41.84  &  2.5     &  0.13   &  118  \\                                  
58 &  104.76166 &  -55.96578 &  0.2997  &  2  &  IRAC    &  ALG  &                                   &                               &  A    &  3.98    &       &           &        \\                                          
59 &  104.78922 &  -55.81475 &  0.291   &  4  &  IRAC    &  ELG &                                    &  {\it Candidate} AGN         &  C    &  9.373   &       &           &   &  & 15   \\                                          
60 &  104.79439 &  -56.05934 &  0.304   &  2  &  IRAC    &  ELG &                                    &  Galaxy                      &  C    &  8.261   &       &           &    &  & 1  \\                                          
61 &  104.79490 &  -56.08235 &  0.296   &  2  &  Chandra &  ELG &  {\it Bona fide} AGN               &  {\it Bona fide} AGN                            &  C    &  9.411   &  42.36  &  $>$7.0  &  0.03   &  392  & 0\\                                  
62 &  104.86802 &  -55.93734 &  0.291   &  3  &  WFI     &  ELG &                                    &  Galaxy                      &  B    &  7.463   &       &           &    &  & 20  \\                                          
63 &  104.97185 &  -55.94513 &  0.294   &  3  &  WFI     &  ALG  &                                   &                               &  C    &  10.92   &       &           &        \\                                          
64 &  104.97657 &  -55.80897 &  0.29    &  3  &  WFI     &  ELG &                                    &  {\it Candidate} AGN         &  C    &  13.93   &       &           &   &   & 147  \\
  \hline
  \end{tabular}
\end{small}

$^{a}$ Source identification number; $^{b}$ redshift quality flag, ranging from 1 for very insecure redshifts to 4 corresponding to secure ones; $^{c}$ instrument used to select the source; $^{d}$ galaxy classification: according to the optical spectral features, objects have been classified in emission line galaxies (ELG) and absorption line galaxies (ALG); $^e$ classification based on X--ray emission (see Sect. 6.3); $^f$ classification  based on emission line optical spectra, using diagnostic diagrams (BPT: Baldwin, Phillips \& Terlevich, 1981, Veilleux \& Osterbrock, 1987, see Fig. \ref{BPT}); $^g$ region where the source lands (A, B, C in Fig. \ref{image}, the symbol * indicates the red elliptical region in Fig. \ref{image}, which has been excluded by the X-ray analysis). Note that at the Bullet's redshift the flux limits the A, B and C regions, correspond to a 2--10 keV luminosity of $\sim 1.5 \times 10^{41}$ erg s$^{-1}$, $\sim 2 \times 10^{41}$ erg s$^{-1}$, $\sim 3.5 \times 10^{41}$ erg s$^{-1}$, respectively; $^h$ distance from the cluster centre in arcmin; $^i$ log(2-10 keV unabsorbed luminosity); $^l$ N$_H$ in unit of 10$^{22}$ cm$^{-2}$, evaluated by the (H-S)/(H+S) ratio using an absorbed power$-$law spectrum with $\Gamma=$1.9;$^m$ X/O ratio: the optical flux is computed by converting R magnitudes in specific fluxes and then multiplying by the width of the Rc filter ($f_R(0)=2.15\times10^{-9}$  erg cm$^{-2}$ s$^{-1}$ $\AA^{-1}$, $\Delta\lambda_R=1568 \AA$, Fukugita et al. (1995)).$^{n}$ star formation rate in M$_\odot$/yr, evaluated by the 2--10 keV luminosity using the relation from Ranalli et al. (2003); $^{o}$ star formation rate in M$_\odot$/yr, evaluated by the H${\alpha}$ luminosity using the relation from Kennicutt et al.(1994); $^{p}$ redshift from a literature (see text);$^q$ the X-ray spectrum of the source \#13 (see Sect. 4.4, Fig. \ref{spettri2b} and Tab. \ref{tabfit}) shows a strong iron emission, which suggests a column density larger than 10$^{24}$ cm$^{-2}$.
\label{souzbullet}
  \end{table} 
\end{landscape}

\subsection{Redshift distribution of the sources contributing to the observed over--densities.}

The 127 X--ray sources with measured redshift are distributed almost uniformly in and around the Bullet cluster: 40 are located in region A, 39 in region B, 44 region in C. 4 are located in the central area with the strongest ICM emission at distance less than $< 0.3\times R_{200}$ from the cluster centre (red elliptical region in Fig. \ref{image}). 28\% of the X-ray sources in region A and B are at the Bullet cluster redshift, while only 13.6\% of the X-ray sources in region C belong to the cluster.

The observed over--density at bright fluxes in region B is totally due to {\it bona fide} AGN at the Bullet redshift.  The situation in region C is different.  67\% of the brightest H band sources (4 out of 6) in the region C are field AGN, and 100$\%$ of the S and HH brightest sources in region C are luminous AGN at redshift in the range 0.21--1.54. Therefore in this region the observed over--density at the brightest S and HH fluxes is due to field AGN, whereas the excess at the brightest H fluxes could be only in part due to field AGN. The 16 X--ray emitting normal galaxies/LLAGN contribute to the excess densities observed near the survey flux limits.

We analysed the spatial distribution around the cluster of the {\it bona fide} AGN, {\it candidate} AGN, X--ray normal/LLAGN, ELG and ALG galaxies. To this aim, we applied the Kolmogorov Smirnov test to the distributions of the projected distance of each source from the cluster centre. We found no statistically significant difference between the spatial distribution of two types of AGN and normal galaxies in the region A, B and C. However, in the core of cluster there are only ALG galaxies, as expected.

\begin{figure}
\begin{center}
\includegraphics[width=6.5cm]{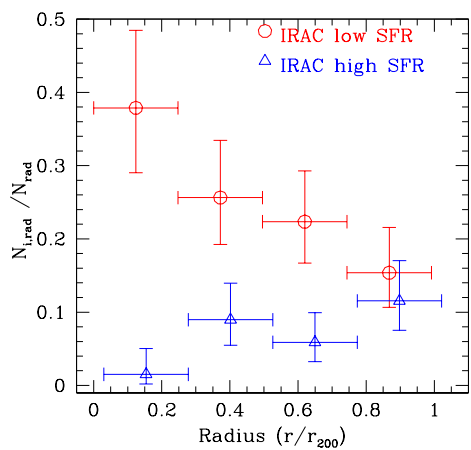}
\caption{ Ratio between the number of IRAC {\it LSFRG} and {\it HSFRG} to the total samples as a function of the distance from the cluster centre for the 308 cluster members with IRAC fluxes.  Red open dots = IRAC {\it LSFRG}; blue open triangles = IRAC {\it HSFG}.
}
\label{irac2}
\end{center}
\end{figure}

We found 308 sources detected in all the four IRAC channels at the Bullet cluster redshift. Using these sources, we recalculated the fraction of high and low star--formation galaxies in the Bullet cluster field as done in Sect. 5.2. We confirm that the fraction of IRAC {\it LSFRG} is maximum near the cluster centre and decreases with the distance from the centre, and that the fraction of {\it HSFRG} is roughly constant with the radius and remains smaller than the fraction of {\it LSFRG}. The fractions evaluated using only the sources with a redshift consistent with that of the Bullet cluster are higher than the one evaluated by using the total IRAC sample, see Figs. \ref{iracfra} and \ref{irac2}.

\subsection{AGN fraction in the Bullet cluster}

\begin{figure}
\begin{center}
\includegraphics[width=6.5cm]{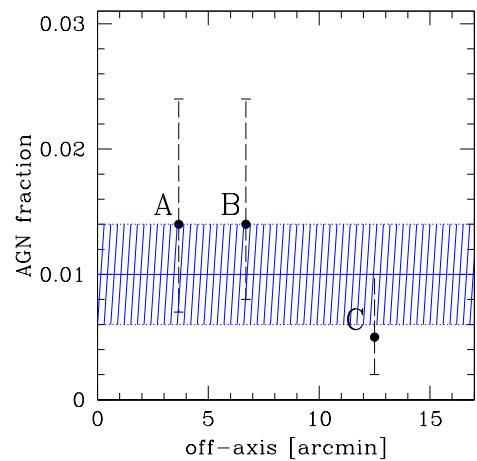}
\caption{The fraction of cluster members hosting an X--ray AGN versus the distance from the cluster centre, in units of R$_{200}$ (black solid circles). The average X--ray AGN fraction in the A$+$B$+$C total region is marked by the blue solid line, with 1$\sigma$ uncertainties identified by the dashed regions.
}
\label{AGNfracz}
\end{center}
\end{figure}

The availability of a large sample of galaxies with measured redshift and the identification of 13 AGN associated to the cluster ({\it bona fide} + candidates) allows us to study the fraction of cluster optical galaxies hosting an X--ray AGN. To this aim, we first selected galaxies with optical magnitude $R\leq$22 mag, a sample including most of the X-ray source counterparts. The fraction of X--ray AGN at the Bullet's redshift is then given by

\begin{equation}
Fraction = {(N_{XRz}/ Comp_{XRz}) \over (N_{Rz} /Comp_{Rz}}
\end{equation}

\noindent
where N$_{XRz}$ is the number of galaxies at the cluster redshift hosting an X--ray AGN, which are in regions A, B and C, and most of them have spectroscopic redshift (see Sect. 6), while  $N_{Rz}$ is the total number of R band selected galaxies  at the cluster redshift (most of them are photometric, see Sect. 6.2) in the same regions. Comp$_{XRz}$ and Comp$_{Rz}$ are correction factors, which take into account the redshift incompleteness of N$_{XRz}$ and N$_{Rz}$, respectively. 
The upper ($+\Delta$ Fraction) and lower ($-\Delta$ Fraction) uncertainties are evaluated using the Gehrels (1986) formulae for Poisson statistics for N$_{XRz}$ and N$_{Rz}$. Fig. \ref{AGNfracz} shows the AGN fraction in the Bullet cluster as a function of the distance from the cluster core.  The average fraction of X-ray AGN in the total A$+$B$+$C region is 1.0$\pm{0.4}\%$.

We have also estimated the fraction of X--ray AGN at the cluster redshift with L(2--10 keV) $ \ge 10^{43}$ erg s$^{-1}$. This is  0.25$\pm^{+0.32}_{-0.12}\%$ and 0.5$\pm^{+0.6}_{-0.2}\%$ in the  in the total A$+$B$+$C region, and within R$_{200}$, respectively. This difference is due to the lack of luminous X--ray AGN at distances larger than R$_{200}$.

\section{Discussion and conclusions}

The analysis of cluster environments is a valuable instrument to investigate the origin of the galaxy nuclear and star-formation activities trigger mechanisms. To this aim we performed a detailed analysis of point X--ray sources in the Bullet cluster field. Thanks to deep and wide {\it Chandra} observations accumulating a total of $\sim600$~ks, we were able to produce a catalogue of 381 X--ray point sources up to a distance of $\sim 2 \times R_{200}$ (1.5 virial radii). We reached a flux limit of $\sim 1\times 10^{-16}$\cgs (corresponding to a luminosity of $\sim$a few $10^{40}$ erg/s at the cluster redshift), an unprecedented sensitivity in cluster analysis.

Using ESO/MPG 2.2m WFI and Spitzer/IRAC imagers we identified optical and near infrared counterparts of $\sim$84\% and $\sim$48\% of the X--ray sources, respectively.

We obtained new spectroscopic redshifts for 106 X-ray sources using ESO VLT/VIMOS. We collected from the literature additional 21 redshifts of the X-ray sources (8 photometric and 13 spectroscopic). 29 of the 127 X-ray sources with a measured redshift are associated with the Bullet cluster, in the redshift range 0.28--0.31. Additional 525 infrared selected sources and 113 optical selected sources have a photometric or spectroscopic redshift in the same range, and are thus likely to be associated to the cluster.

\subsection{X--ray over--density}

Point-like X-ray sources in the Bullet cluster are denser than in the field by a factor 1.5--2 in all S, H and HH energy bands, as found in most cluster in the past twenty years. 
The over--densities correspond to $\sim$ 100, 70, 50 excess sources in the S, H and HH bands, respectively, higher than what reported by Gilmour et al. (2009, $7.7_{-4.6}^{+5.7}$) in the central Mpc.  Limiting the analysis at the flux limits and area covered by Gilmour et al (2009) we found an over--density identical to their evaluation, showing that deep and wide observations are truly mandatory to probe the full scale of cluster content of point X-ray sources.

We studied the over-density of X-ray sources as a function of the distance from the cluster core to highlight any radial variation. This is of particular interest in this case because, unlike the majority of clusters of galaxies, in the Bullet cluster the peak of the dark matter distribution is significantly shifted eastward from the peak of the X-ray emission (Clowe et al. 2004, 2006). Magnification of background sources is therefore probably higher outside the cluster core and this may produce an enhancement of X-ray point source detection eastward of the core, where the ICM X-ray emission is fainter than in the core, thus limiting less our ability to detect point sources (both cluster members, and background or foreground sources). We studied the three regions A, B and C, located at increasing distances from the cluster core: A, $R\le 0.66\times R_{200}$; B, $0.66\times R_{200}<R<R_{200}$; C, $R_{200}<R<2\times R_{200}$ (see Fig. \ref{image}). The peak of the dark matter distribution lies in region A, eastward from the core, see Fig. 1, where three X-ray sources are detected. Source $\#38$ in Tab. \ref{souzbullet} belong to the cluster, nearby source at R.A.$=$104.65632 and Dec.$=$-55.94062 has a bright point like optical counterpart (R=16, B=15.8) and it is likely a star, source at R.A.$=$104.62821 and Dec.$=$-55.96111 in Tab. \ref{sounozbullet} is at z=0.347, a redshift consistent with the background cluster discovered by Rawle et al. (2010). The density of X-ray sources in the region corresponding to the peak of the dark matter distribution is similar to that of full region A. This is also the region presenting the highest over--density of faint X-ray sources in both S and H bands. Region C presents the highest over--density of bright H and HH sources. These bright sources are distributed uniformly in region C, without significant concentrations.

\subsection{What generates the X--ray over--density?}

In interpreting number counts over--densities we must consider that cosmic variance can produce over--densities, particularly when small regions of the sky are examined. To investigate further this point we evaluated the cosmic variance in regions with dimension and sensitivity similar to those used for the Bullet, based on the C-COSMOS and CDF-S survey data. The values obtained are consistent with those given in the literature, indicating that the cosmic variance cannot exceed $\sim$ 25$\%$ in regions of the same size as region A, and is $\sim$ 10$\%$ in zones as large as region C or the whole Chandra Bullet. These results indicate that the over--densities found in the Bullet cluster field can hardly be explained in terms of cosmic variance.\\
\indent To investigate other scenarios for the over-density of X-ray sources (cluster AGN, cluster SFG, magnified background sources, etc), we have obtained redshifts for $\sim 33\%$ of the X--ray sources, that is 127 sources: 119 spectroscopic redshifts and 8 photometric redshifts. These sources are uniformly distributed in the three regions A, B and C (see Fig. \ref{image}). The redshift distribution of the X--ray sources shows a statistical significant peak at the Bullet cluster redshift (see Fig. \ref{isredshift})  suggesting an excess of X--ray sources belonging to the cluster. We found a similar behaviour in the redshift distributions in each of the regions A, B and C. Thus, a total of 29 X--ray sources likely belong to the Bullet cluster. We identified seven {\it bona fide} AGN, those with L$_{2-10~keV}\gtrsim 10^{42}$ erg s$^{-1}$, and six {\it candidate} AGN, those with $3\times 10^{41}\lsimeq$ L$_{2-10~keV} \lsimeq 10^{42}$ erg s$^{-1}$. The remaining sixteen X-ray sources have L$_{2-10~keV} \lsimeq 3\times 10^{41}$ erg s$^{-1}$, and were classified as normal galaxies/LLAGN. For comparison, Haines et al. (2012) found a total of 28 AGN with L$_{0.3-7 keV}\gtrsim 10^{42}$ erg s$^{-1}$ and 48 AGN with L$_{0.3-7 keV} \gtrsim 2\times 10^{41}$ erg s$^{-1}$, in a sample of 26 clusters at z=0.15--0.3. This corresponds to an average number of 1 and 2 AGN per cluster, respectively, with an upper limit of 6 AGN per cluster. Gilmour et al. (2009) found similar numbers. Therefore, we find in the Bullet Cluster at least three times more AGN than the average value of Gilmour et al. (2009) and Haines et al. (2012), and possibly twice their upper limit.\\
\indent The 13 {\it bona fide} or {\it candidate} AGN contribute to the observed over-density of X-ray source with respect to the field at medium/high H band fluxes. In particular, the observed over--density at bright fluxes in region B is totally due to Bullet cluster {\it bona fide} AGN. In region C 2/3 of the bright H band sources have an estimated redshift and are field AGN, as well as all bright HH sources. Therefore in the C region part of the observed over--density in the H band could be due to field AGN, even if the remaining sources belonged to the Bullet cluster. Spectroscopic identification of more X-ray bright sources in region C is needed to definitely assess whether sources associated to the Bullet cluster contributes to the over--density in this region. The 16 X--ray emitting normal galaxies/LLAGN contribute to the over-density observed near the survey flux limits. The analysis of X-ray hardness ratios suggests that region A contains a higher concentration of obscured AGN. Moreover, in region A we found a candidate Compton thick AGN at the redshift of the Bullet cluster.\\
\indent In conclusion, our analysis suggest that the over-density of bright H and HH sources in regions B and C is mostly due to AGN belonging to the Bullet cluster and that the over-density of faint sources X-ray sources in region A is mostly due to star-forming galaxies and low luminosity and/or highly obscured AGN belonging to the Bullet cluster.\\
\indent It is known that the Bullet cluster has high star-formation activity (Rawle et al. 2010), therefore we can speculate that the over-densities observed in all regions A, B and C can be explained if the same mechanism can trigger both star-formation and nuclear accretion. Magnification of background sources has likely little effect in producing the observed over-densities, otherwise we would have observed an an higher over-density in the inner regions, closer to the peak of the dark matter distribution.

\subsection{Fraction of active, X-ray galaxies} 

Cluster of galaxies are over-dense regions of the Universe, thus it is not surprising that excess X-ray sources with respect to the field are found. Haines et al. (2012) and Ehlert et al. (2014) found that the average fraction of bright, active galaxies ($F>10^{-14}$\cgs) in cluster cores is 2--3 times lower than in the field, while it is similar to that in the field at radii above $R_{200}$. In the Bullet cluster we find a fraction of bright, active galaxies consistent with that in the field at all radii but the statistical uncertainty at $R<0.66R_{200}$ are too big to draw strong conclusions (see Fig. \ref{frax}). 

Finally, we estimated the fraction of galaxies belonging to the Bullet cluster hosting an AGN (we use for this estimate both "bona fide" and "candidate" AGN). We found a fraction of $1.0 \pm 0.4 \%$ constant, within the rather large error-bars, with the distance from the cluster core.  This AGN fraction is consistent with the results of Haines et al. (2012, $0.73 \pm 0.14\%$) and Martini et al. (2007, 1$\%$) for AGN of similar luminosities in clusters at redshift similar to that of the  Bullet cluster, and also with the results of Haggard et al. (2010, 1.19$^{+0.11}_{-0.08}\%$) for AGN in field. Limiting the analysis to AGN with L$_{2-10keV}> 10^{43}$ erg s$^{-1}$ reduces the AGN fraction to $0.25^{+0.32}_{-0.12}\%$ in the total A+B+C region and to $0.5^{+0.6}_{-0.2}\%$ in the region A+B, within $R_{200}$ (no luminous AGN is present in region C). At these luminosities Martini et al. (2013) find a slightly smaller fraction, 0.107$^{+0.057}_{-0.039}\%$. Haggard (2012, private communication in Martini et al. 2013) reports a field AGN fraction of  0.64$^{+0.04}_{-0.05}\%$. In conclusion, the higher luminosity AGN fraction in the Bullet cluster is consistent with the fraction found in the field and thus slightly higher than the average fraction found in clusters. This result is however based on only three AGN, and thus we cannot exclude that it is due to a statistical fluctuation.

\subsection {Spatial distribution of active galaxies}
 
We analysed the spatial distribution around the cluster of the {\it bona fide} AGN, {\it candidate} AGN, X–ray normal/LLAGN, star-forming (ELG) and passive (ALG) galaxies at the redshift of the Bullet cluster. The spatial distribution of the different samples of sources in region A, B and C, is not statistically different. Nevertheless, as expected, in the core of cluster (the red elliptical region in Fig. \ref{image}), there are only ALG galaxies.  The similar spatial distribution of active galaxies suggests that the trigger of gas accretion on SMBHs and star-formation is also similar. 

The conclusions in Section 7.3 are based on galaxies with a measured redshift, which are however a minority population. To increase the sample and therefore avoid selection effects which can bias the results, we considered also the sample of all galaxies with a measured IRAC colour. This includes of course both cluster members and foreground or background galaxies. We assume that the distribution of the latter galaxies do not depend on the cluster geometry outside the cluster core because outside this region the magnification of background galaxies is likely negligible, as found for the AGN sample.

  We considered samples of high and low star-formation galaxies ({\it HSFRG} and {\it LSFRG}), selected by comparing their IRAC colours to those of \textsc{M82}, a local star-forming galaxy, and \textsc{NGC 4429} a passive galaxy. The fraction of {\it LSFRG} is higher near the cluster centre, and decreases at increasing radii. Conversely, the fraction of {\it HSFRG} is constant with the radius, and remains at all radii smaller than the fraction of {\it LSFRG}.  {\it Chandra} sources with IRAC counterparts, follow a similar behaviour, thus suggesting that AGN and star-formation activities are triggered by the same mechanism.

We finally find a high concentration of X--ray sources with IRAC colours near zero, in particular in the C region. We obtained redshifts for 8 of these 29 sources: 7 are stars and 1 is not a cluster member, suggesting that most of these sources are stars.

\begin{acknowledgements}
  This work is based on observations collected at the European Southern Observatory under ESO programme 090.A-0267. We are grateful to Andrea Grazian for useful discussions on WFI and IRAC data reduction and to Chiara Feruglio for comments that helped to improve the presentation. We acknowledge the anonymous referee for her/his comments, that helped improving the quality of the manuscript. FF and AB acknowledge support from contract PRIN-INAF-2016 FORECAST, and contract ASI/INAF I/037/12/0.
\end{acknowledgements}

\begin{appendix}

\section{VIMOS optical spectra}
The figures in these section show the VIMOS optical spectra of 49
cluster members (see Tab. \ref{souzbullet}).

\begin{figure*}[ht!]

\centering
\includegraphics[width=6cm]{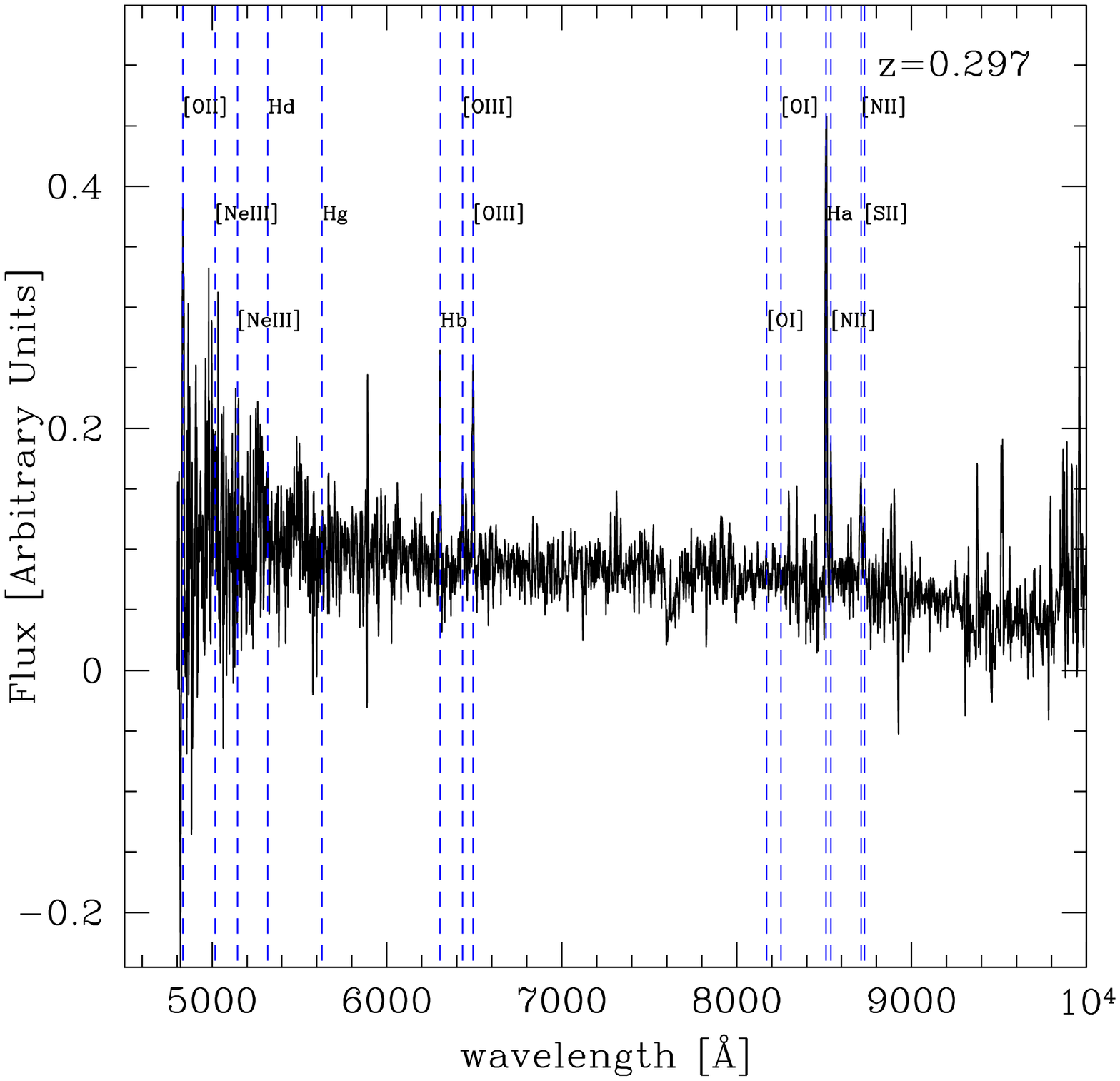}\includegraphics[width=6cm]{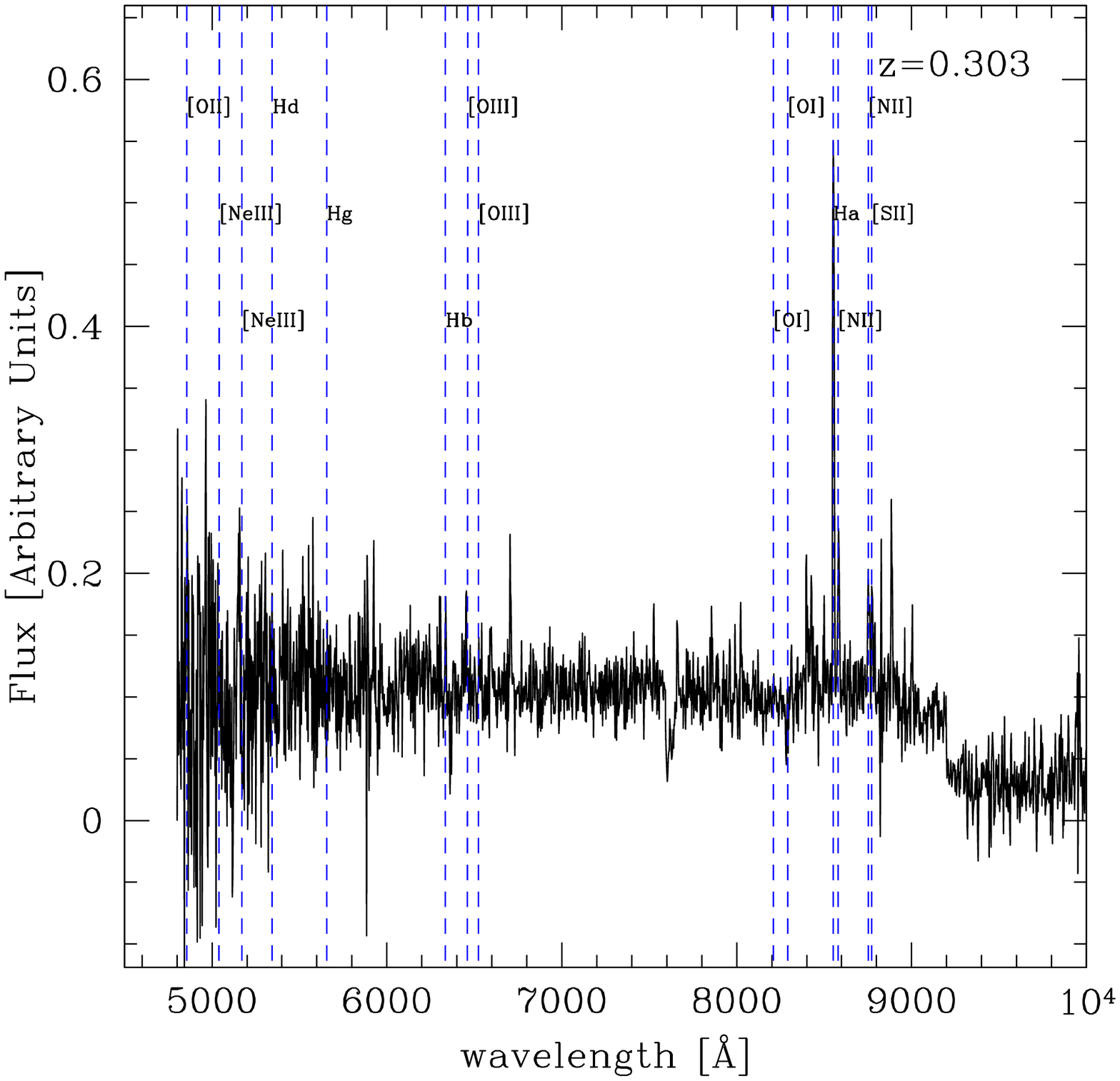}\includegraphics[width=6cm]{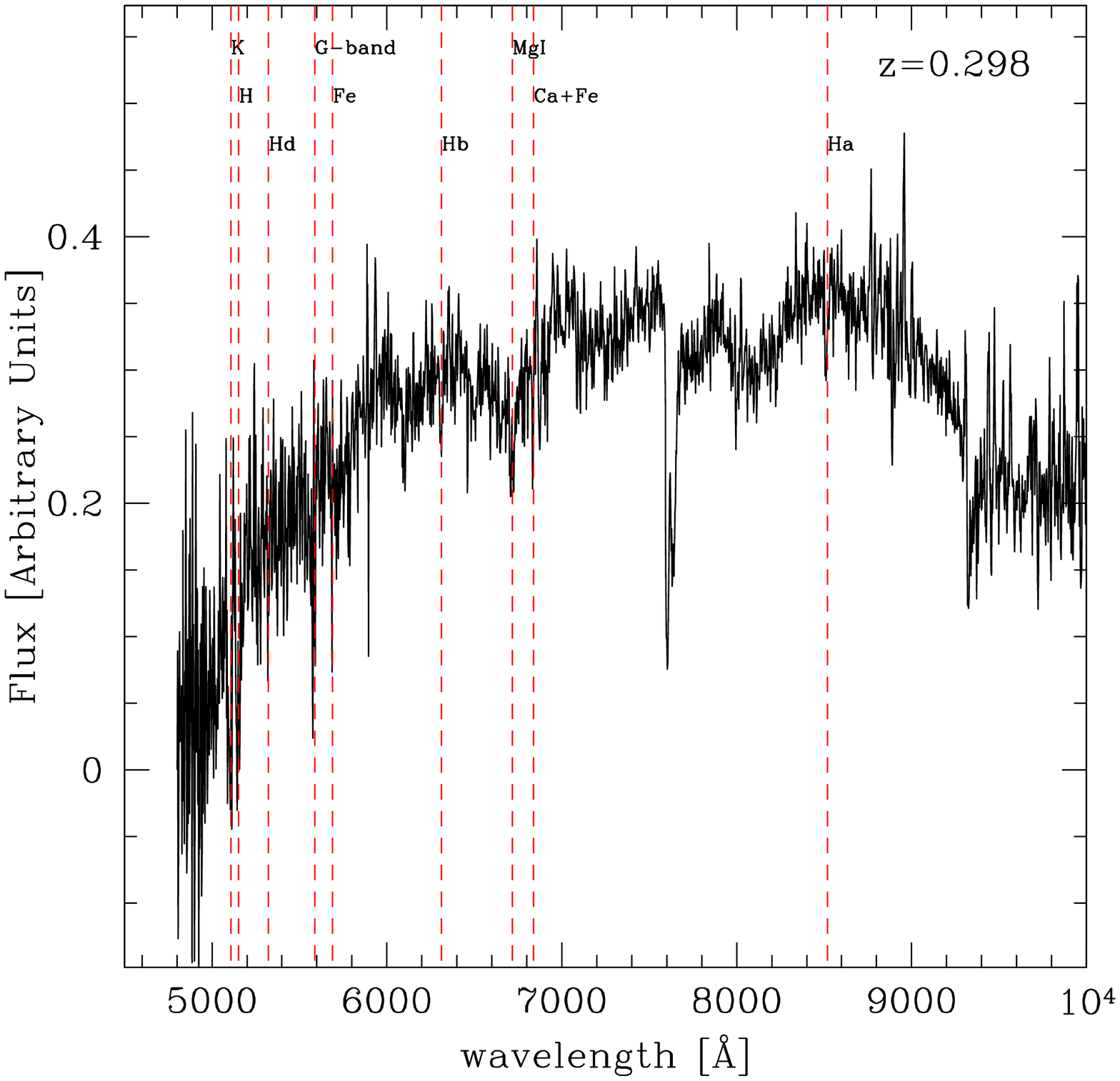}

\includegraphics[width=6cm]{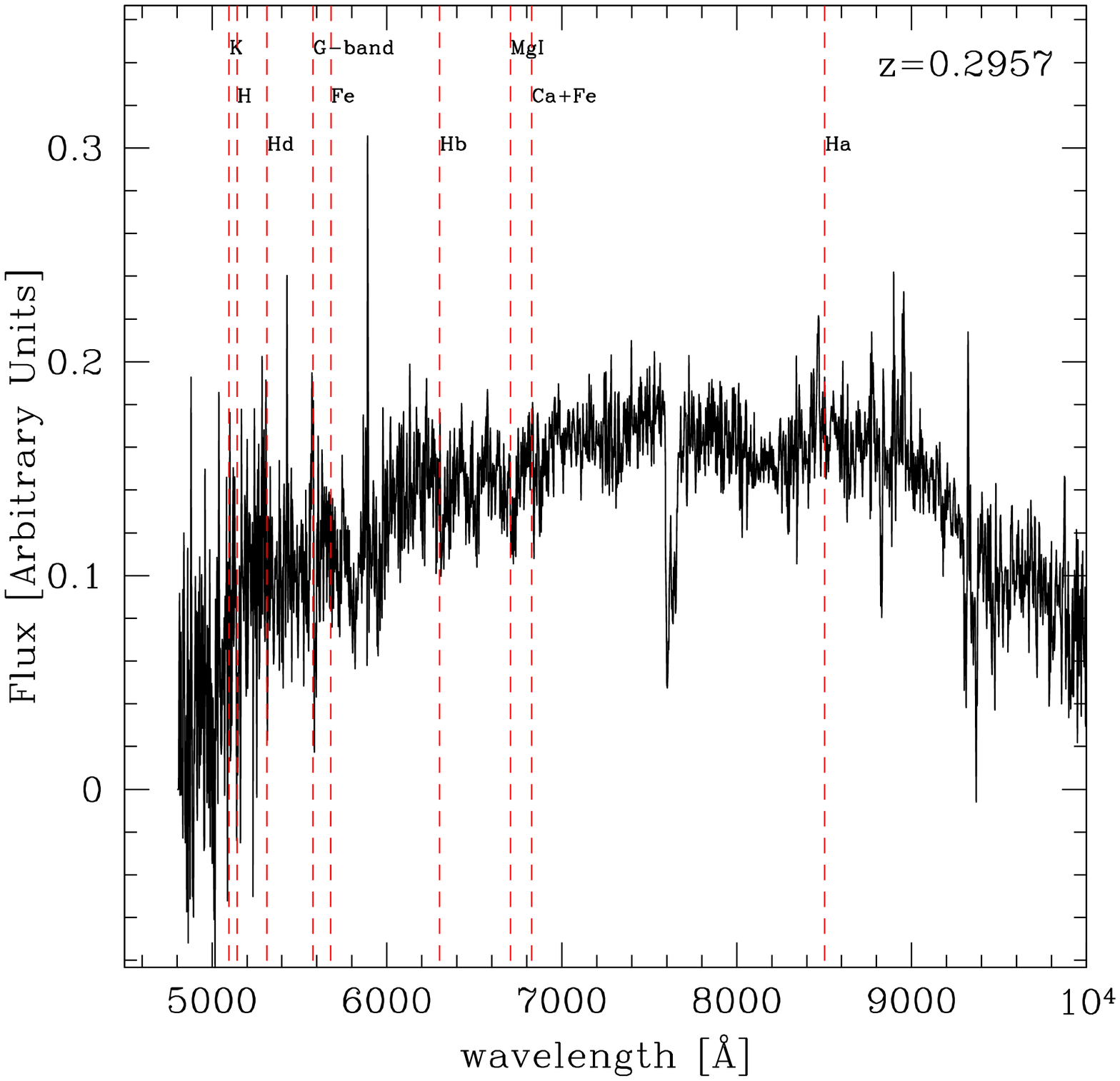}\includegraphics[width=6cm]{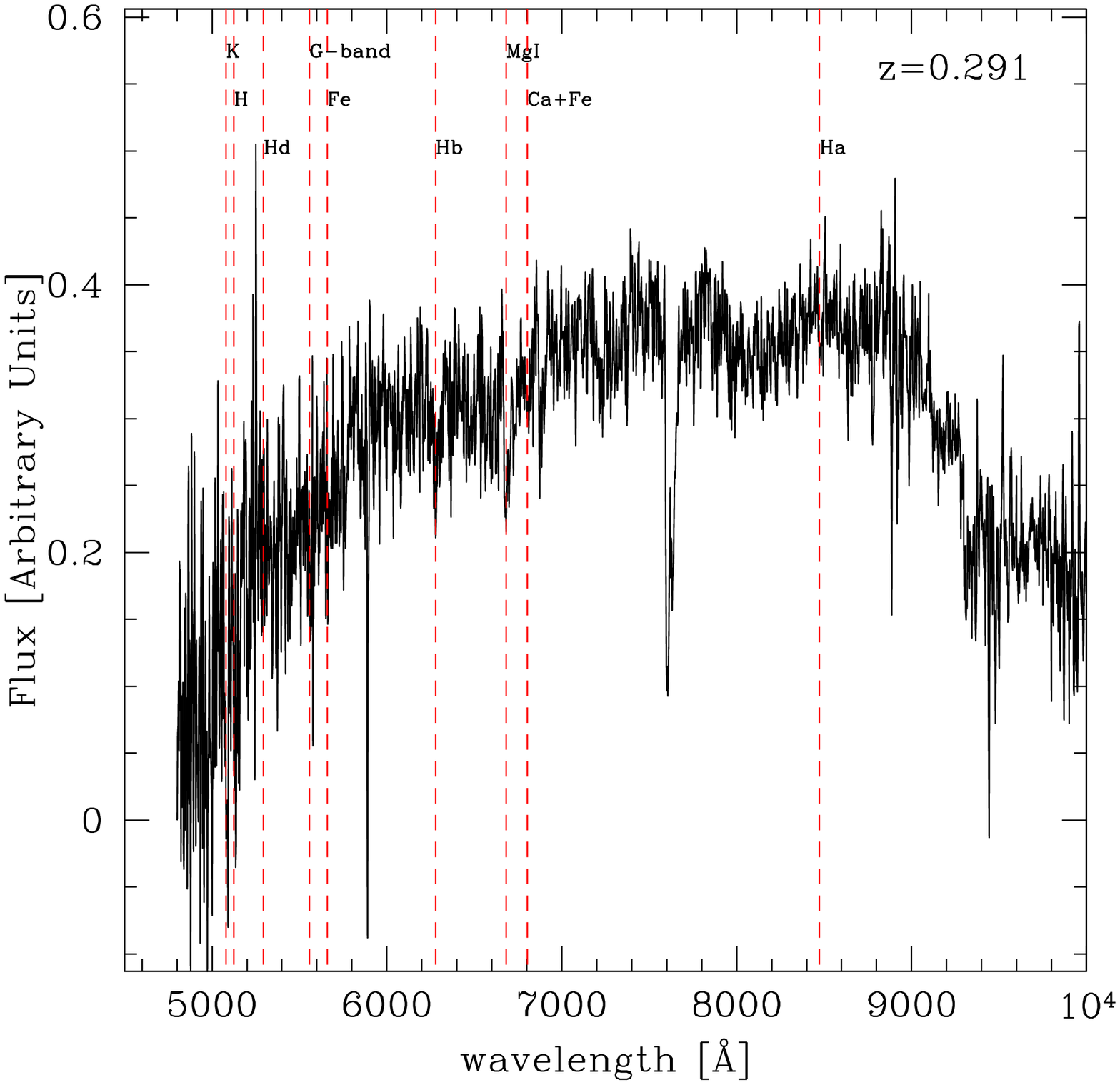}\includegraphics[width=6cm]{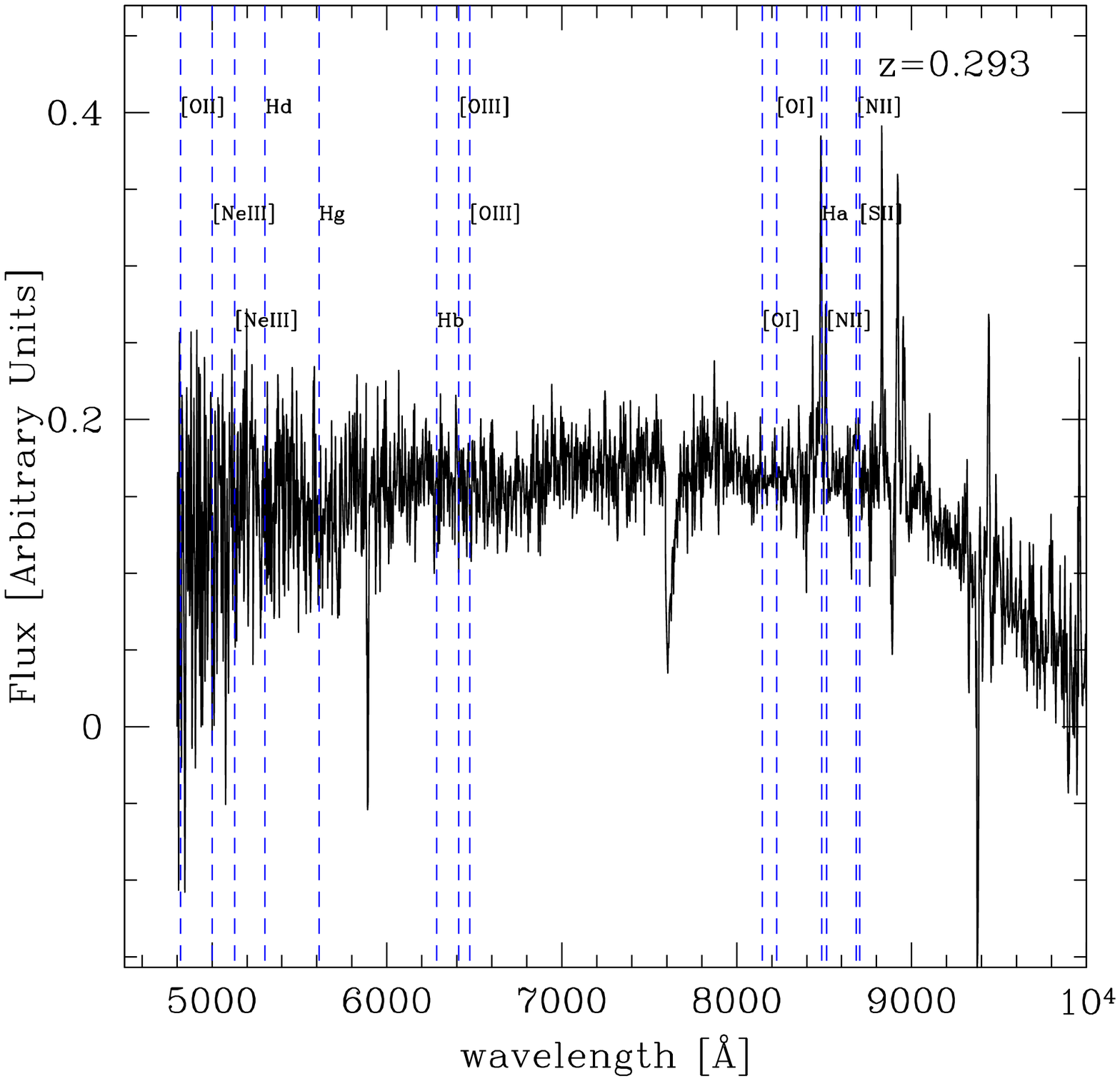} 

\includegraphics[width=6cm]{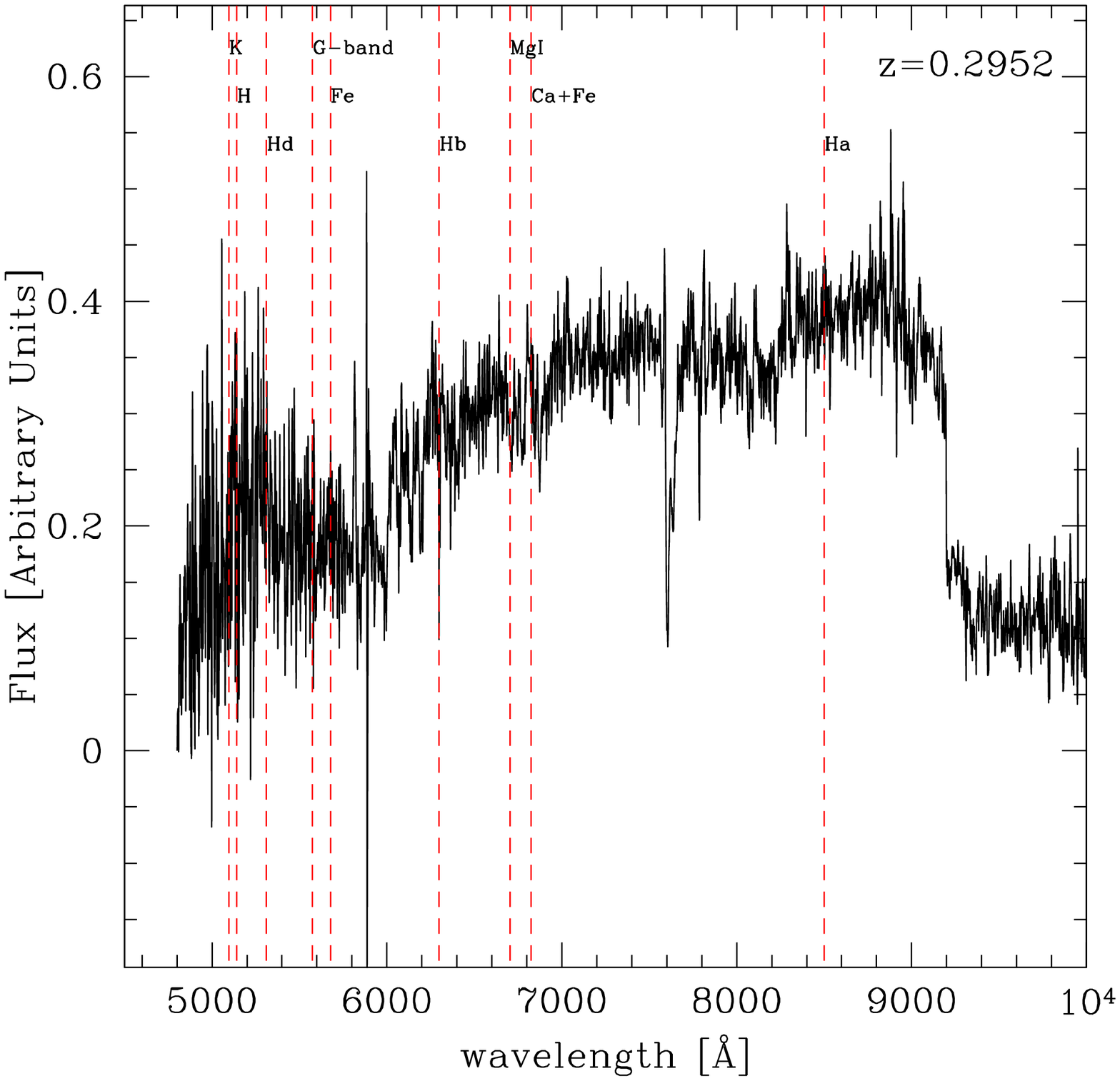}\includegraphics[width=6cm]{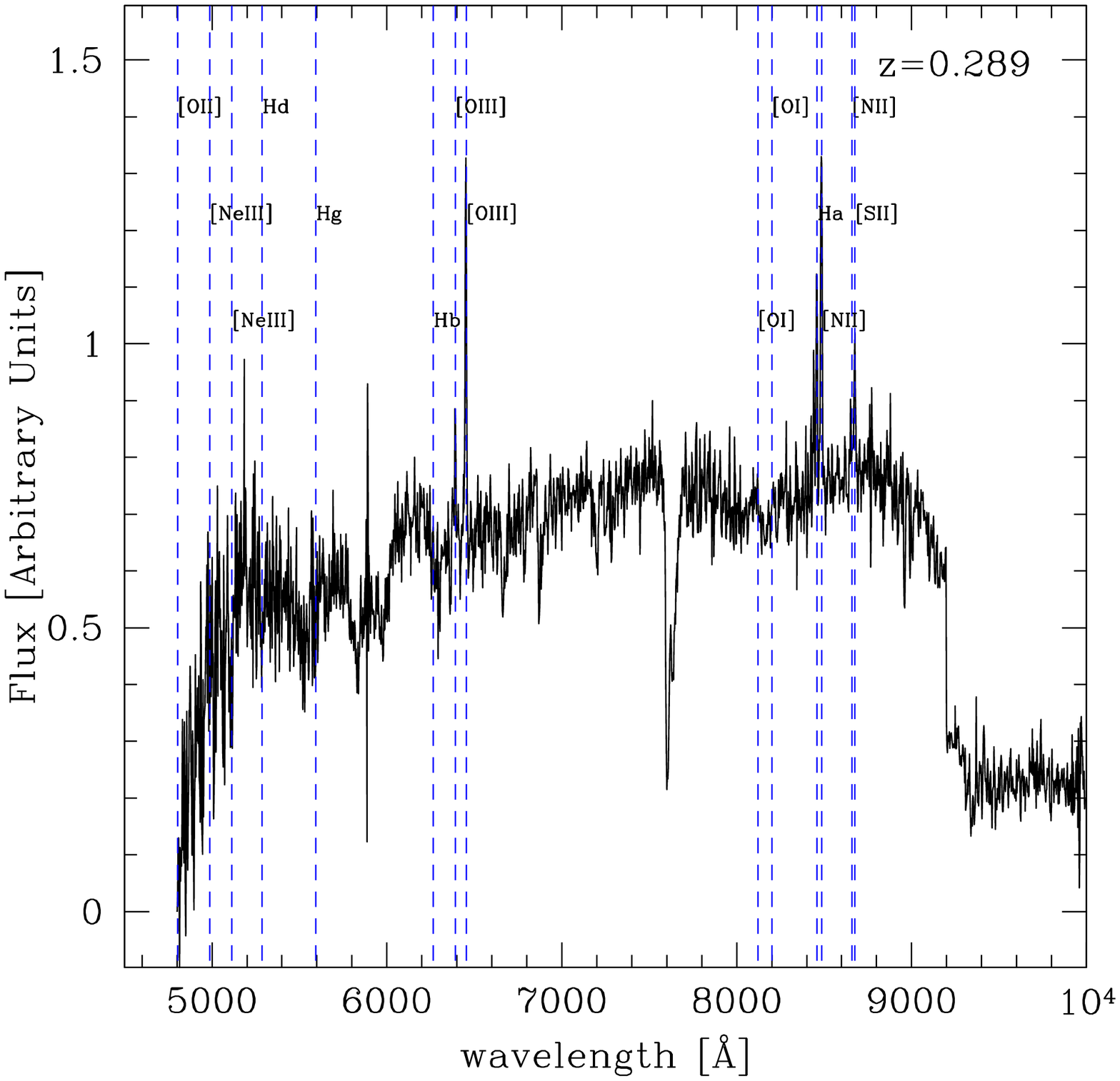}\includegraphics[width=6cm]{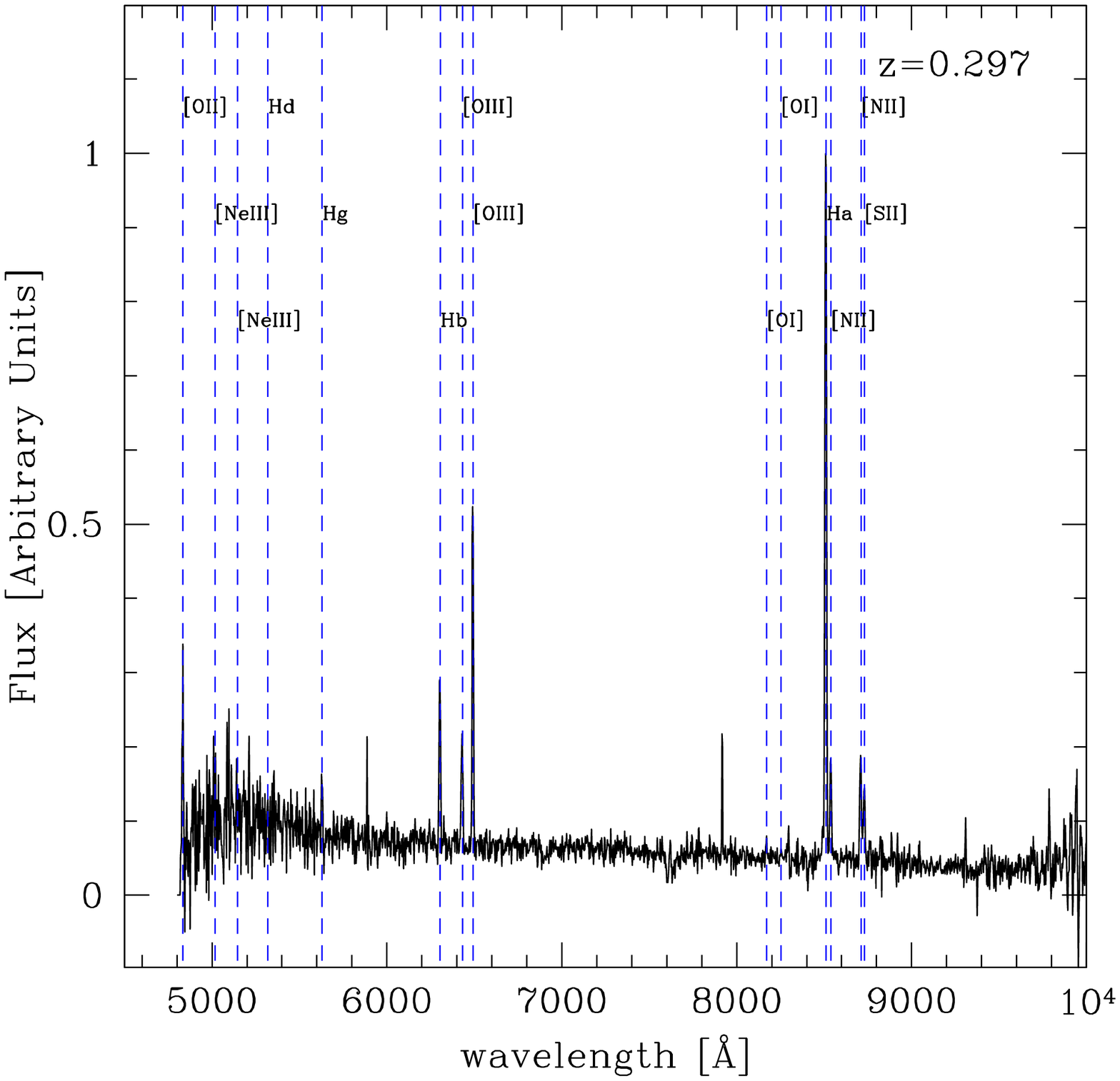} 

\includegraphics[width=6cm]{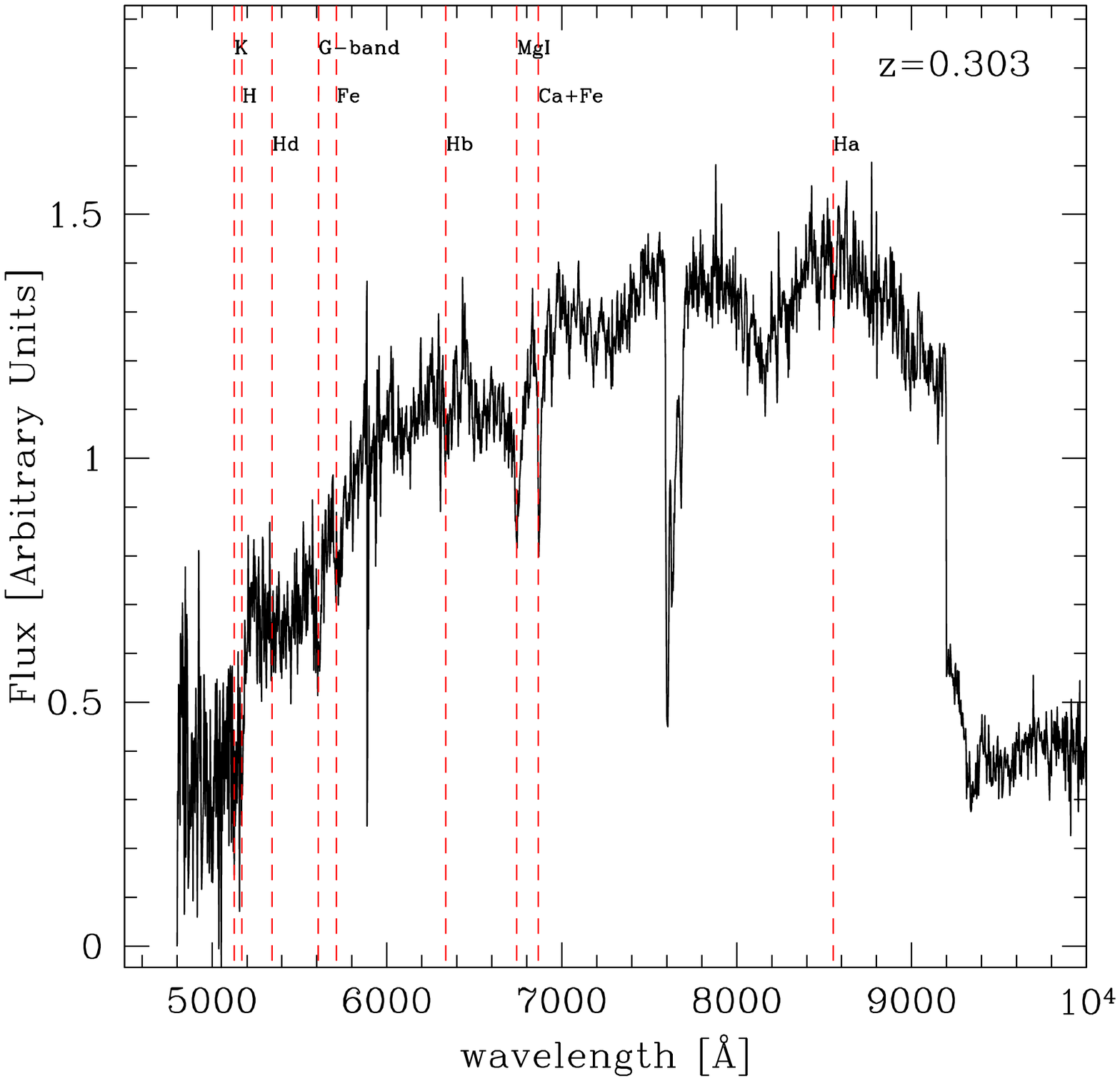}\includegraphics[width=6cm]{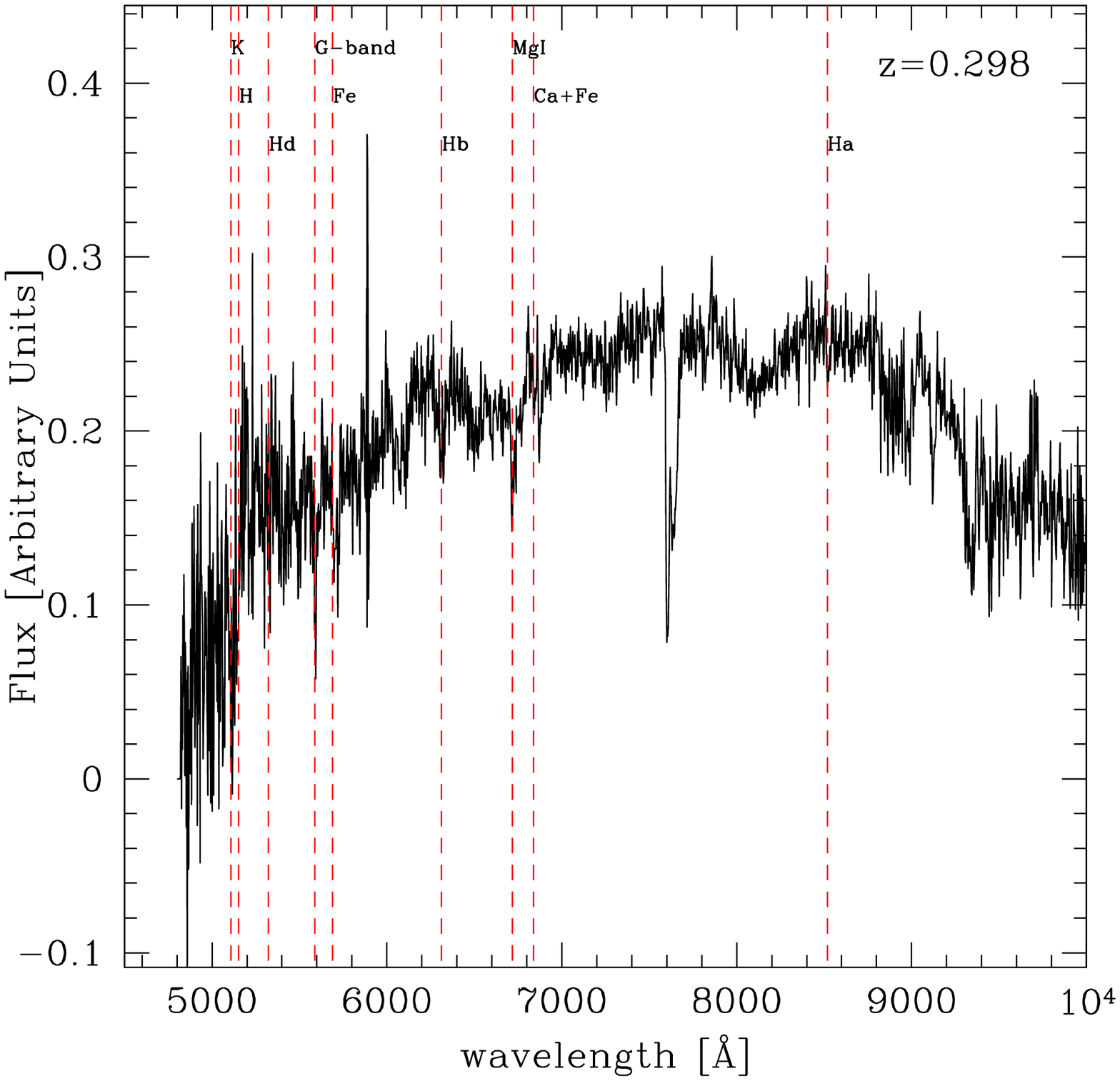}\includegraphics[width=6cm]{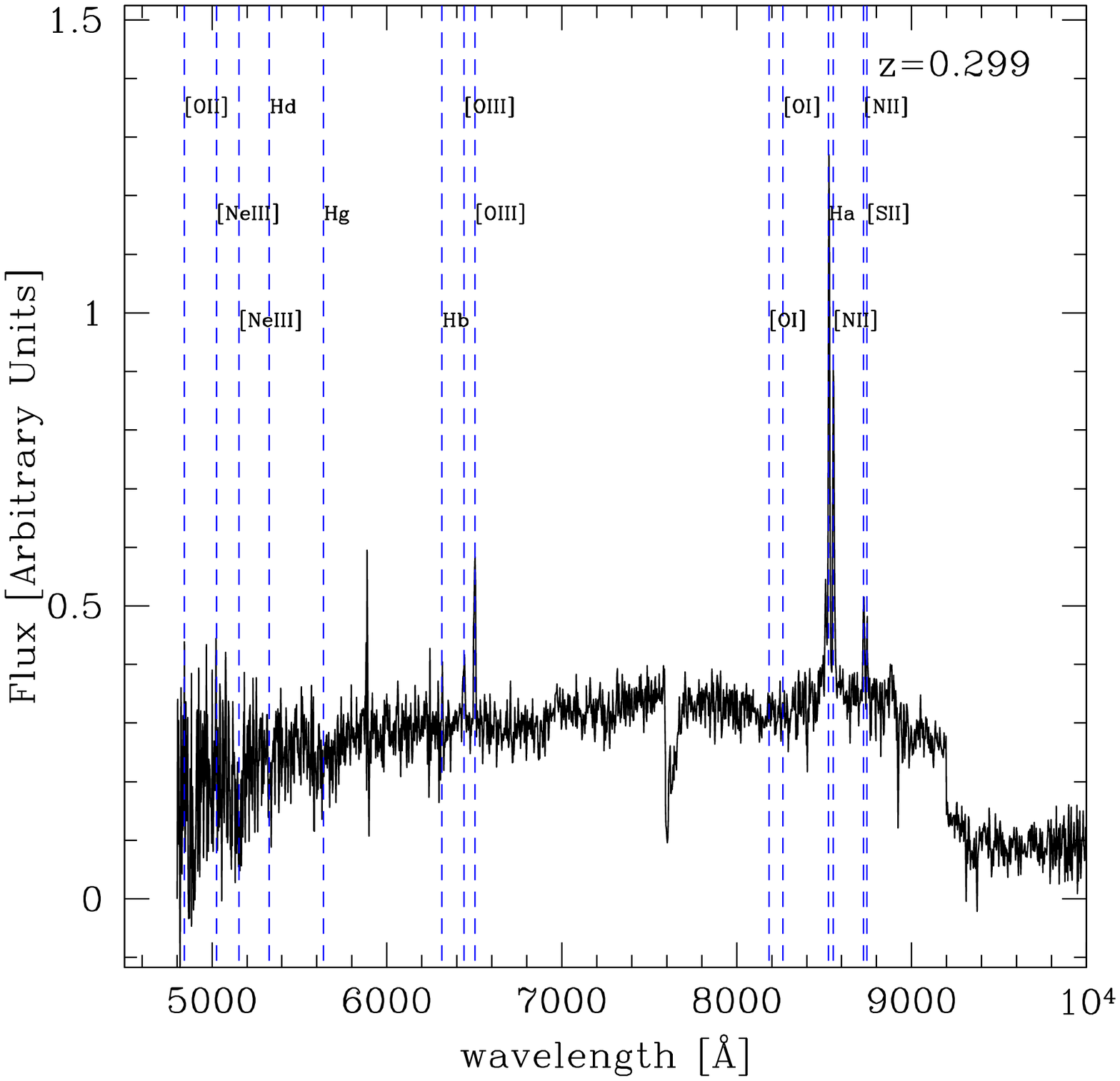} 

\caption{VIMOS optical spectra of 49 cluster members (see Tab.
  \ref{souzbullet}). From top to bottom and from left to right
  spectrum of sources \#2, 3, 4, 5, 6, 7, 10, 13, 16, 17, 19, 20.
\label{spettri}}
\end{figure*}

\begin{figure*}
\ContinuedFloat
\centering
\includegraphics[width=6cm]{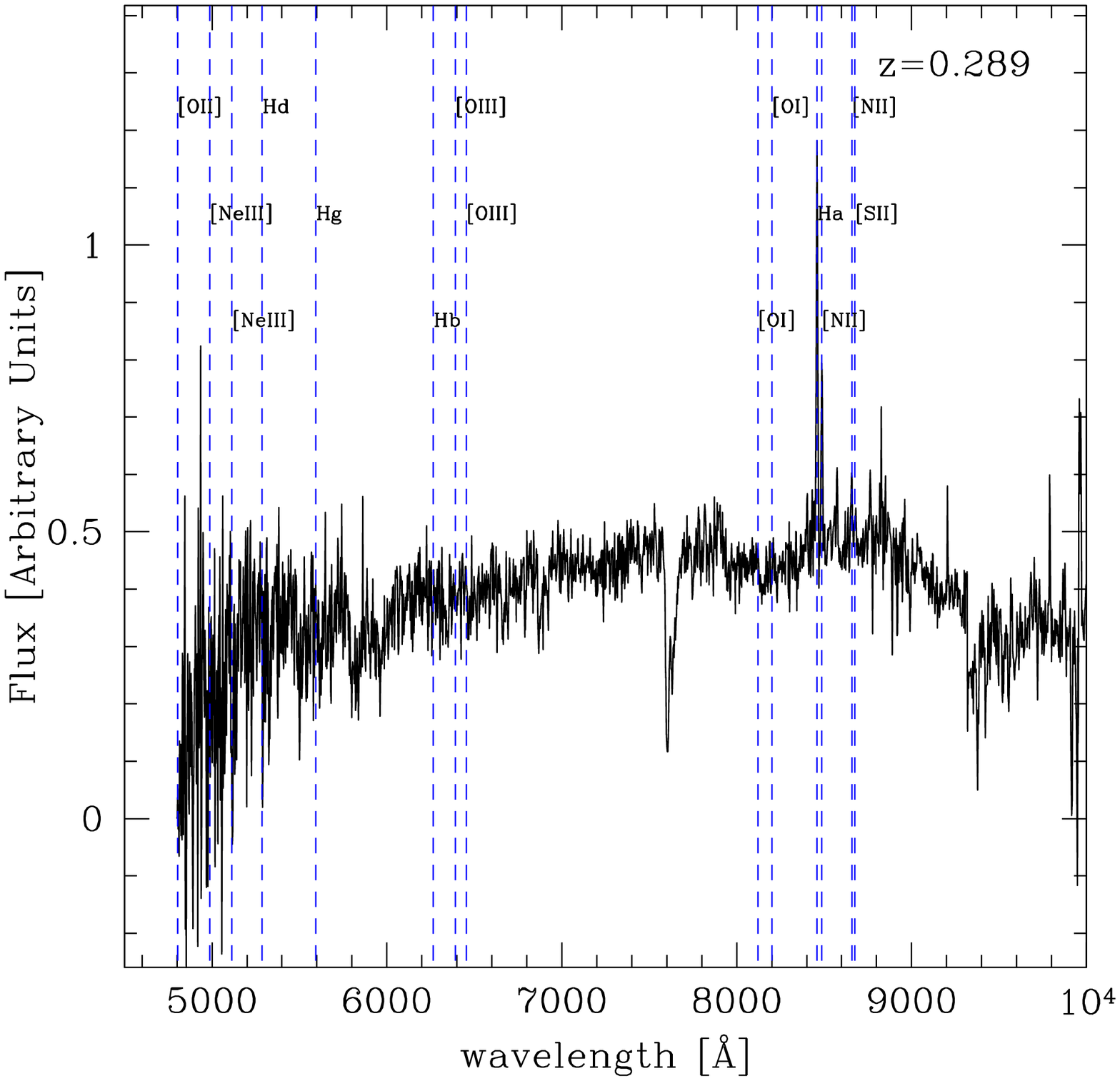}\includegraphics[width=6cm]{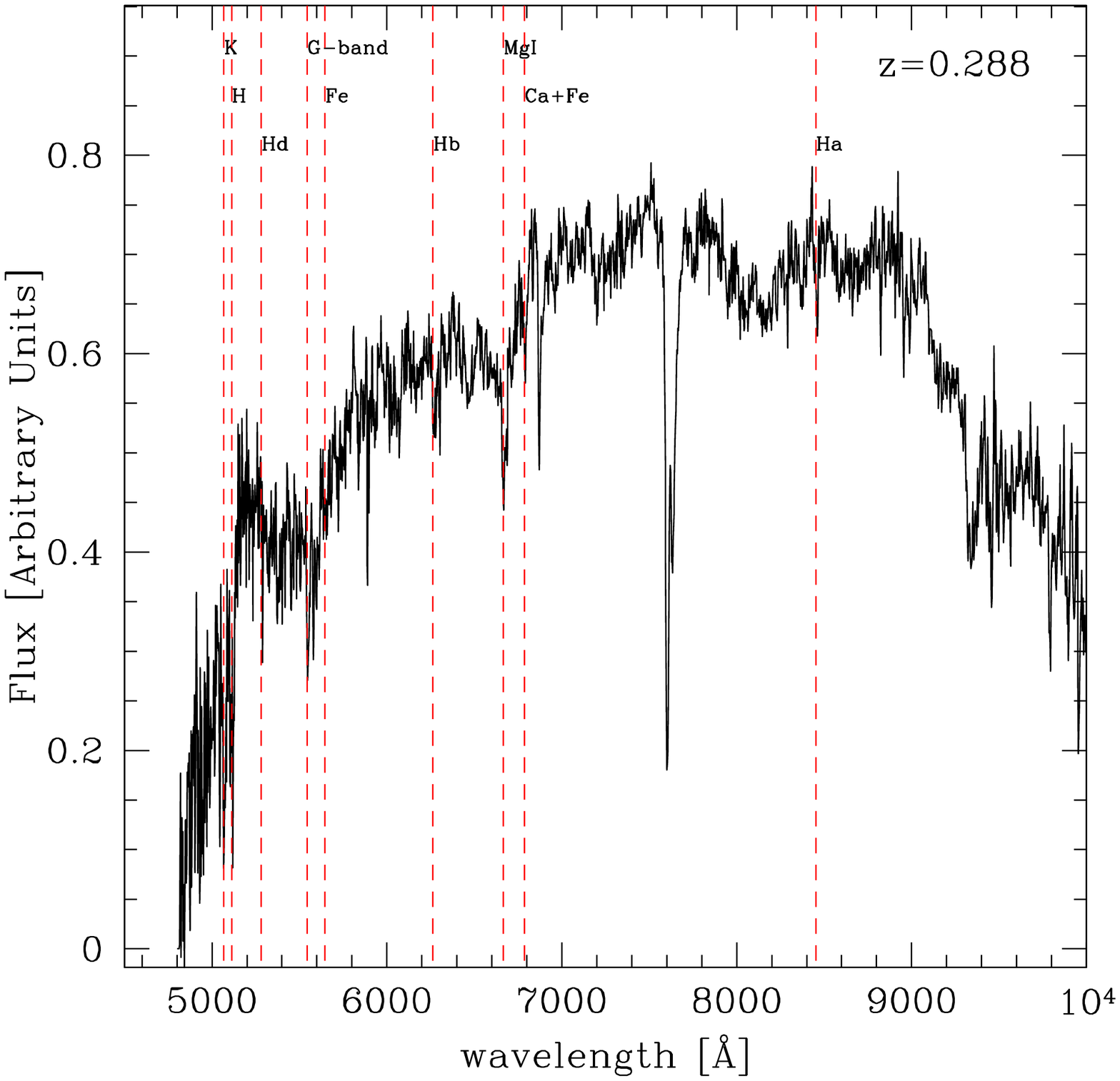}\includegraphics[width=6cm]{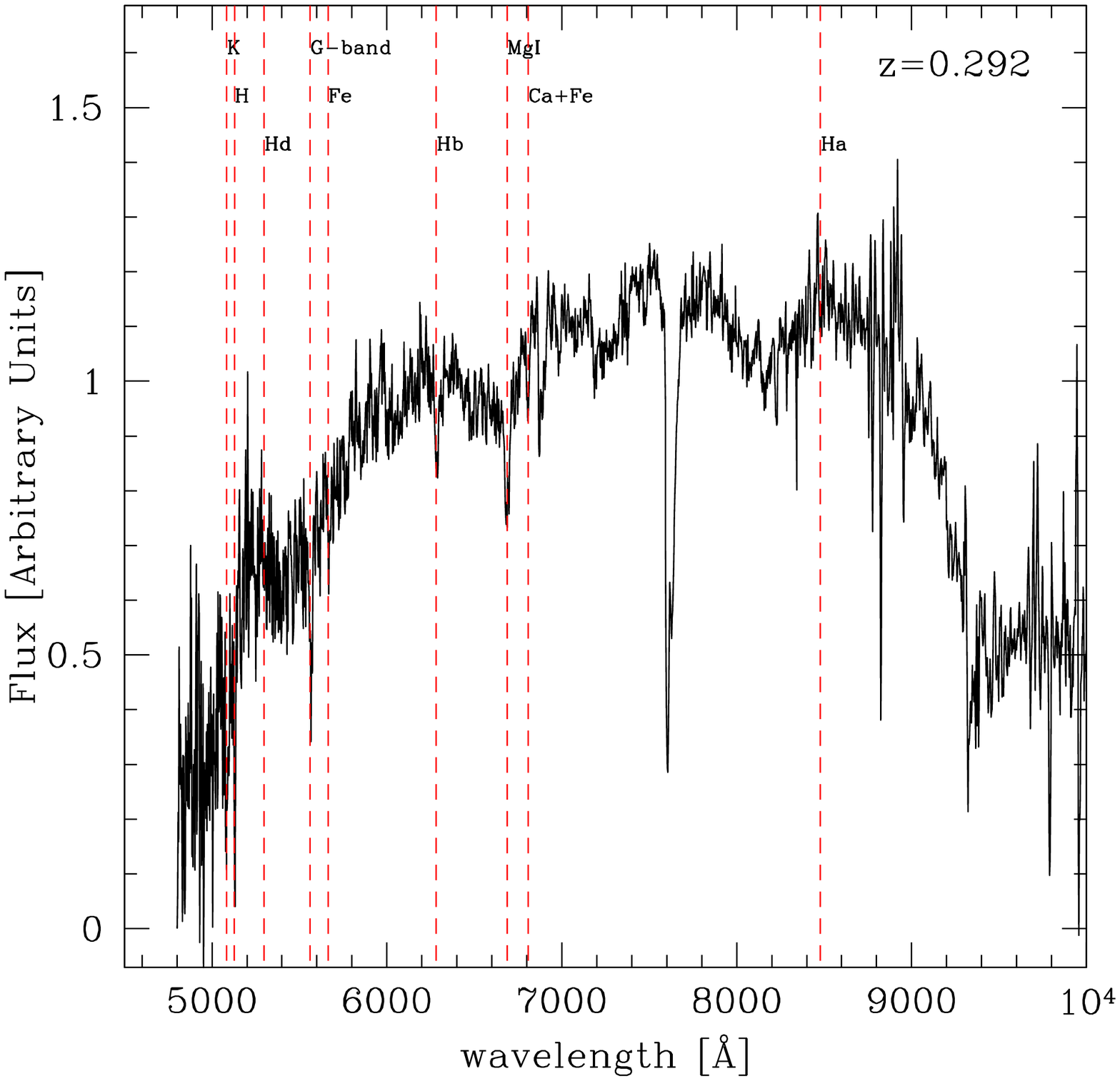}

\includegraphics[width=6cm]{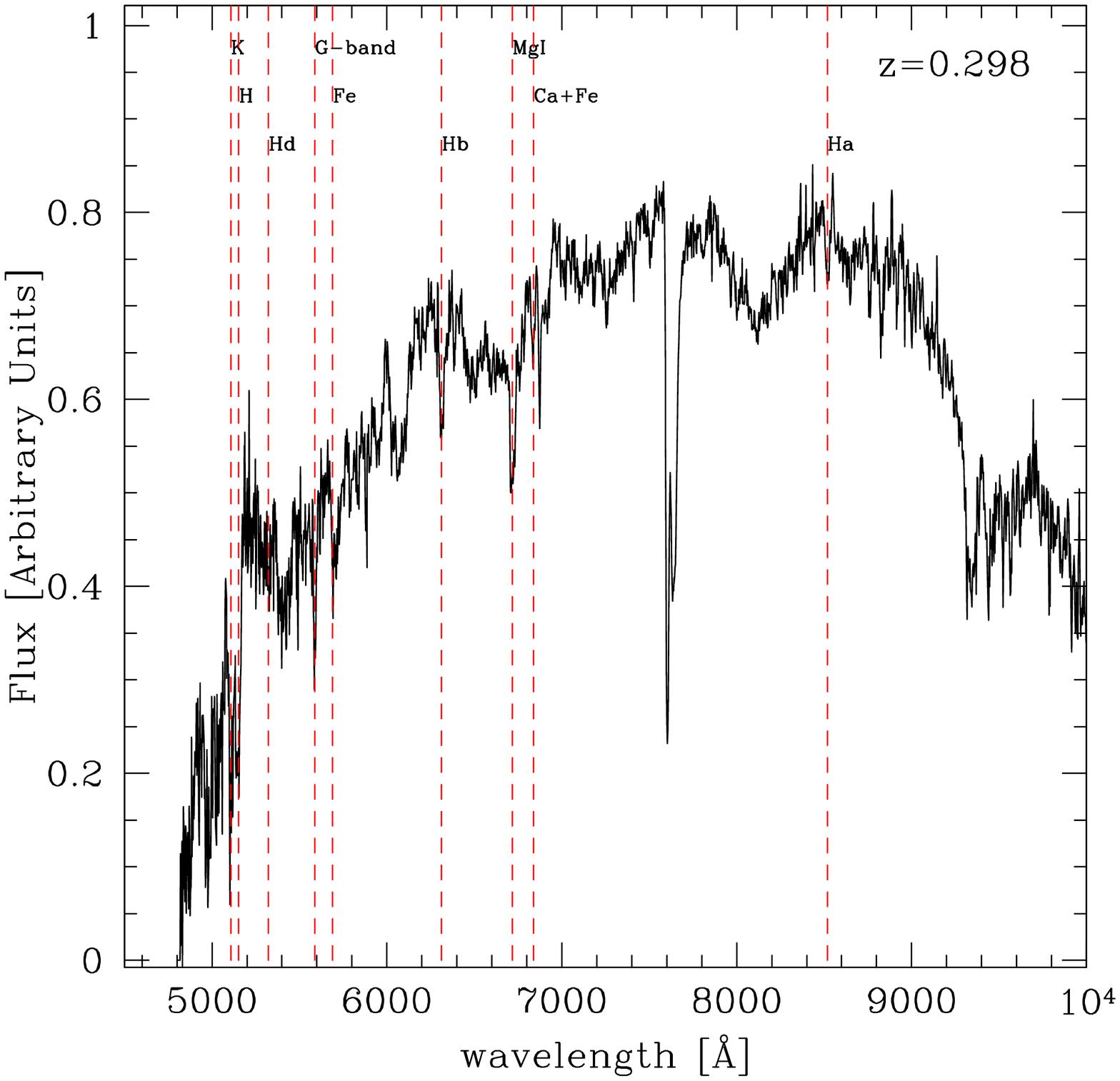}\includegraphics[width=6cm]{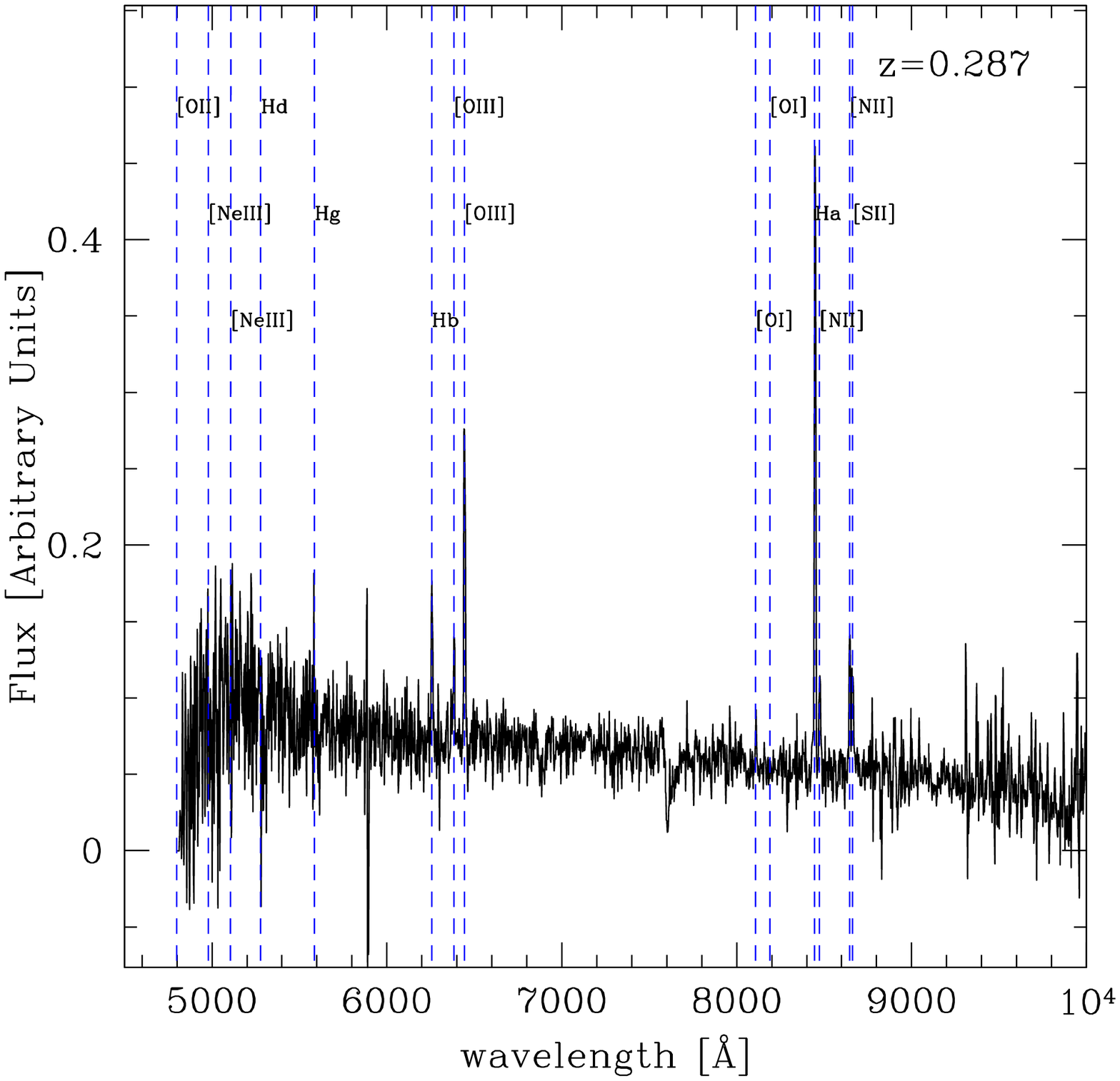}\includegraphics[width=6cm]{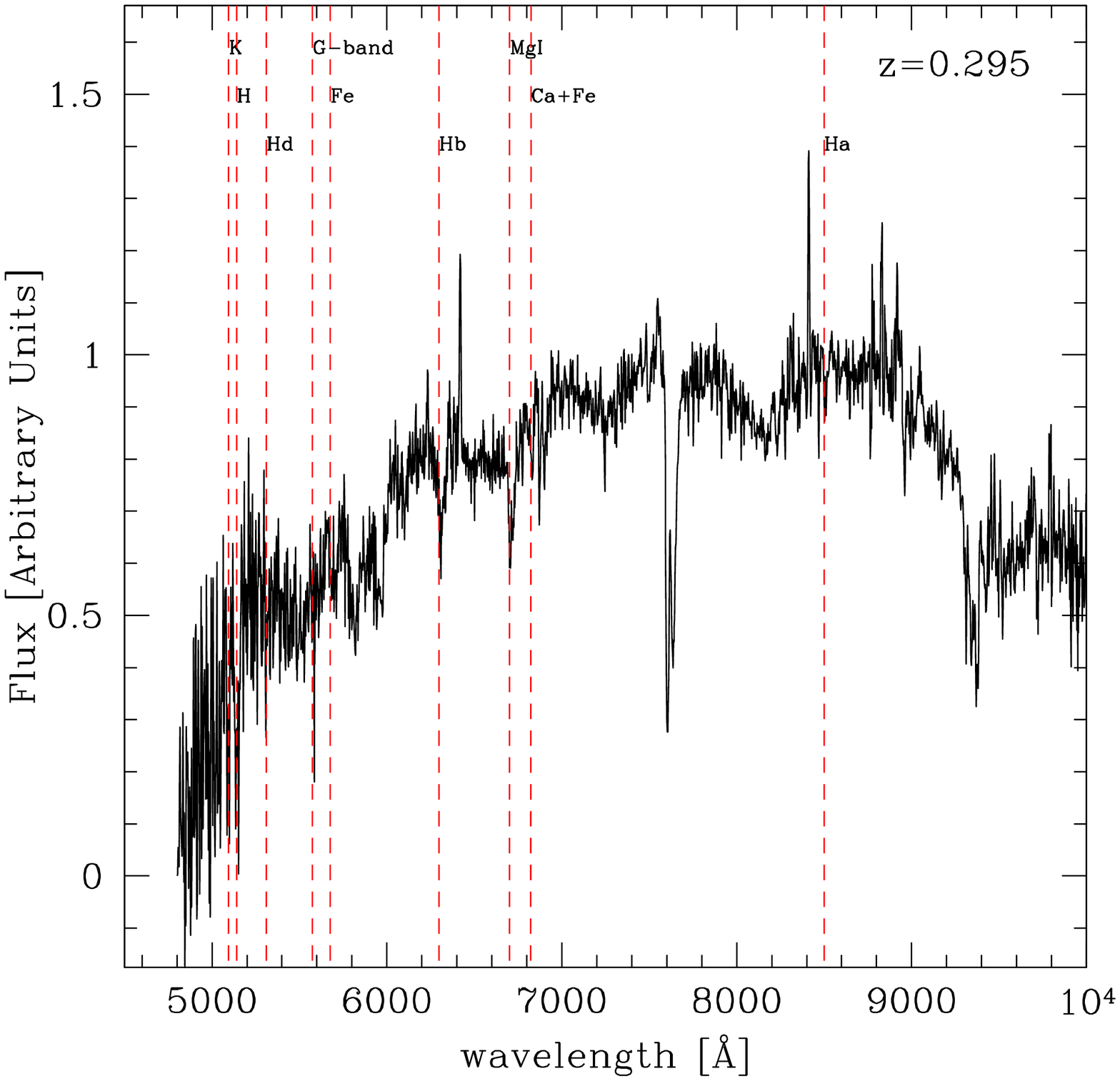} 

\includegraphics[width=6cm]{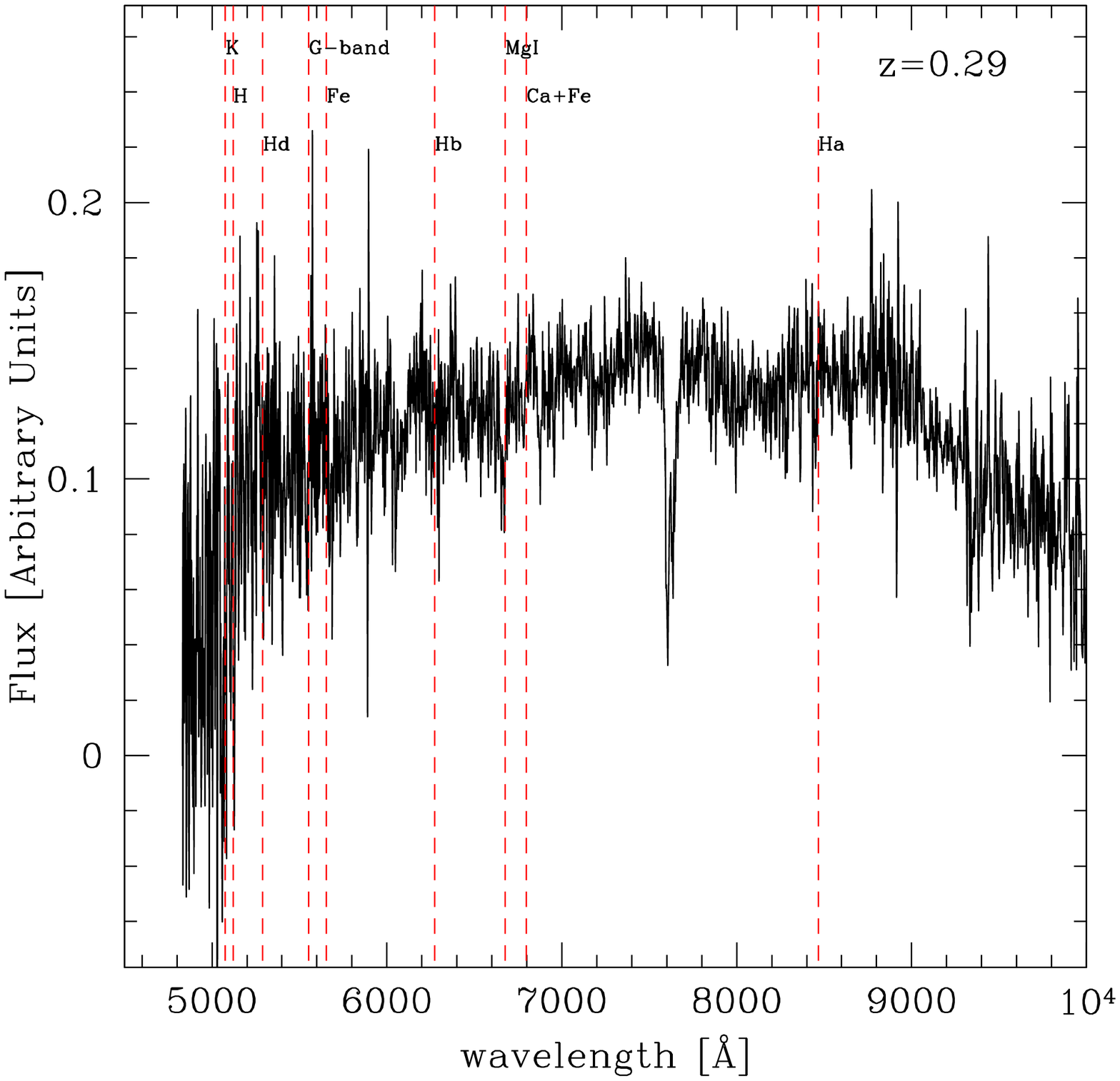}\includegraphics[width=6cm]{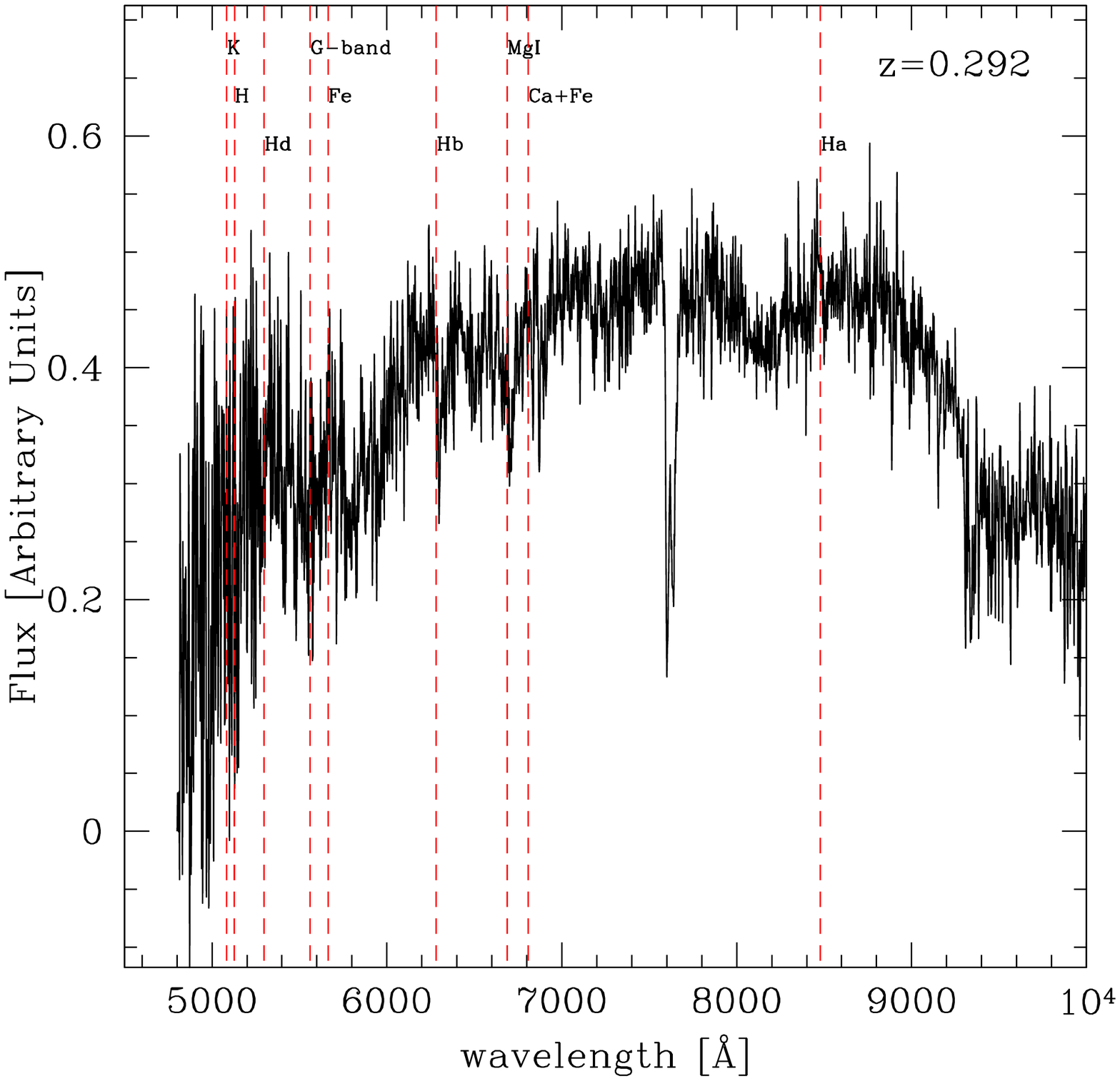}\includegraphics[width=6cm]{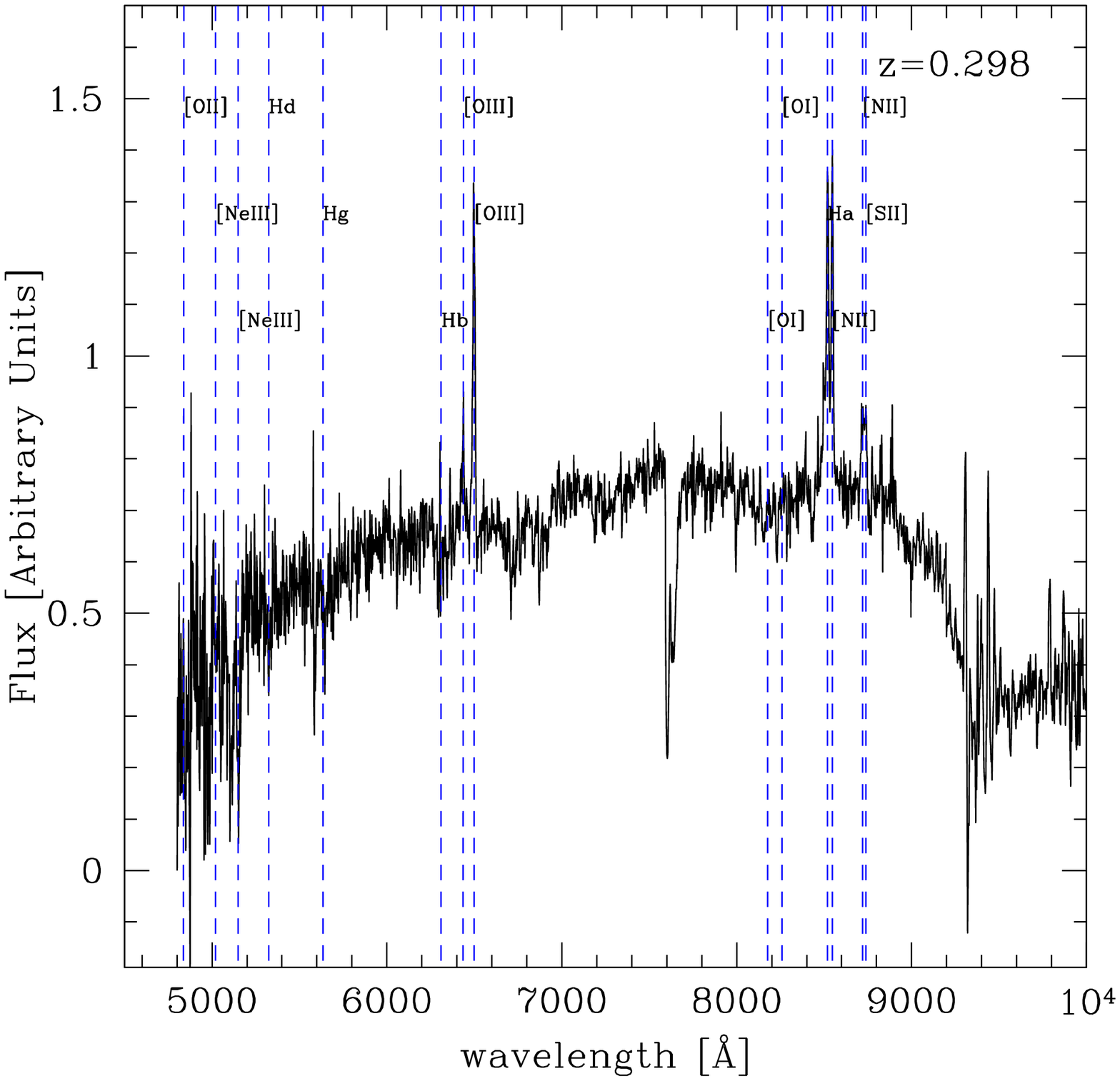} 

\includegraphics[width=6cm]{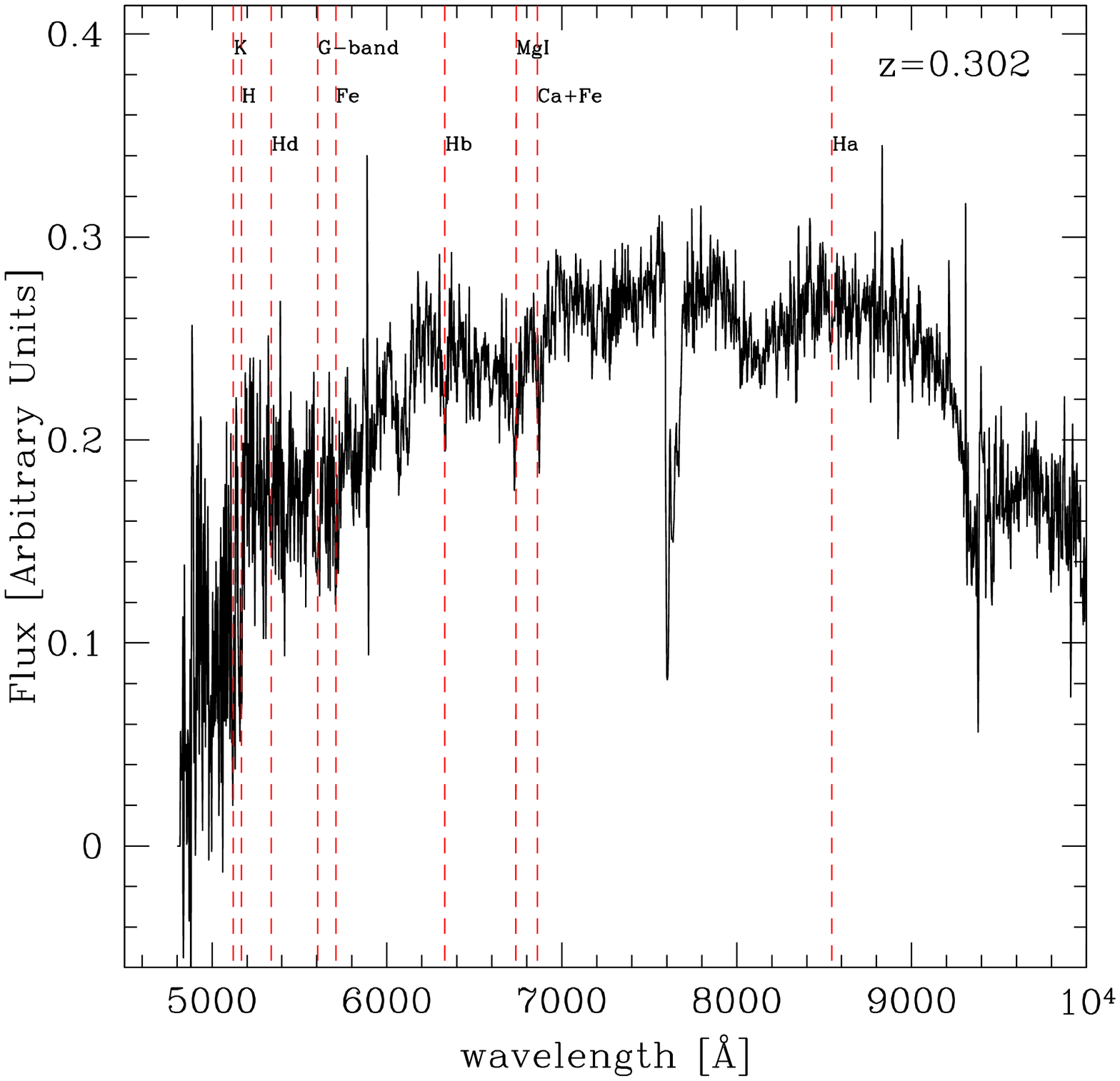}\includegraphics[width=6cm]{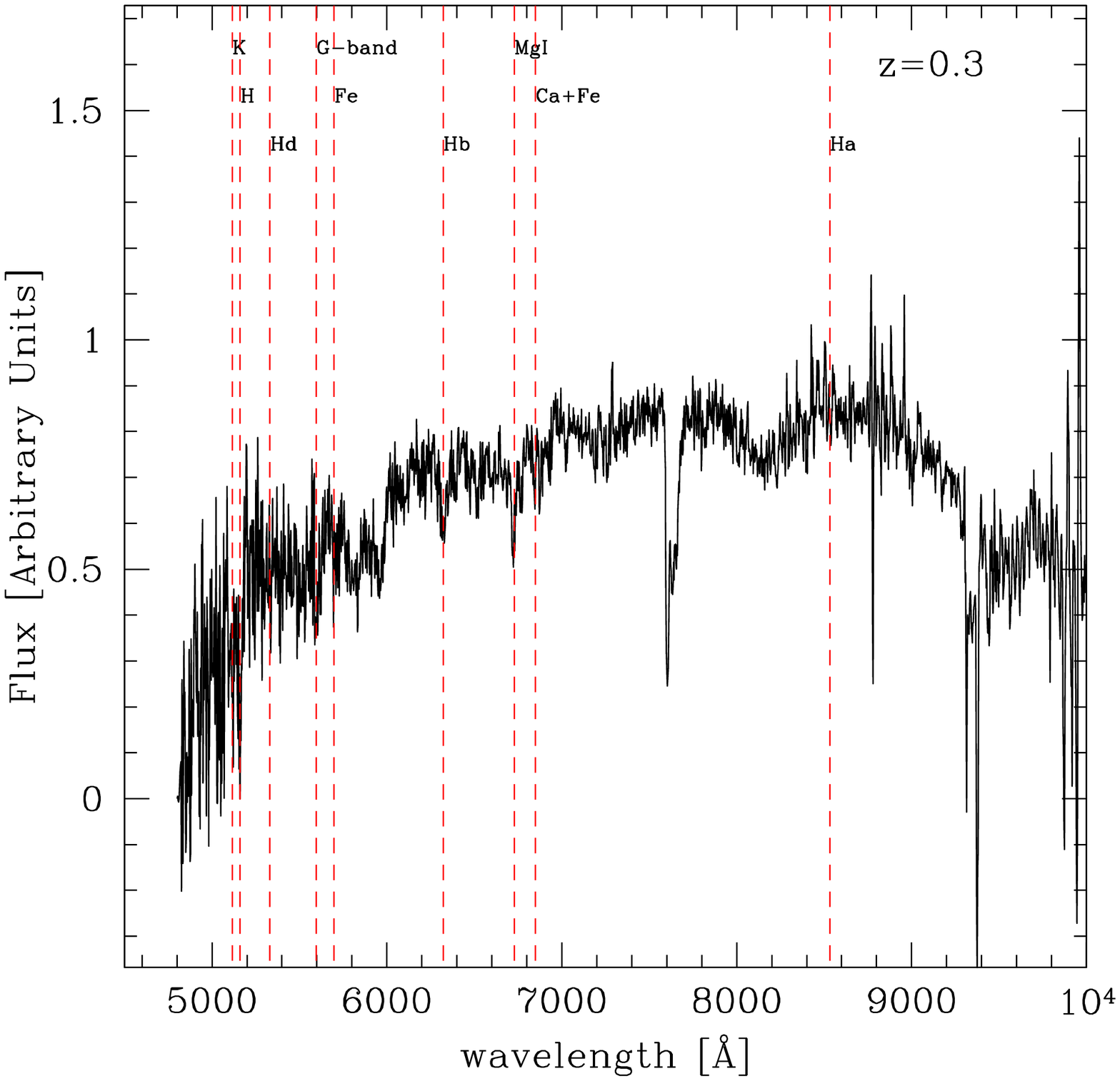}\includegraphics[width=6cm]{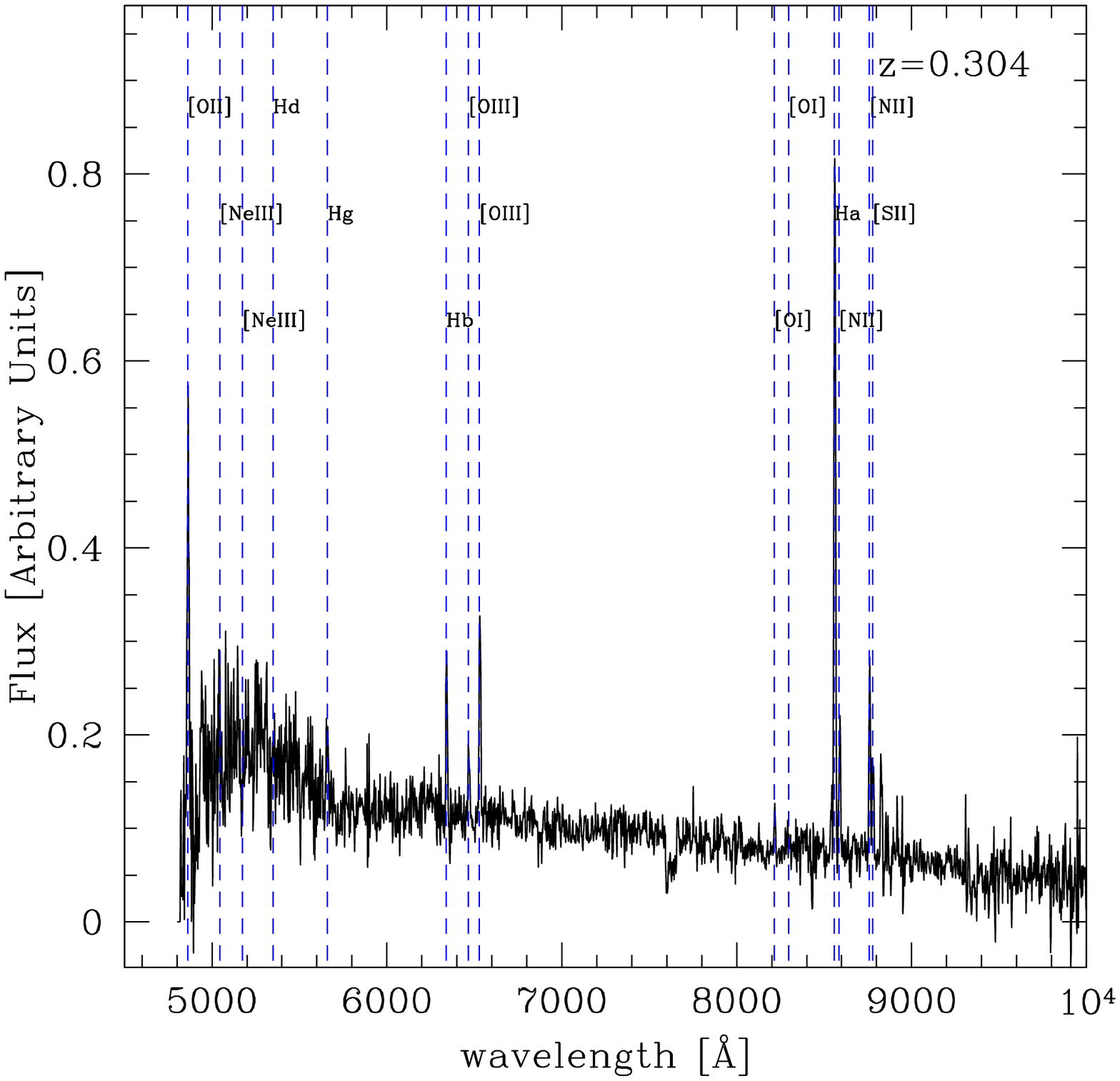} 

\caption{VIMOS optical spectra of 49 cluster members (see
  Tab. \ref{souzbullet}). From top to bottom and from left to right
  spectrum of sources \#21, 22, 23, 24, 25, 26, 29, 31, 32, 33, 34, 35.
\label{spettri}}
\end{figure*}

\begin{figure*}
\ContinuedFloat
\centering
\includegraphics[width=6cm]{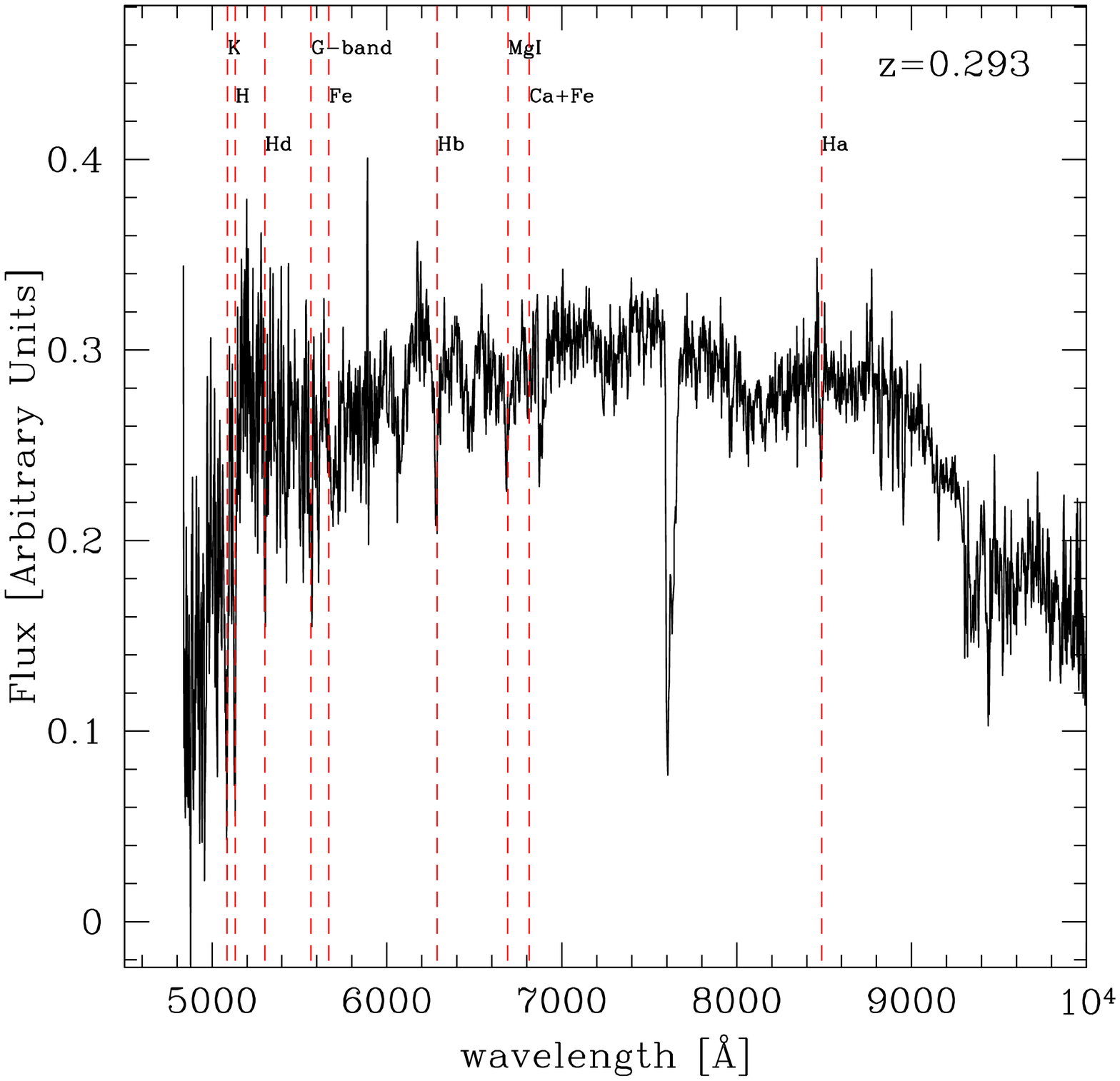}\includegraphics[width=6cm]{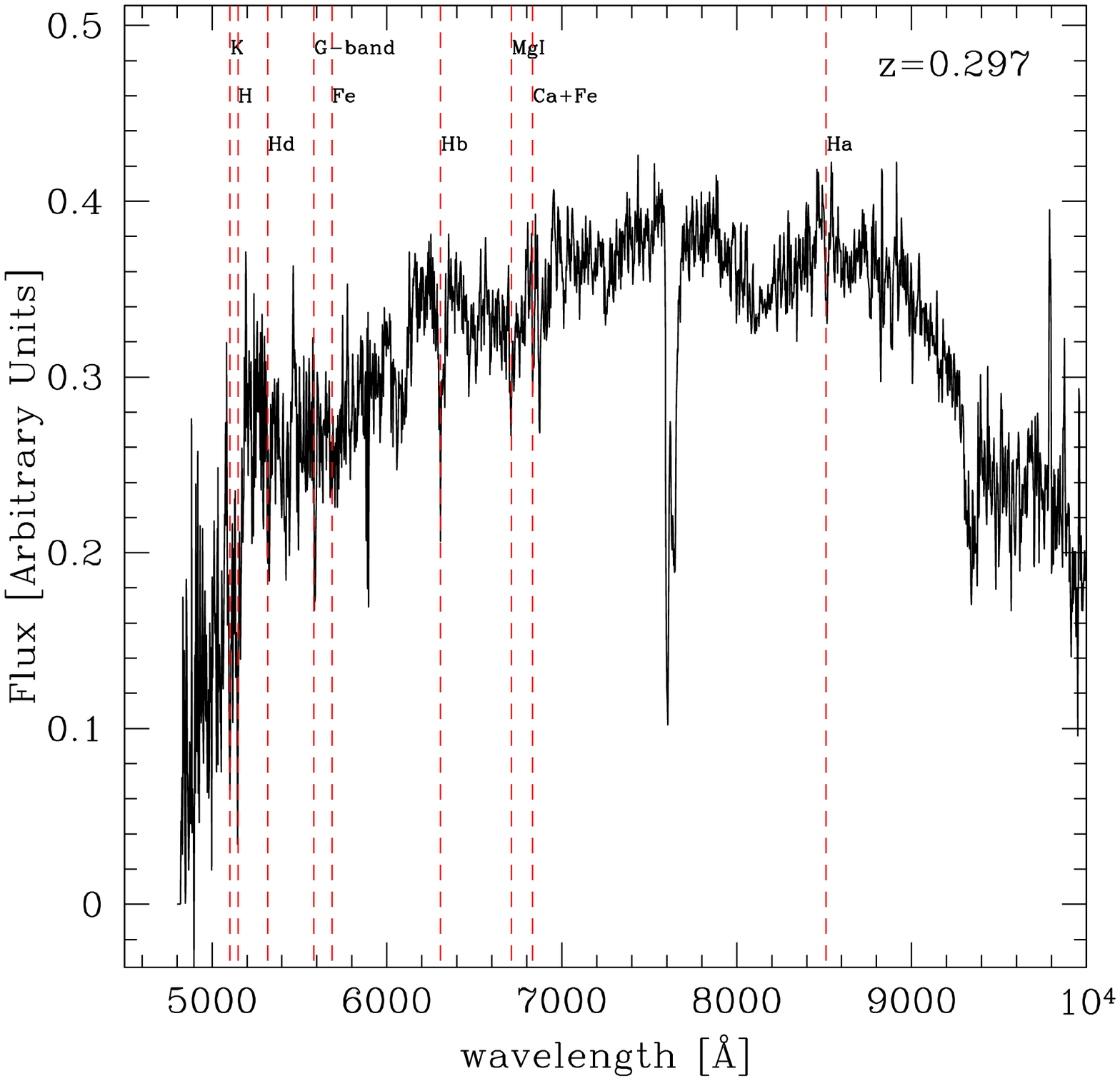}\includegraphics[width=6cm]{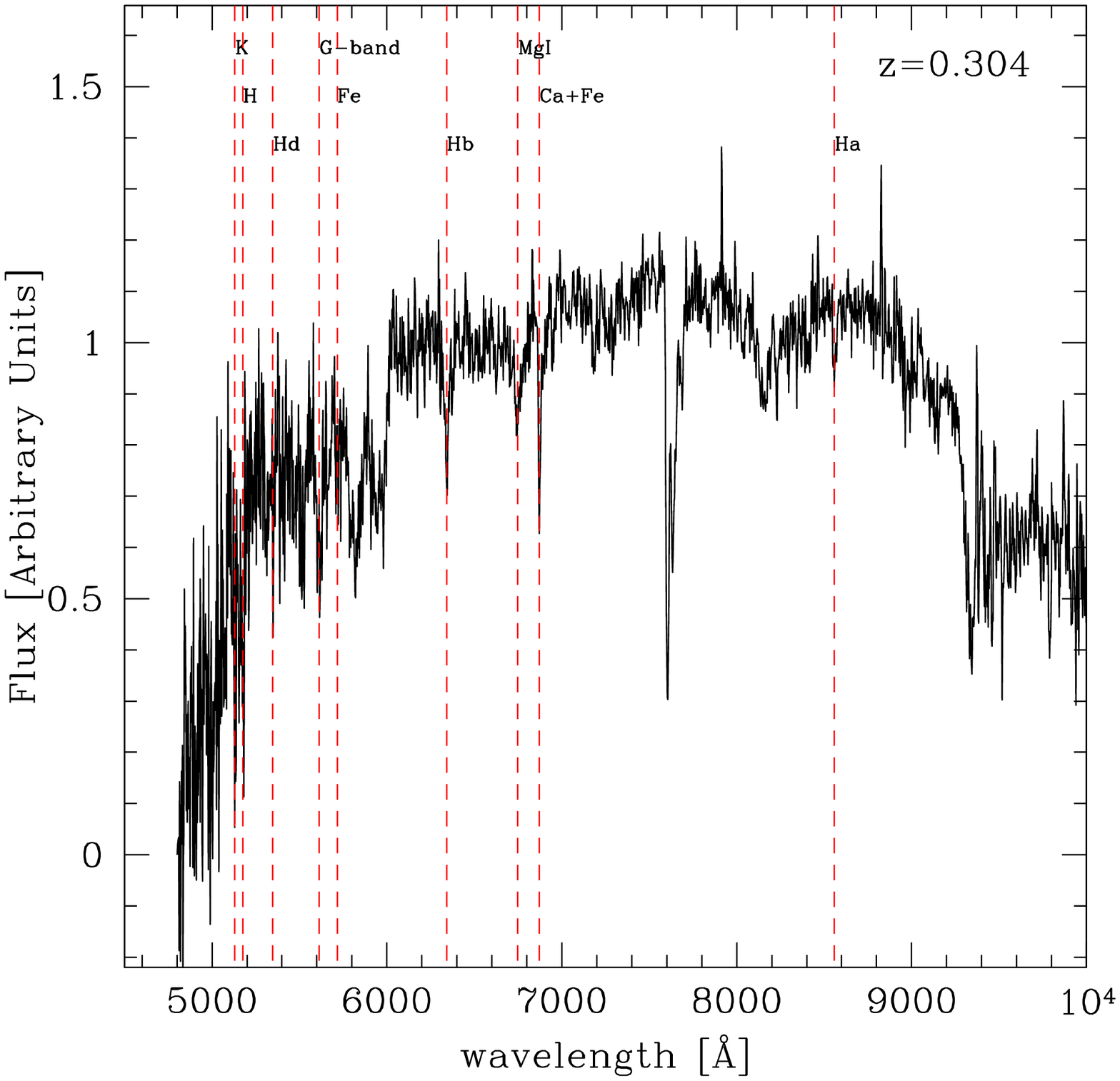}

\includegraphics[width=6cm]{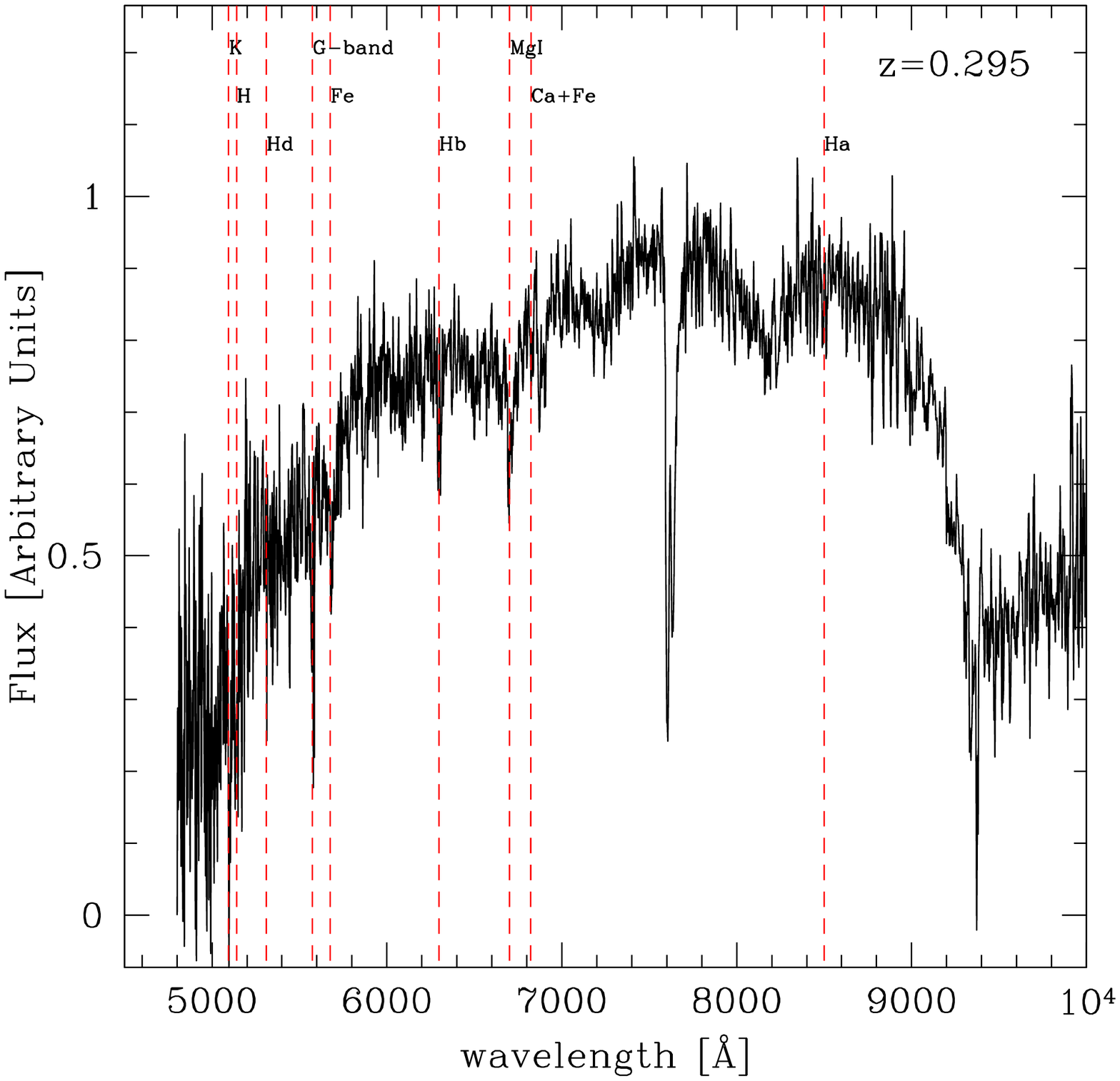}\includegraphics[width=6cm]{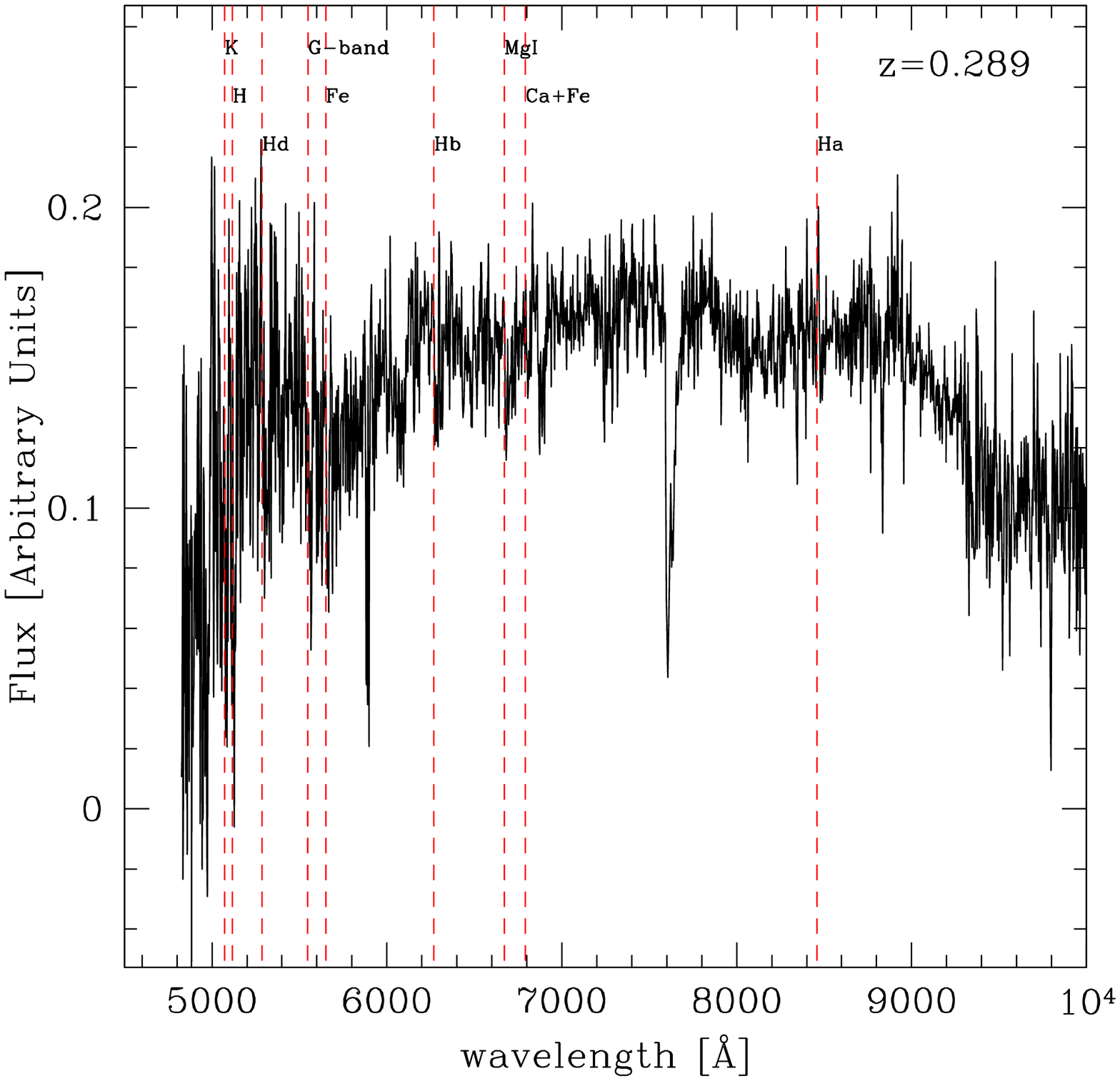}\includegraphics[width=6cm]{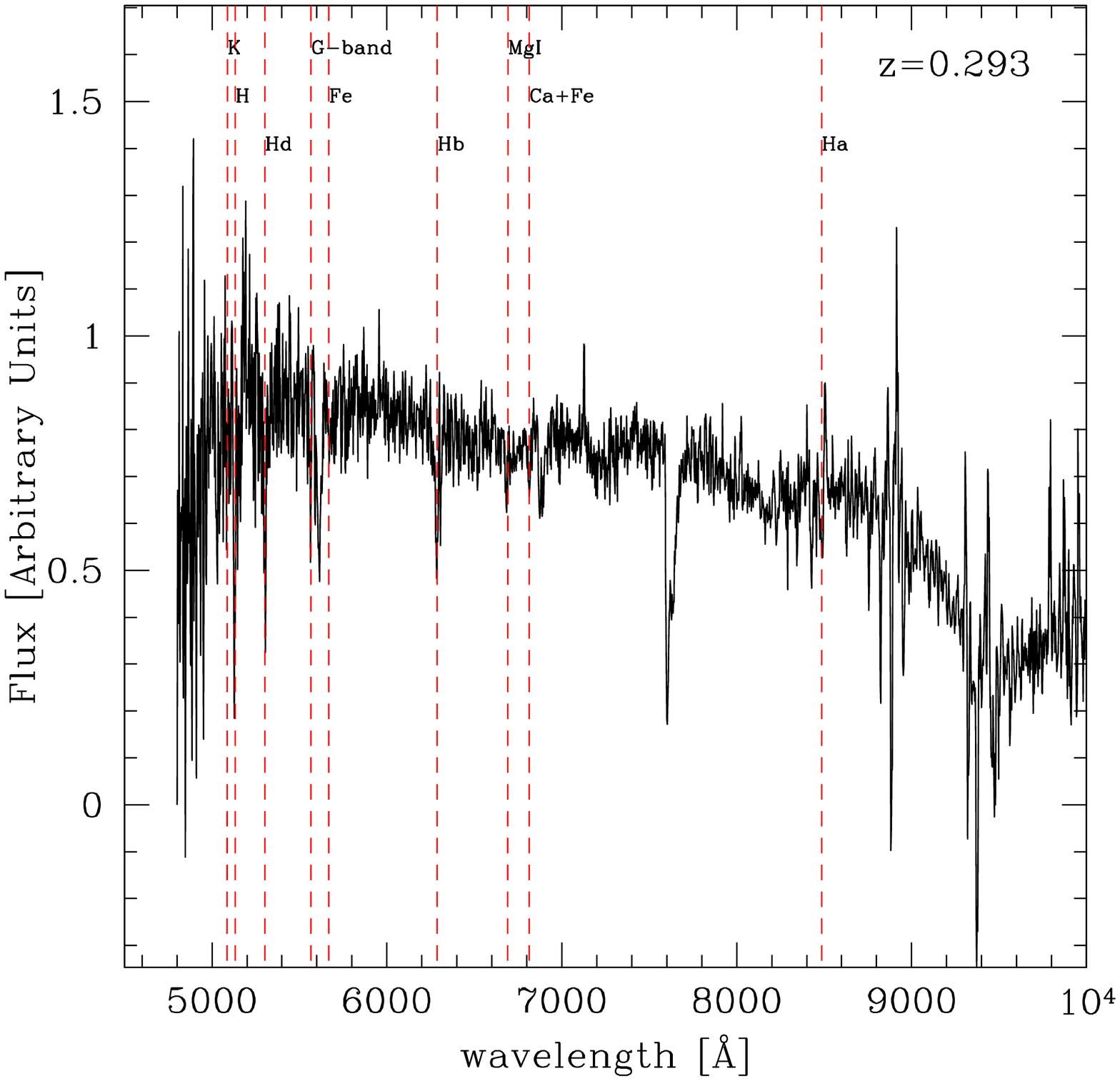} 

\includegraphics[width=6cm]{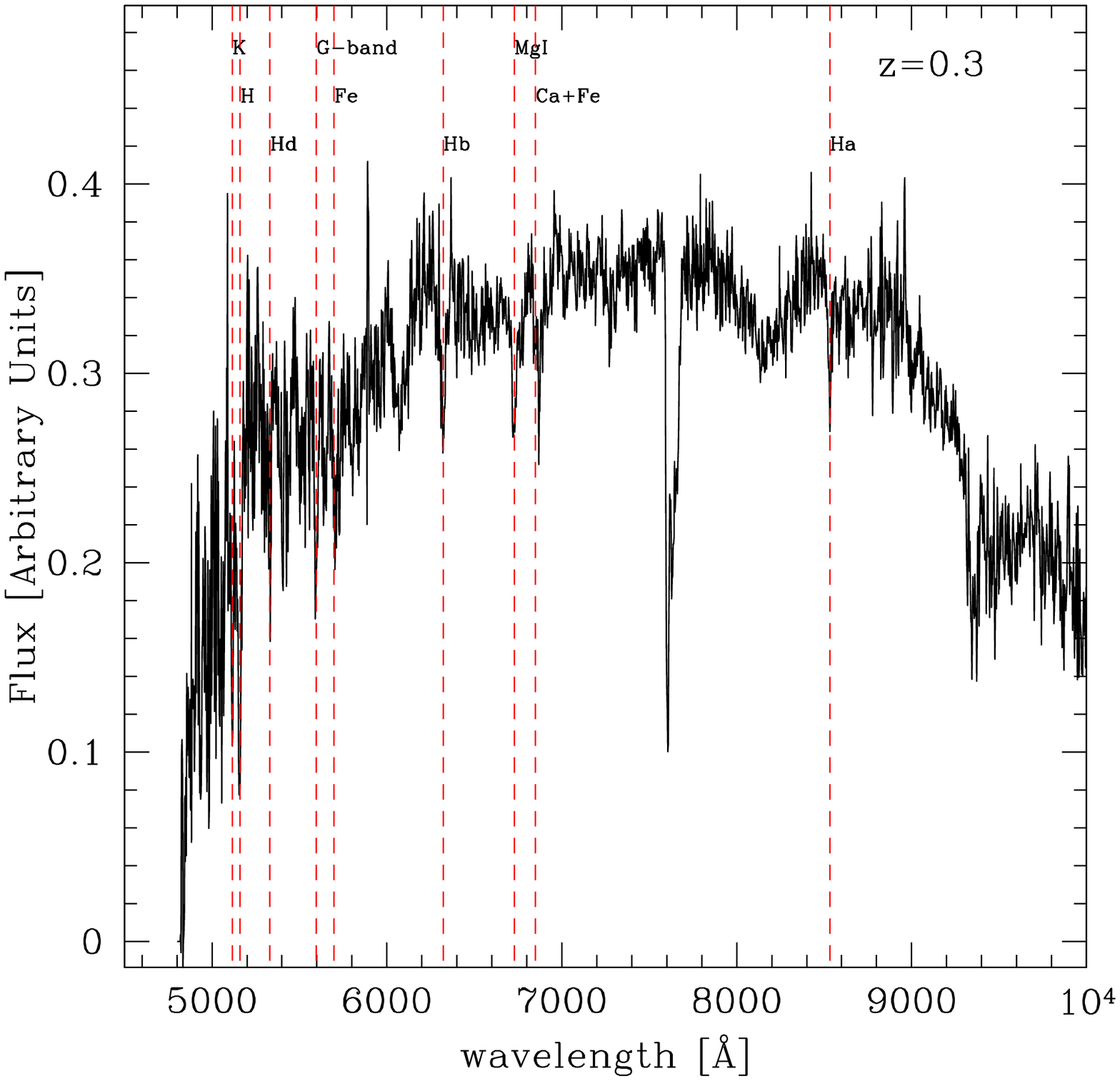}\includegraphics[width=6cm]{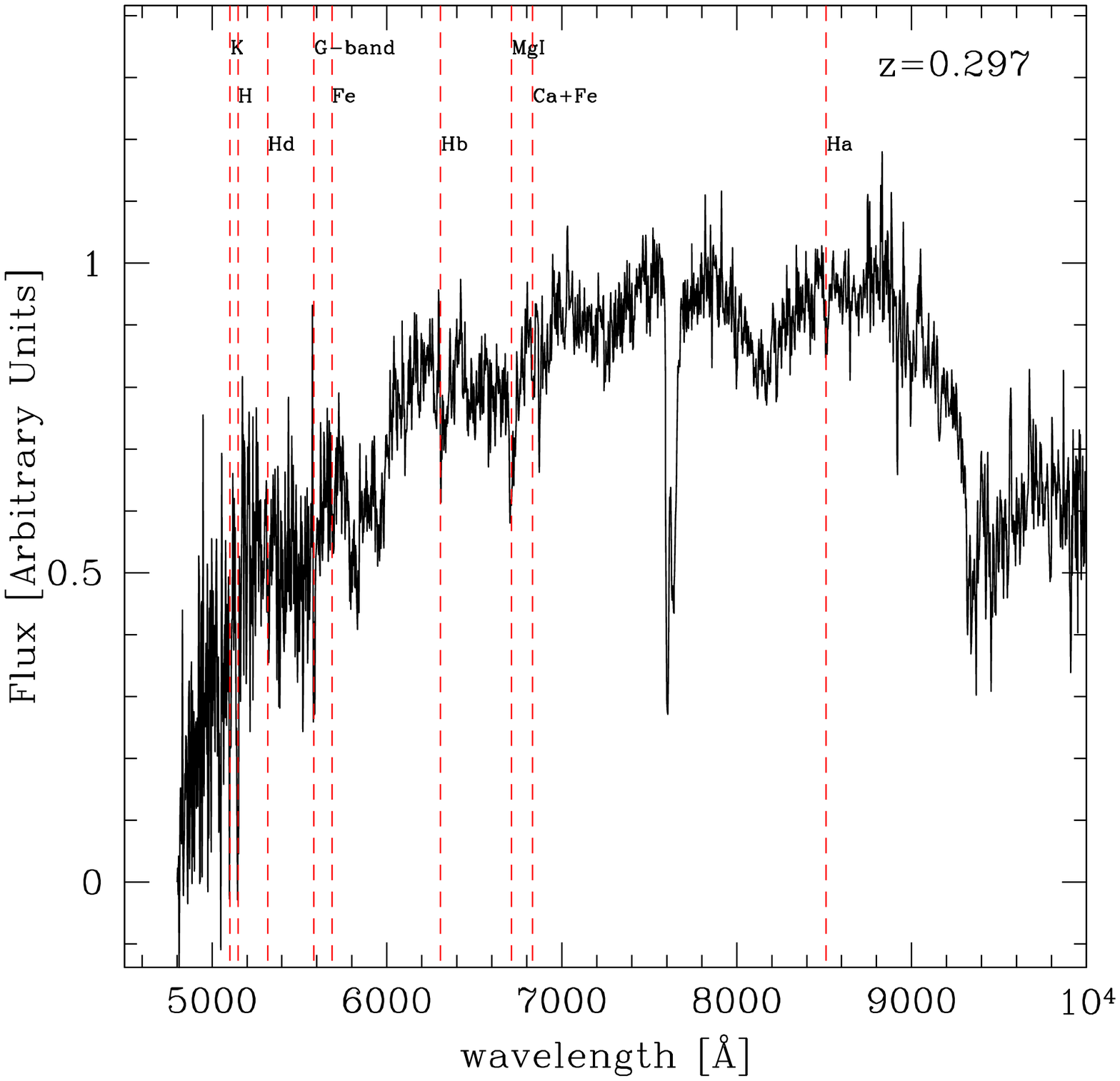}\includegraphics[width=6cm]{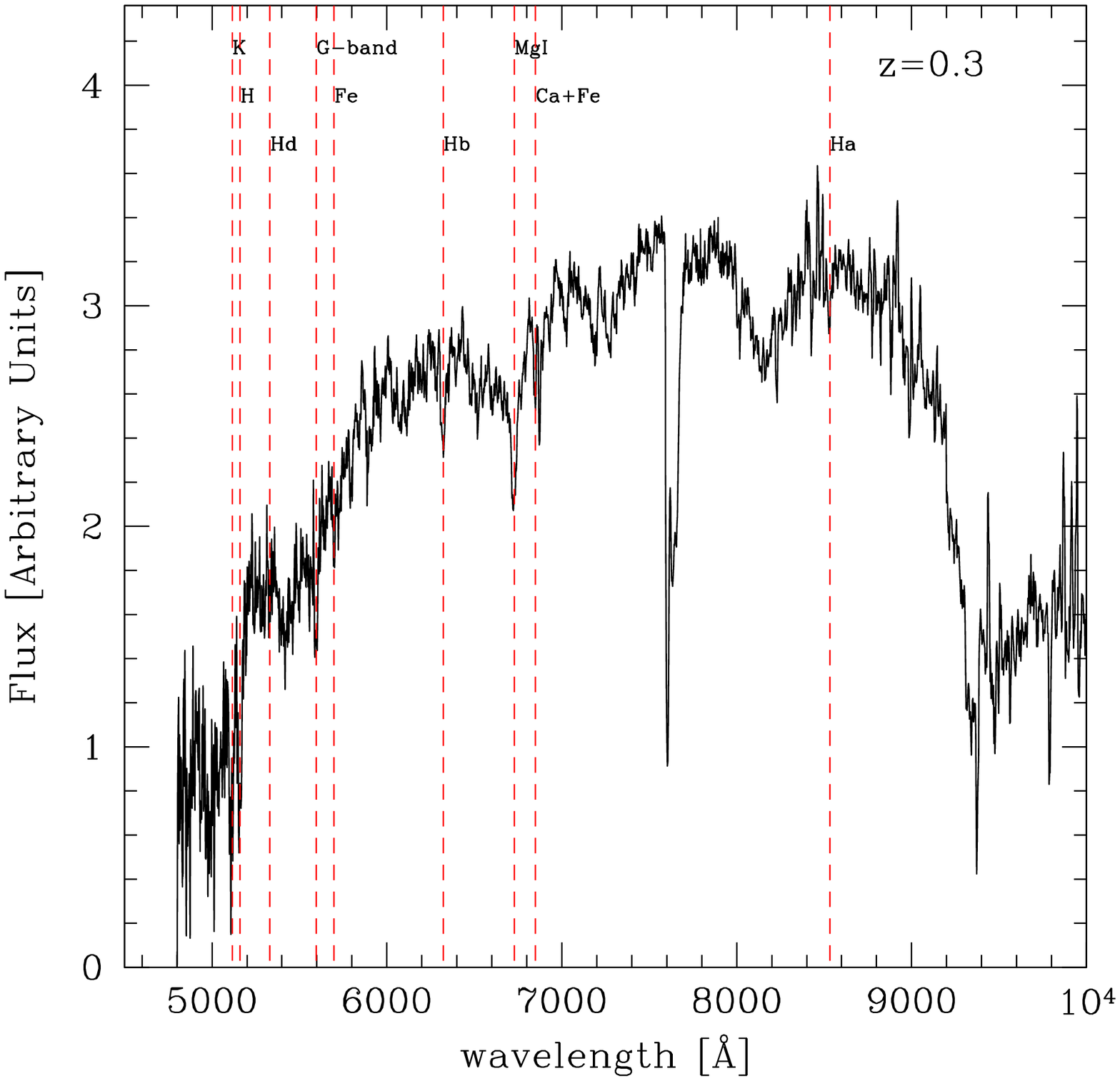} 

\includegraphics[width=6cm]{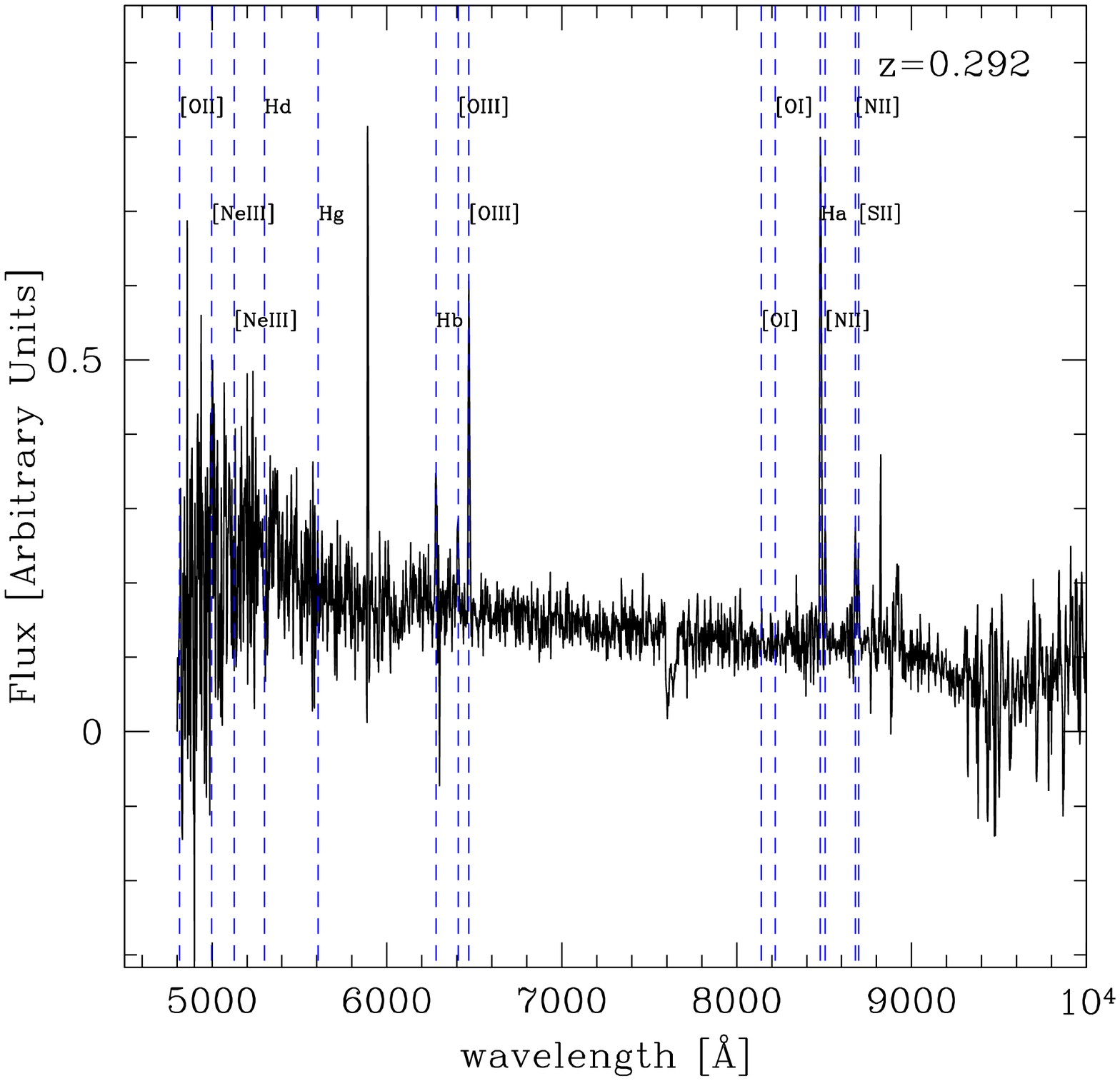}\includegraphics[width=6cm]{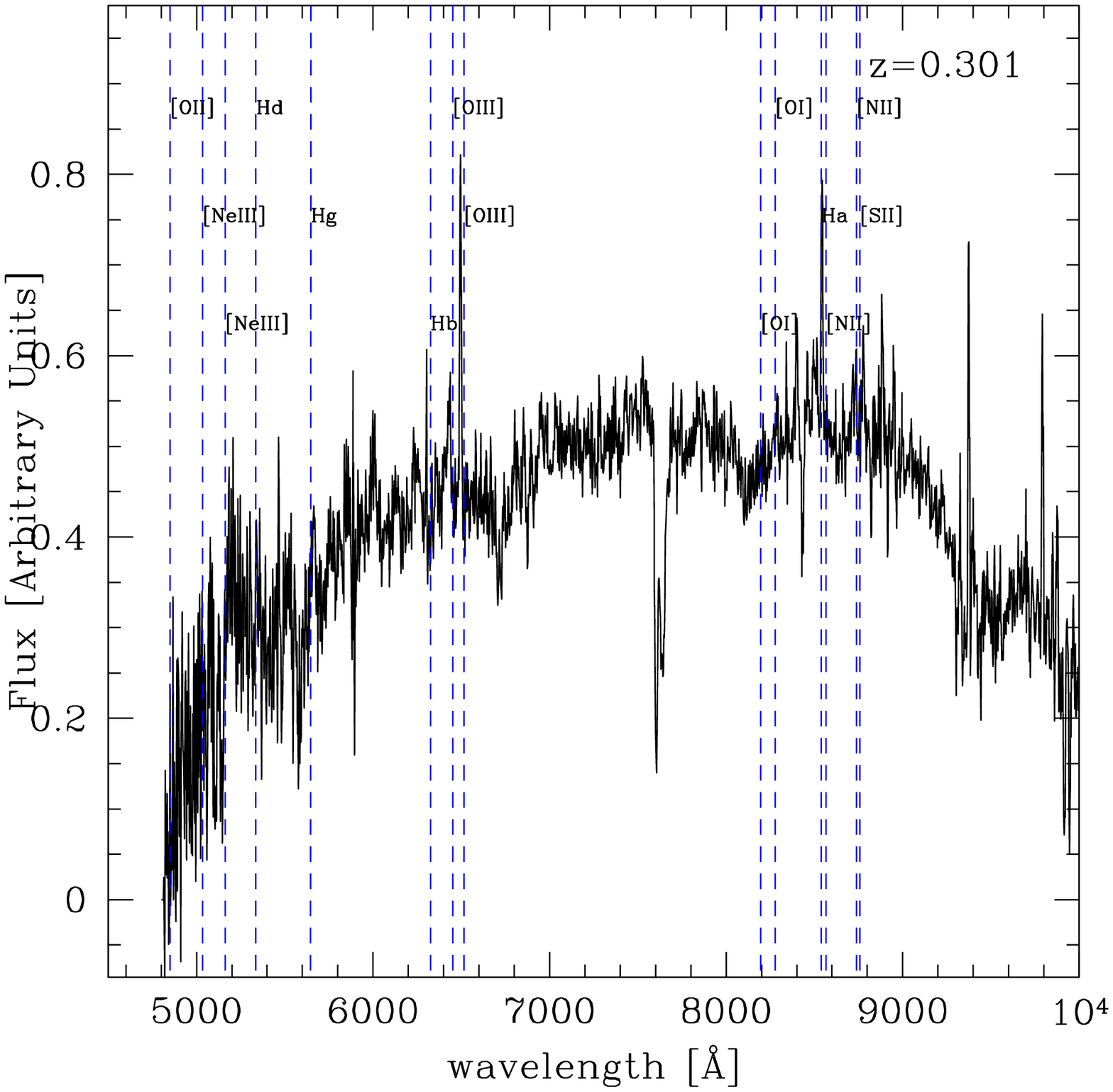}\includegraphics[width=6cm]{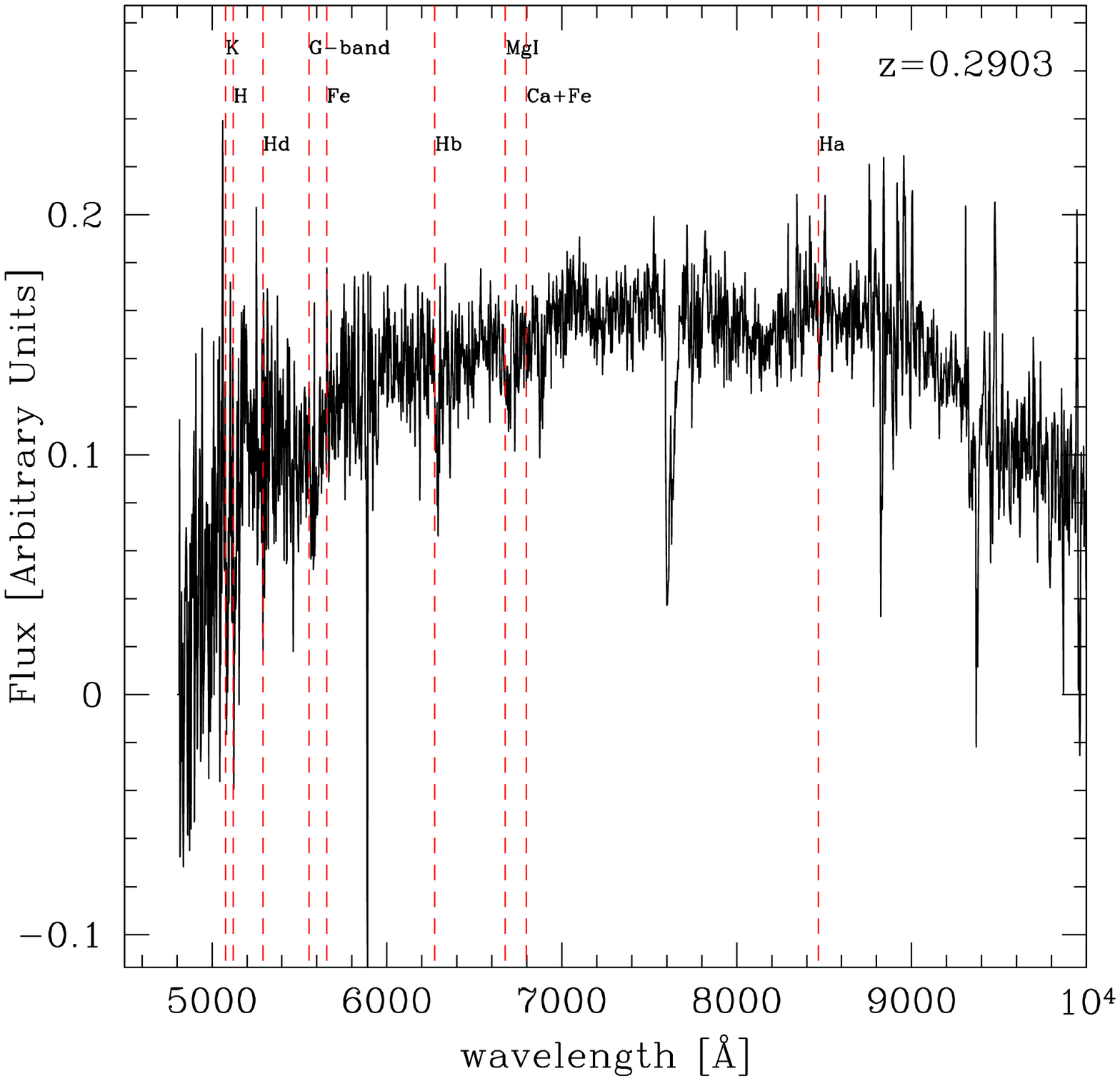} 

\caption{VIMOS optical spectra of 49 cluster members (see
  Tab. \ref{souzbullet}).  From top to bottom and from left to right
  spectrum of sources \#37, 38, 39, 40, 41, 42, 44, 45, 46, 47, 48, 50.
\label{spettri}}
\end{figure*}

\begin{figure*}
\ContinuedFloat
\centering
\includegraphics[width=6cm]{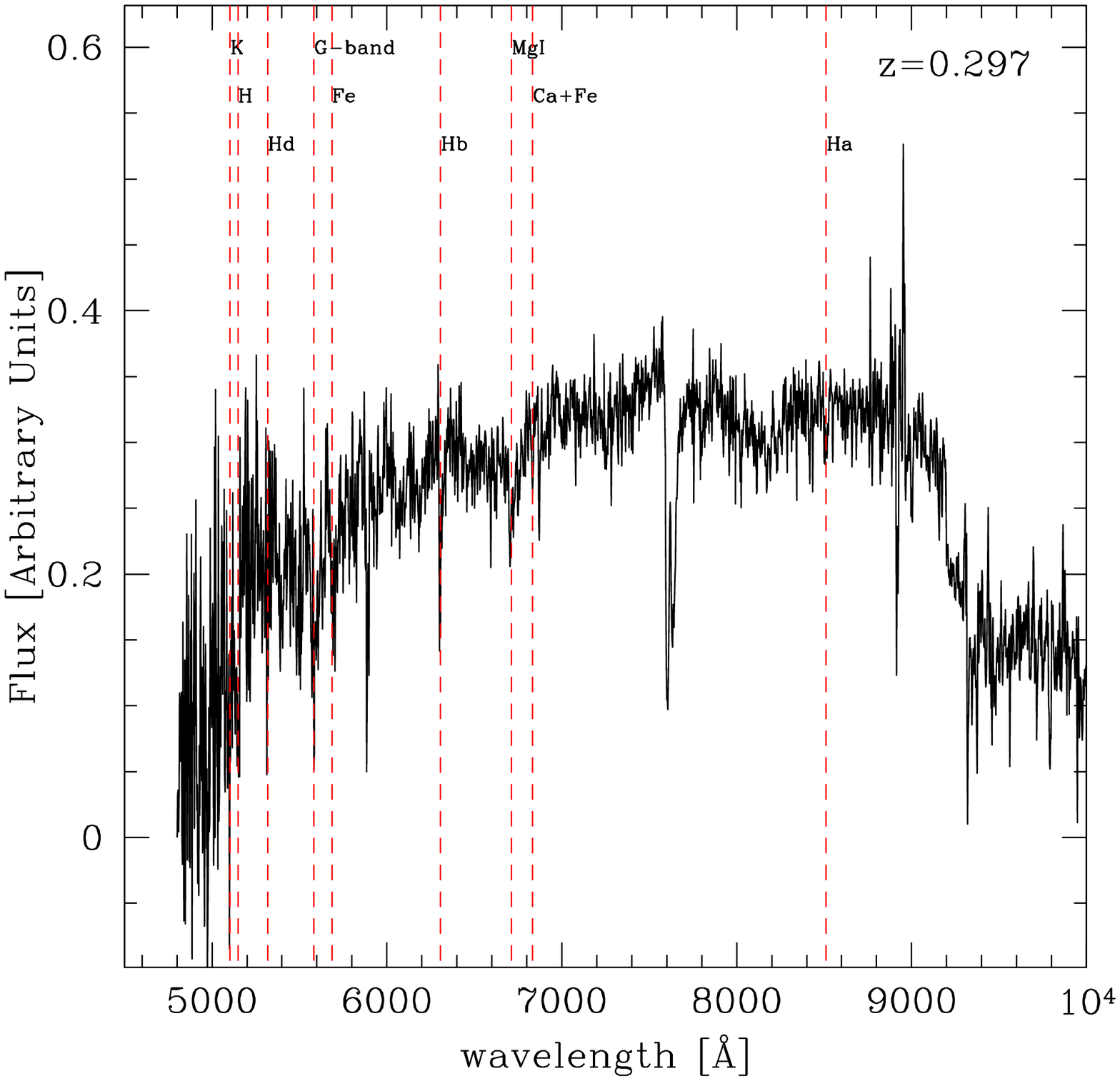}\includegraphics[width=6cm]{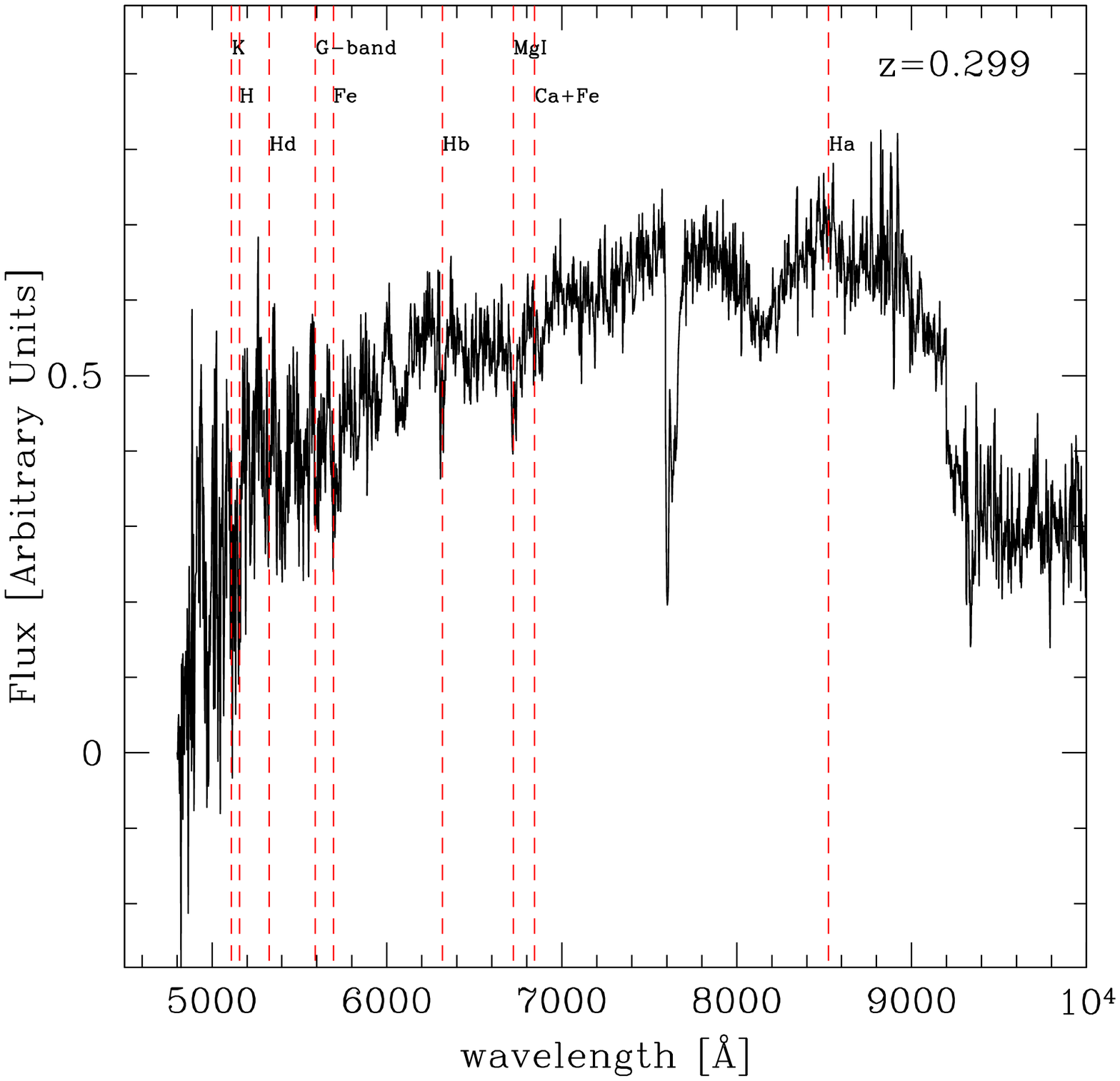}\includegraphics[width=6cm]{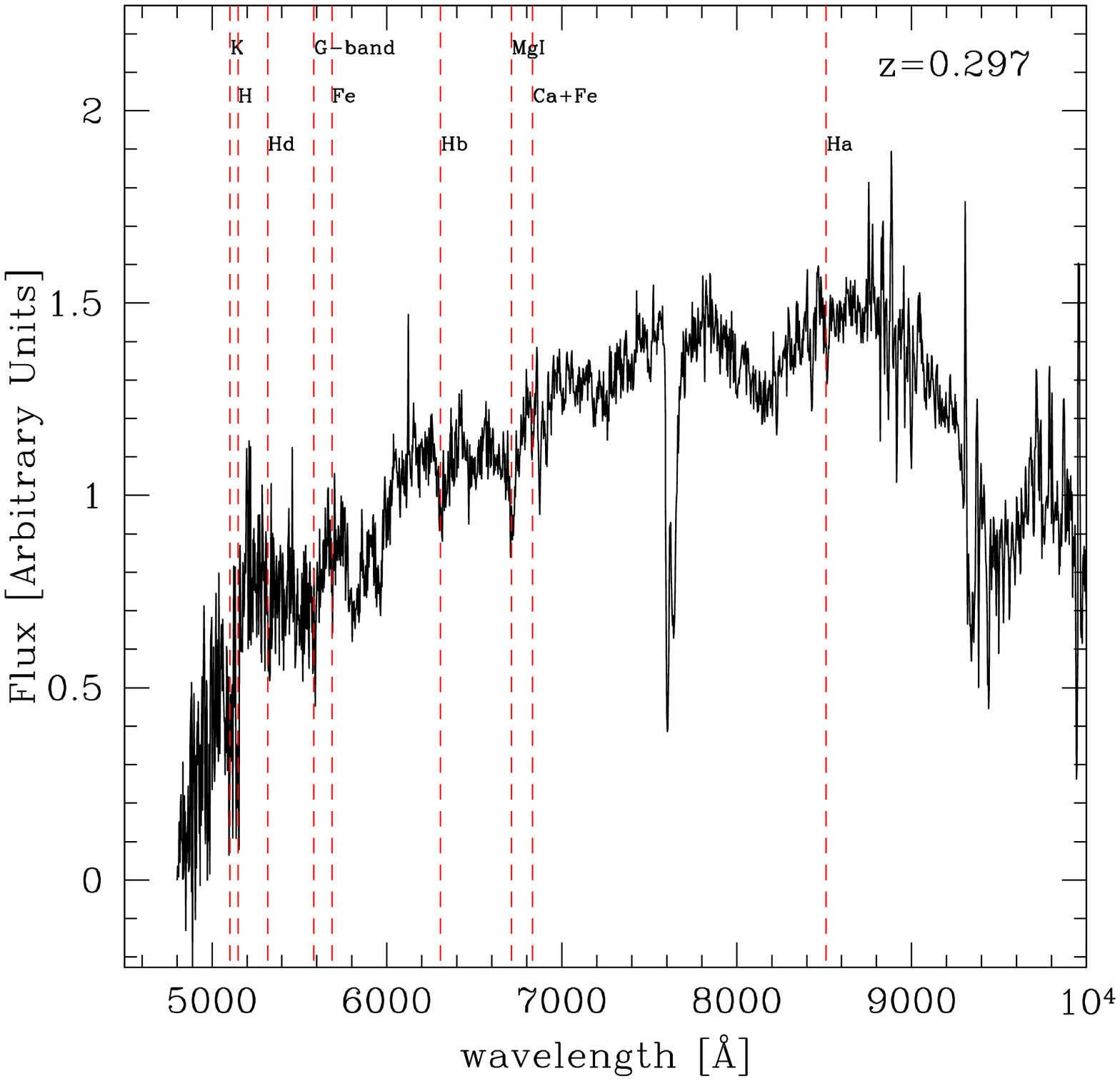}

\includegraphics[width=6cm]{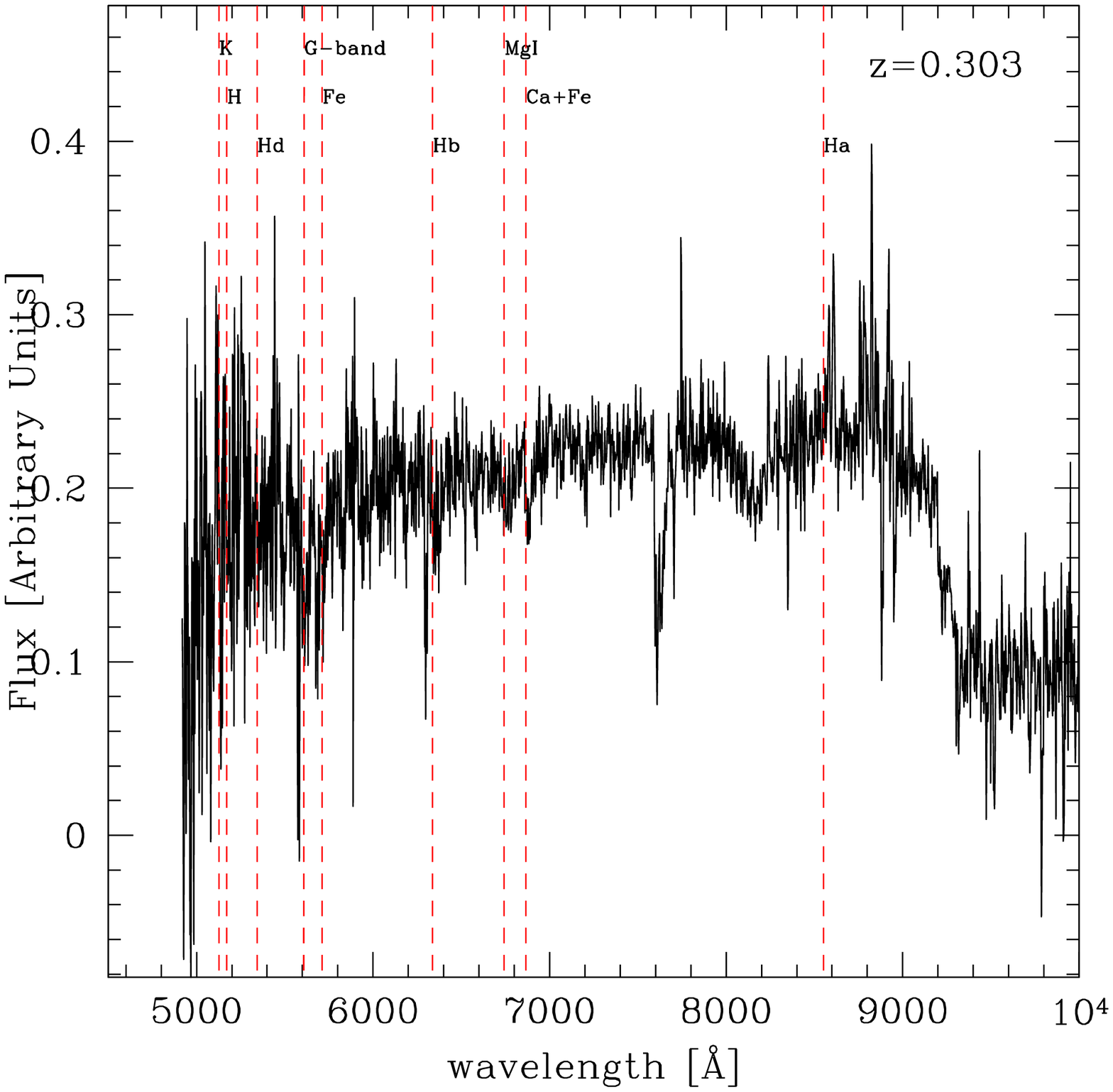}\includegraphics[width=6cm]{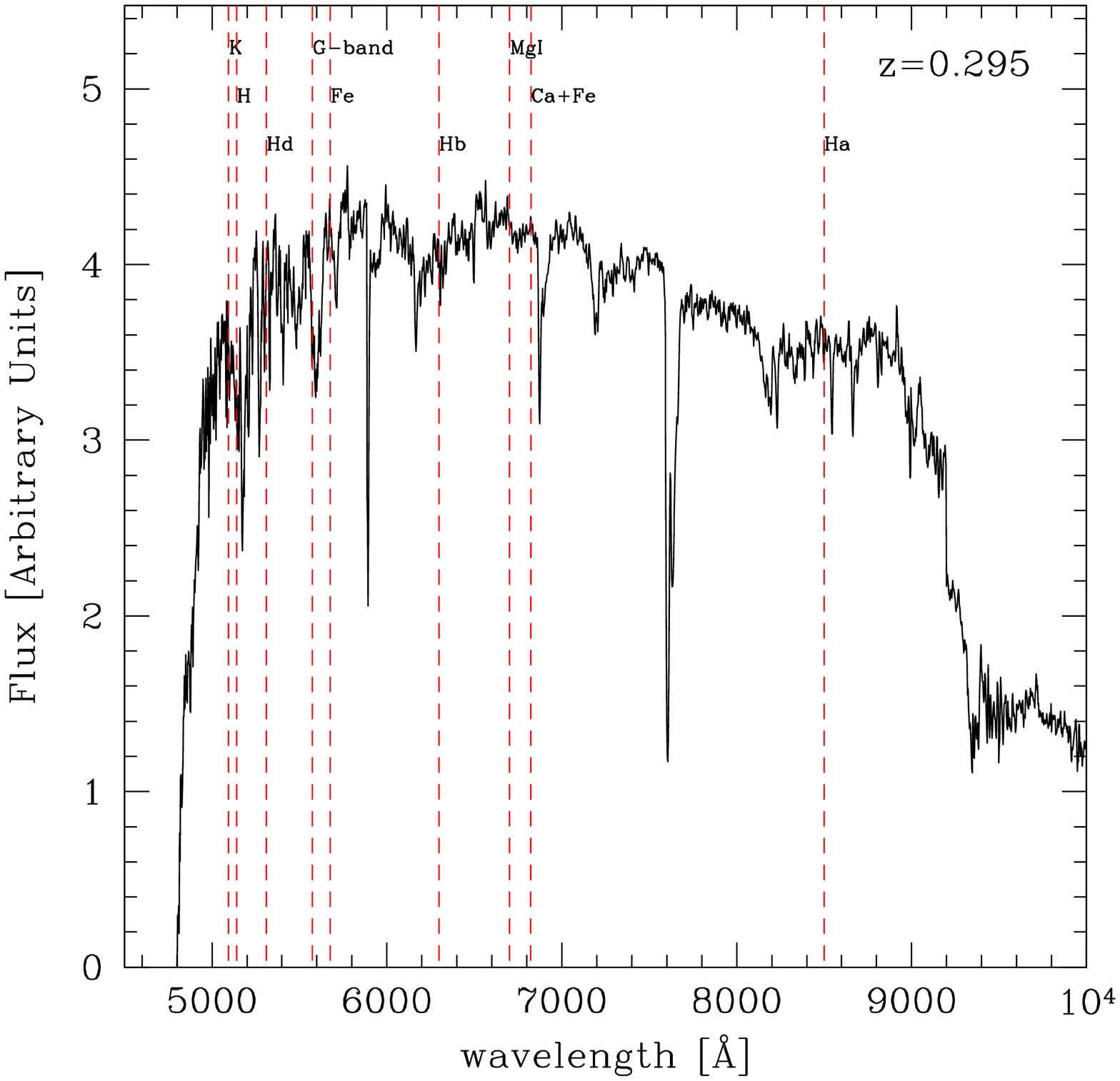}\includegraphics[width=6cm]{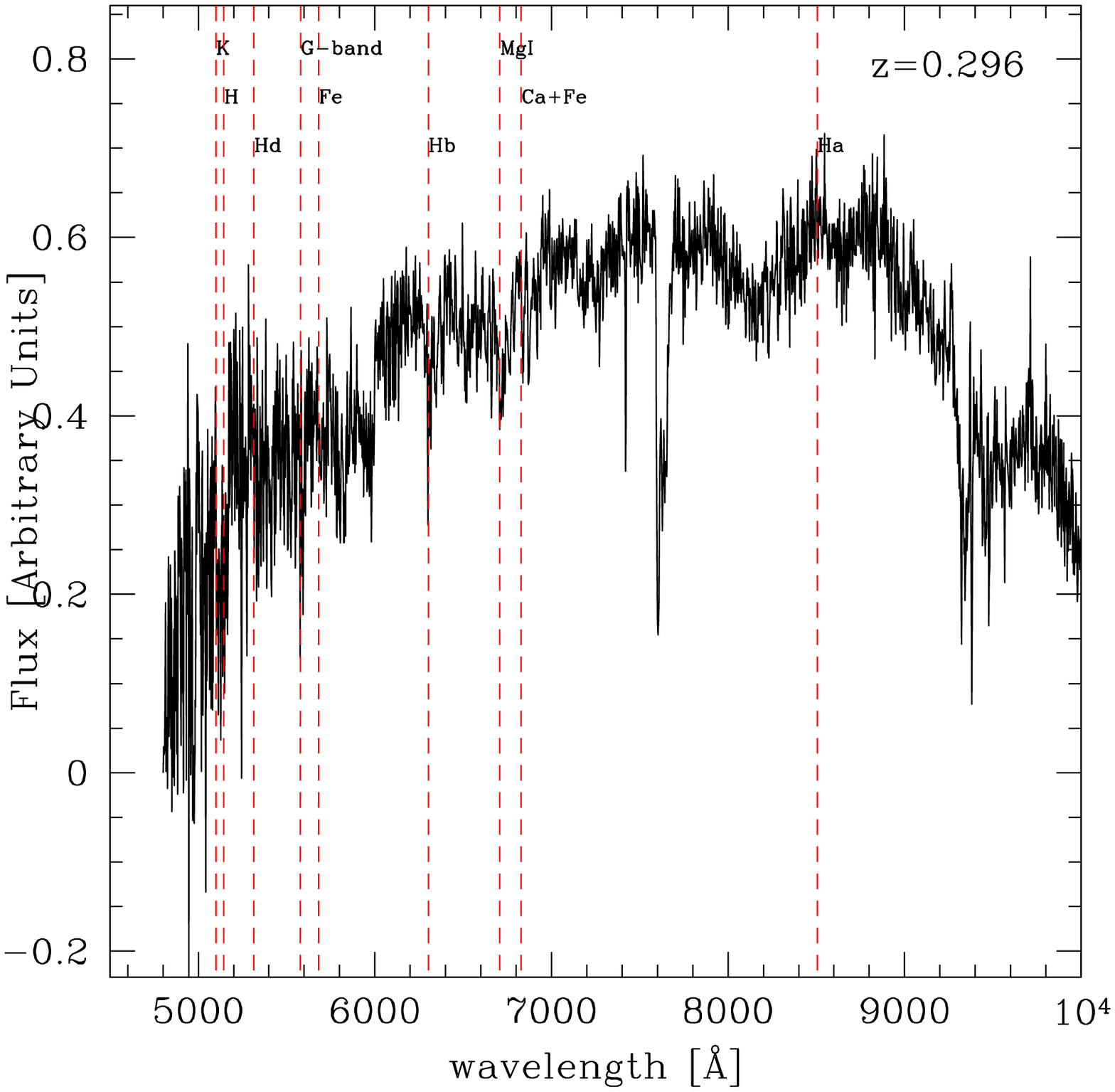} 

\includegraphics[width=6cm]{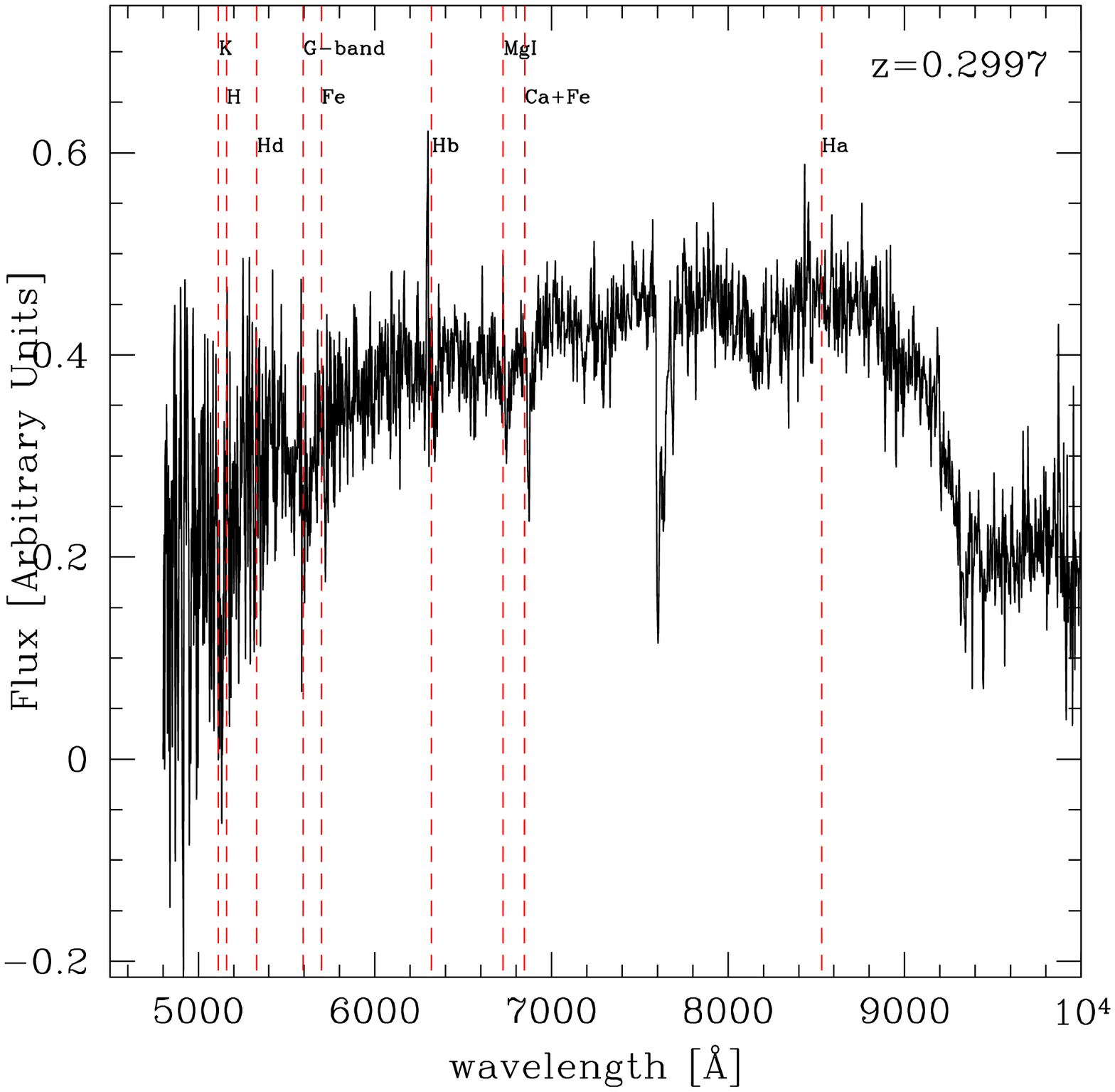}\includegraphics[width=6cm]{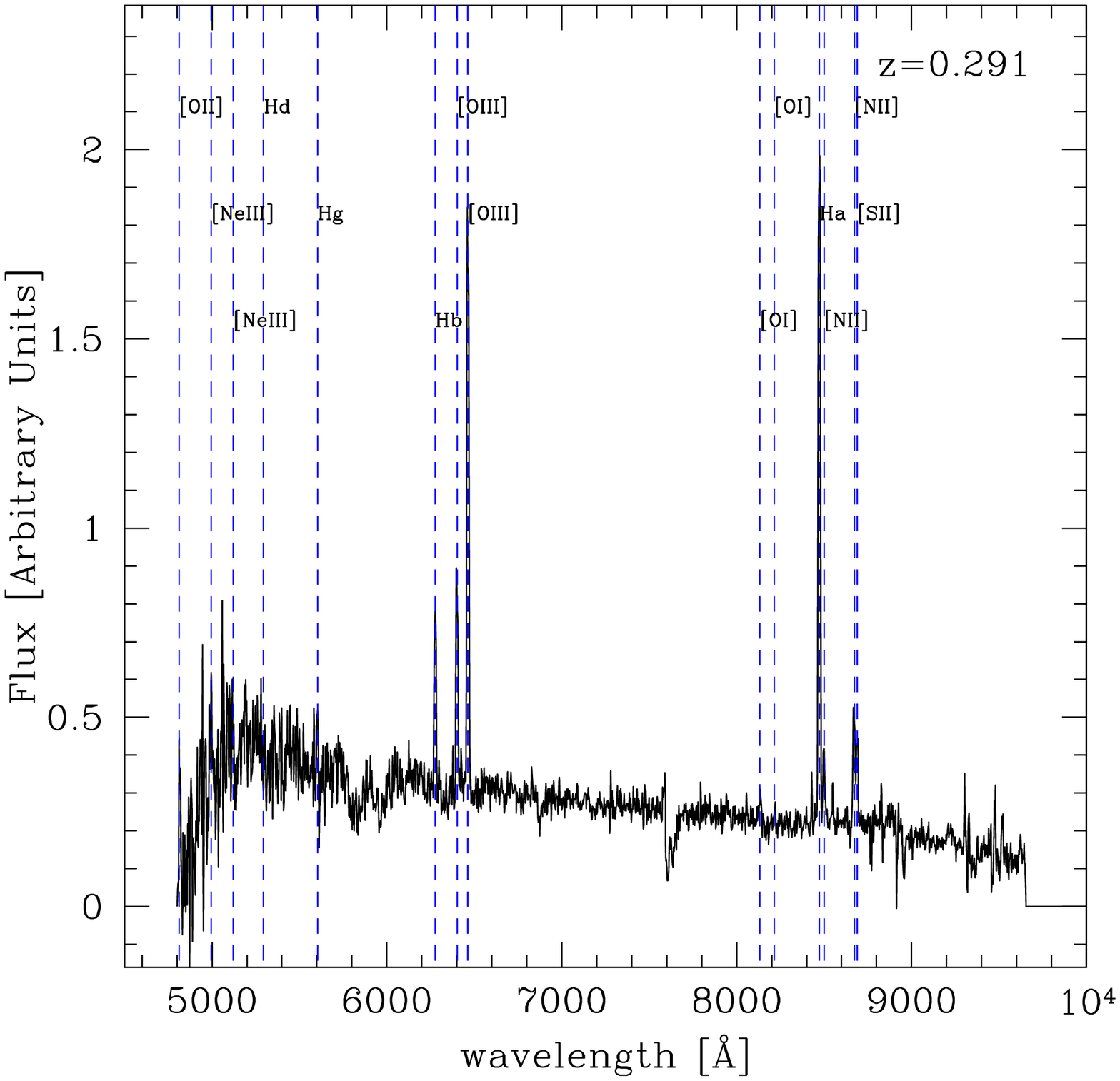}\includegraphics[width=6cm]{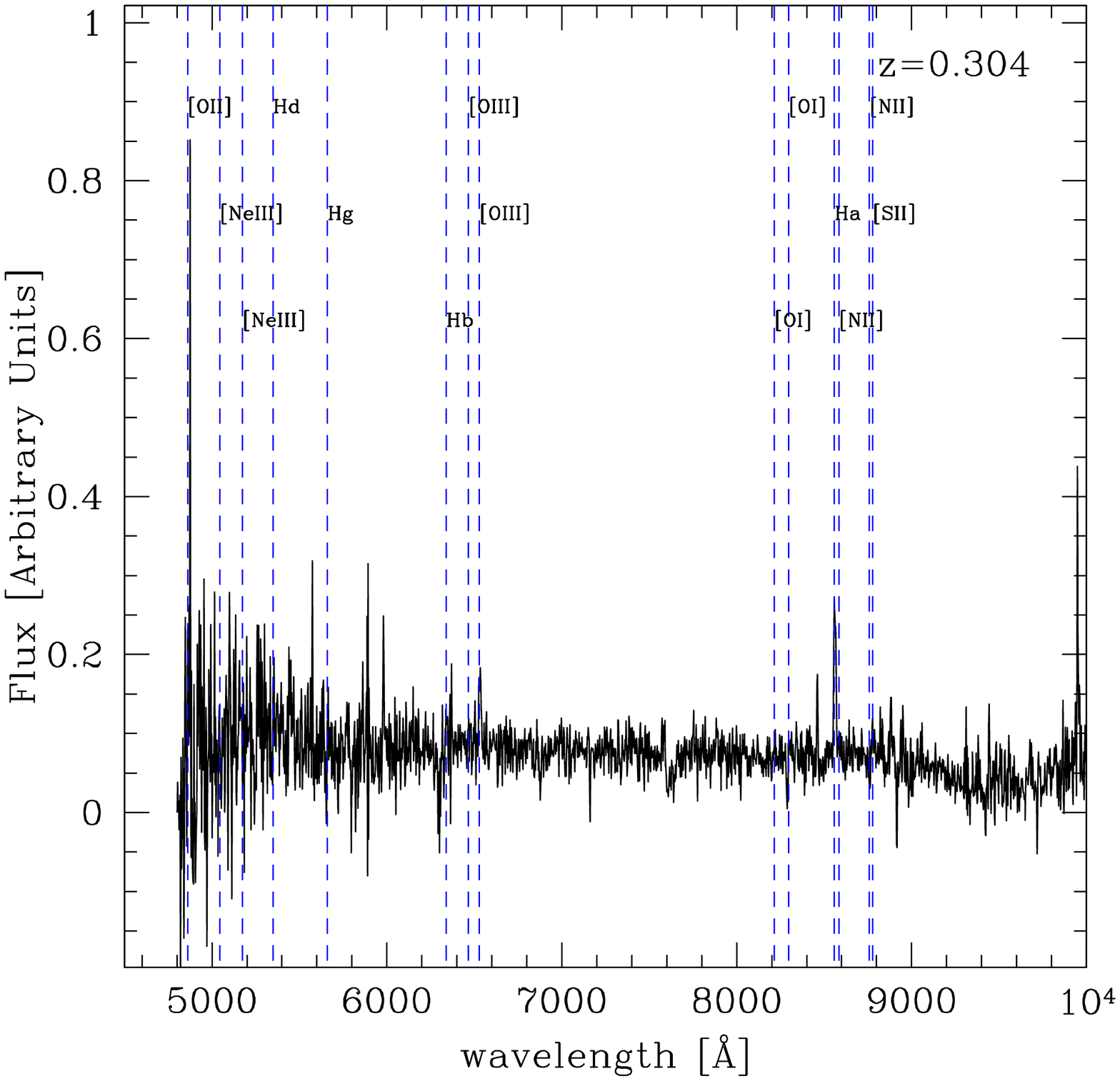} 

\includegraphics[width=6cm]{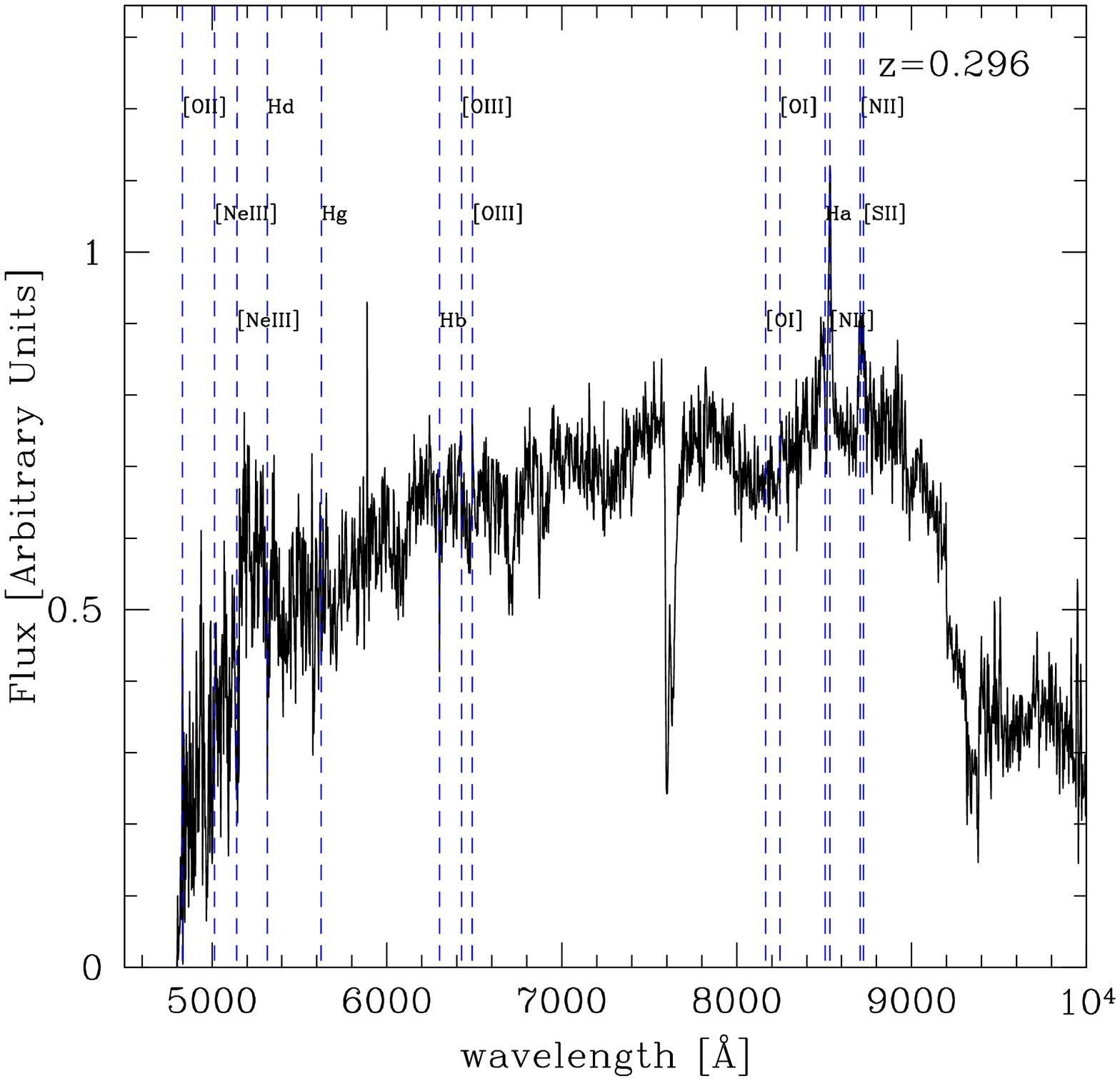}\includegraphics[width=6cm]{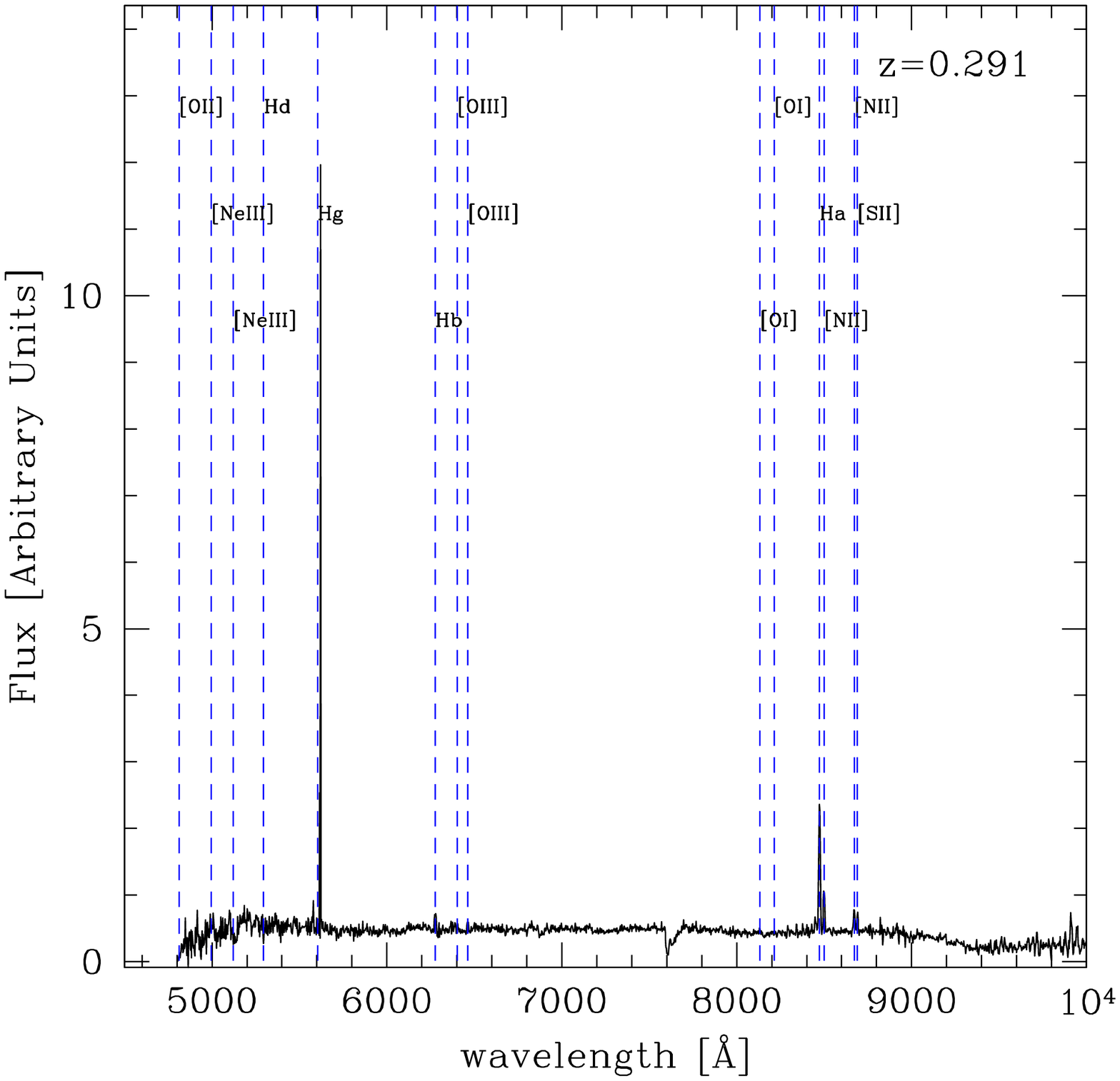}\includegraphics[width=6cm]{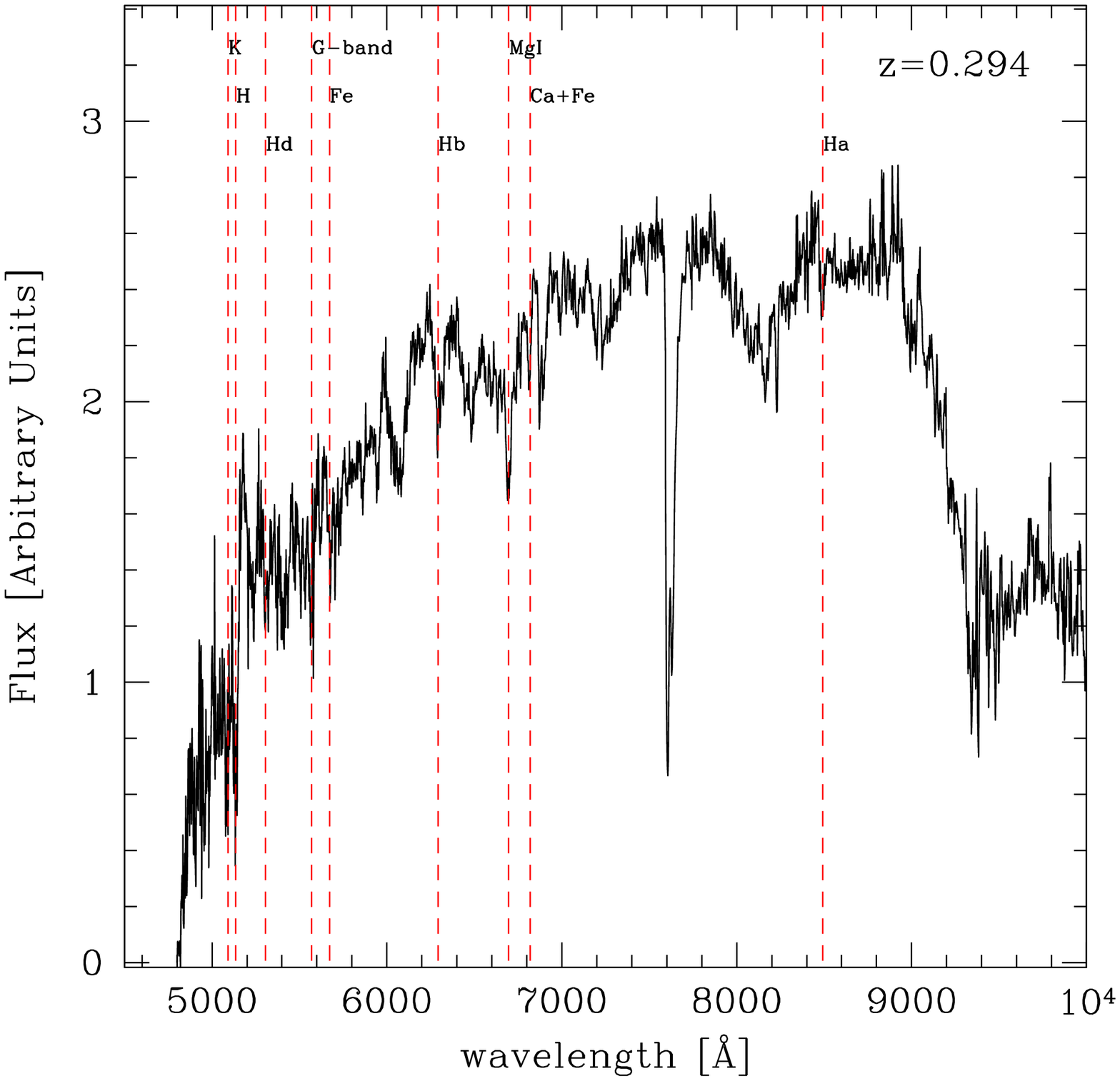}

\caption{VIMOS optical spectra of 49 cluster members (see
  Tab. \ref{souzbullet}). From top to bottom and from left to right
  spectrum of sources \#51, 52, 53, 54, 55, 57, 58, 59, 60, 61, 62, 63.
  \label{spettri}}
\end{figure*}

\begin{figure*}
\ContinuedFloat
  
\centering

\includegraphics[width=6cm]{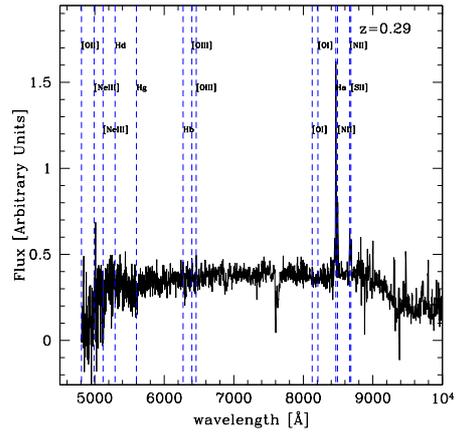}

\caption{VIMOS optical spectra of 49 cluster members (see
  Tab. \ref{souzbullet}). Here source \#64.
\label{spettri}}
\end{figure*}

\section{New spectroscopic redshift of field sources}

\begin{table*}
  \caption{New spectroscopic redshift of sources in the Bullet field}
     \begin{center}
  \begin{small}
    \begin{tabular}{lccccccc}      
      \hline
     R.A. & Dec. & z  &     fl$^{a}$ & Instr.$^{b}$  &  Reg.$^c$ & Dis.$^d$ & Log(Lx$_H$)$^e$ \\
 \hline
 \hline
 104.29484 &    -55.96808 &    0.2215 & 2    & Chandra & C    & 11.89  & 41.94  \\  
  104.29846 &    -55.96422 &    0.222  & 3    & WFI     & C    & 11.75  &    	 \\
  104.31425 &    -56.00233 &    0.961  & 14   & Chandra & C    & 11.62  & 43.99	 \\
  104.33135 &    -55.78235 &    0.116  & 3    & WFI     & C    & 14.61  &    	 \\
  104.33439 &    -55.84008 &    0.0    & 2    & Chandra & C    & 12.4   &    	 \\
  104.3354  &    -55.98148 &    0.326  & 3    & WFI     & C    & 10.65  &    	 \\
  104.34069 &    -55.91983 &    0.6547 & 3    & Chandra & C    & 10.46  & 42.62	 \\
  104.3418  &    -55.77178 &    0.117  & 3    & WFI     & C    & 14.81  &    	 \\
  104.35767 &    -55.83168 &    0.327  & 14   & Chandra & C    & 12.03  & 43.76	 \\
  104.3627  &    -55.85585 &    0.5825 & 2    & Chandra & C    & 11.09  & 42.67	 \\
  104.36485 &    -55.80008 &    0.275  & 13   & Chandra & C    & 13.06  & 42.55	 \\
  104.37303 &    -55.90507 &    0.345  & 3    & WFI     & C    & 9.593  &    	 \\
  104.37357 &    -55.9495  &    0.345  & 1    & IRAC    & C    & 9.195  &    	 \\
  104.38152 &    -55.95877 &    0.479  & 3    & Chandra & C    & 8.945  & 43.41	 \\
  104.38243 &    -55.82737 &    0.118  & 1    & WFI     & C    & 11.53  &    	 \\
  104.3978  &    -55.99726 &    0.6587 & 3    & Chandra & C    & 8.858  & 42.7 	 \\
  104.39784 &    -55.97135 &    0.0    & 3    & Chandra & C    & 8.482  &    	 \\
  104.39837 &    -55.74908 &    1.277  & 19   & Chandra & C    & 14.65  & 44.29	 \\
  104.40704 &    -55.95538 &    0.343  & 1    & IRAC    & C    & 8.078  &    	 \\
  104.40759 &    -55.912   &    0.343  & 1    & IRAC    & C    & 8.359  &    	 \\
  104.41291 &    -55.93628 &    0.346  & 2    & IRAC    & B    & 7.912  &    	 \\
  104.42047 &    -55.90263 &    0.5134 & 2    & Chandra & C    & 8.119  & 42.23	 \\
  104.42352 &    -55.94656 &    0.0    & 3    & Chandra & B    & 7.518  &    	 \\
  104.42357 &    -55.92546 &    0.438  & 2    & Chandra & B    & 7.65   & 42.13	 \\
  104.4248  &    -55.84415 &    0.0    & 2    & Chandra & C    & 9.783  &    	 \\
  104.42571 &    -55.79183 &    0.0    & 2    & Chandra & C    & 12.03  &    	 \\
  104.42809 &    -55.77467 &    0.449  & 2    & IRAC    & C    & 12.81  &    	 \\
  104.4281  &    -55.9309  &    0.0    & 2    & Chandra & B    & 7.445  &    	 \\
  104.43435 &    -55.97685 &    0.234  & 3    & IRAC    & B    & 7.34   &    	 \\
  104.43449 &    -55.75239 &    0.255  & 2    & IRAC    & C    & 13.81  &    	 \\
  104.44178 &    -55.93916 &    0.0    & 4    & Chandra & B    & 6.929  &    	 \\
  104.44379 &    -55.95469 &    0.413  & 3    & Chandra & B    & 6.842  & 42.25	 \\
  104.45172 &    -55.82145 &    0.0    & 2    & Chandra & C    & 10.1   &    	 \\
  104.45254 &    -55.99268 &    0.119  & 3    & WFI     & B    & 7.04   &    	 \\
  104.4532  &    -55.9629  &    0.624  & 3    & Chandra & B    & 6.569  & 42.95	 \\
  104.45782 &    -55.98187 &    1.542  & 13   & Chandra & B    & 6.657  & 44.62	 \\
  104.462   &    -55.88435 &    0.345  & 1    & Chandra & B    & 7.34   & 41.41	 \\
  104.46258 &    -55.97336 &    0.0    & 3    & Chandra & B    & 6.369  &    	 \\
  104.47603 &    -55.97754 &    0.0    & 3    & Chandra & B    & 5.997  &    	 \\
  104.48323 &    -55.98688 &    0.0    & 3    & Chandra & B    & 5.955  &    	 \\
  104.49595 &    -56.08158 &    0.758  & 2    & Chandra & C    & 9.431  & 42.95	 \\
  104.50162 &    -55.90746 &    0.5828 & 1    & Chandra & B    & 5.493  & 42.84	 \\
  104.51277 &    -55.92476 &    0.413  & 3    & Chandra & A    & 4.746  & 41.45	 \\
  104.51501 &    -55.92142 &    0.9824 & 1    & Chandra & A    & 4.741  & 42.53	 \\
  104.5164  &    -56.05258 &    2.146  & 12   & IRAC    & B    & 7.606  &    	 \\
  104.52017 &    -55.84699 &    0.0    & 1    & Chandra & B    & 7.466  &    	 \\
  104.52207 &    -55.93552 &    0.5028 & 1    & Chandra & A    & 4.281  & 41.85	 \\
  104.53542 &    -55.79321 &    0.0    & 3    & Chandra & C    & 10.08  &    	 \\
  104.53557 &    -55.98418 &    0.349  & 4    & IRAC    & A    & 4.3    &    	 \\
  104.53842 &    -56.00427 &    0.098  & 19    & IRAC    & A    & 4.931  &     \\
  104.53938 &    -56.05622 &    0.1492 & 1    & Chandra & B    & 7.379  & 41.39	 \\
  104.54007 &    -55.83098 &    0.7819 & 1    & Chandra & B    & 7.946  & 42.49	 \\
  104.54523 &    -56.12534 &    0.6832 & 2    & Chandra & C    & 11.12  & 43.2 	 \\
  104.5485  &    -55.7807  &    0.174  & 4    & Chandra & C    & 10.63  & 41.59	 \\
  104.55198 &    -55.75025 &    0.211  & 3    & IRAC    & C    & 12.35  &    	 \\
  104.55459 &    -55.80045 &    0.5794 & 1    & Chandra & C    & 9.442  & 42.15	 \\
  104.56466 &    -55.98784 &    0.097  & 4    & Chandra & A    & 3.62   & 40.26	 \\
  104.57104 &    -55.91755 &    0.3278 & 2    & Chandra & A    & 3.178  & 41.31	 \\
  104.57288 &    -56.04702 &    0.3539 & 1    & Chandra & B    & 6.388  & 42.44	 \\
 \hline
  \end{tabular}
  \end{small}
\end{center}

$^{a}$ redshift quality flag, ranging from 1 for very insecure
     redshifts to 4 corresponding to secure ones. 1 in front of the
     flags, indicates objects classified as QSO. The flag 19 indicates
     spectra with a single isolated emission line, which are 100\%
     reliable; $^{b}$ instrument used to select the source; $^c$
     region where the source lands (A, B, C in Fig. \ref{image}, the
     symbol * indicates the red elliptical region in Fig. \ref{image},
     which has been excluded by the X-ray analysis); $^d$ distance
     from the cluster centre in arcmin; $^e$ log(2-10 keV observed
     luminosity, which has been evaluated using an absorbed power$-$law spectrum with
     $\Gamma=$1.9.
\label{sounozbullet}
\end{table*} 

\begin{table*}
  \ContinuedFloat
  \caption{New spectroscopic redshift of sources in the Bullet field}
     \begin{center}
  \begin{small}
    \begin{tabular}{lccccccc}      
      \hline
     R.A. & Dec. & z  &     fl$^{a}$ & Instr.$^{b}$  &  Reg.$^c$ & Dis.$^d$ & Log(Lx$_H$)$^e$ \\
 \hline
 \hline
  
  104.57409 &    -55.92671 &    0.7799 & 2    & Chandra & A    & 2.795  & 42.38	 \\
  104.57942 &    -55.89782 &    0.7114 & 2    & Chandra & A    & 3.822  & 42.73	 \\
  104.58056 &    -55.97838 &    0.097  & 3    & Chandra & A    & 2.847  & 40.38	 \\
  104.58611 &    -55.8789  &    0.136  & 3    & Chandra & A    & 4.679  & 40.81	 \\
  104.59541 &    -55.75976 &    0.396  & 3    & Chandra & C    & 11.49  & 42.44	 \\
  104.5968  &    -55.9042  &    0.78   & 4    & Chandra & A    & 3.174  & 43.39	 \\
  104.59803 &    -55.75567 &    0.271  & 1    & IRAC    & C    & 11.72  &    	 \\
  104.61816 &    -55.91006 &    0.328  & 3    & Chandra & *    & 2.529  & 42.05	 \\
  104.62383 &    -55.85702 &    0.174  & 1    & IRAC    & B    & 5.572  &    	 \\
  104.62821 &    -55.96111 &    0.347  & 2    & Chandra & *    & 0.9657 & 42.53	 \\
  104.62891 &    -55.78974 &    0.368  & 4    & Chandra & C    & 9.573  & 42.27	 \\
  104.63384 &    -55.85087 &    0.348  & 3    & IRAC    & B    & 5.903  &    	 \\
  104.63775 &    -55.76893 &    0.327  & 2    & IRAC    & C    & 10.81  &    	 \\
  104.6397  &    -55.81648 &    0.619  & 1    & Chandra & B    & 7.953  & 42.92	 \\
  104.64069 &    -55.97303 &    0.351  & 3    & IRAC    & *    & 1.46   &    	 \\
  104.65137 &    -55.873   &    0.109  & 4    & Chandra & A    & 4.56   & 40.7 	 \\
  104.65225 &    -55.84737 &    0.356  & 3    & IRAC    & B    & 6.098  &    	 \\
  104.66145 &    -55.96805 &    0.325  & 2    & IRAC    & *    & 1.243  &    	 \\
  104.66775 &    -56.10359 &    0.876  & 1    & Chandra & C    & 9.303  & 43.25	 \\
  104.67078 &    -56.06422 &    0.216  & 1    & IRAC    & B    & 6.961  &    	 \\
  104.68223 &    -55.95154 &    0.351  & 2    & Chandra & *    & 1.192  & 41.85	 \\
  104.6866  &    -55.92378 &    0.4798 & 1    & Chandra & A    & 2.013  & 41.69	 \\
  104.68729 &    -55.90688 &    0.353  & 2    & Chandra & A    & 2.865  & 42.42	 \\
  104.69197 &    -55.78604 &    0.369  & 3    & IRAC    & C    & 9.893  &    	 \\
  104.69354 &    -55.81759 &    0.512  & 3    & IRAC    & B    & 8.037  &    	 \\
  104.69592 &    -55.84348 &    0.2652 & 11   & Chandra & B    & 6.54   & 42.35	 \\
  104.70166 &    -55.95094 &    2.15   & 14   & Chandra & A    & 1.839  & 44.06	 \\
  104.70307 &    -56.06774 &    0.32   & 3    & IRAC    & B    & 7.369  &    	 \\
  104.70411 &    -55.77522 &    0.368  & 3    & IRAC    & C    & 10.6   &    	 \\
  104.70916 &    -56.05041 &    1.599  & 12   & Chandra & B    & 6.433  & 44.03	 \\
  104.71108 &    -55.97227 &    0.326  & 2    & IRAC    & A    & 2.565  &    	 \\
  104.71113 &    -55.80706 &    0.0    & 3    & Chandra & C    & 8.784  &    	 \\
  104.71347 &    -55.92693 &    0.8797 & 2    & Chandra & A    & 2.596  & 42.7 	 \\
  104.71365 &    -56.0895  &    0.34   & 2    & IRAC    & C    & 8.722  &    	 \\
  104.71987 &    -55.94819 &    0.345  & 1    & IRAC    & A    & 2.448  &    	 \\
  104.72316 &    -55.98171 &    1.678  & 19   & Chandra & A    & 3.223  & 43.08	 \\
  104.72377 &    -55.95946 &    0.9856 & 2    & Chandra & A    & 2.653  & 43.37	 \\
  104.72429 &    -55.88401 &    0.0    & 1    & Chandra & A    & 4.685  &    	 \\
  104.72623 &    -56.02138 &    0.3935 & 2    & Chandra & A    & 5.092  & 41.74	 \\
  104.72773 &    -55.92093 &    0.326  & 2    & IRAC    & A    & 3.193  &    	 \\
  104.73515 &    -55.92275 &    0.222  & 2    & Chandra & A    & 3.355  & 41.29	 \\
  104.7395  &    -56.01251 &    0.26   & 13   & Chandra & A    & 4.915  & 42.44	 \\
  104.74231 &    -55.80296 &    0.589  & 3    & IRAC    & C    & 9.332  &    	 \\
  104.74493 &    -55.84026 &    0.428  & 4    & Chandra & B    & 7.309  & 42.0 	 \\
  104.747   &    -56.02988 &    1.525  & 12   & Chandra & B    & 5.9    & 44.01	 \\
  104.74816 &    -56.08019 &    0.234  & 3    & Chandra & C    & 8.571  & 41.07	 \\
  104.75086 &    -56.02563 &    0.0    & 2    & Chandra & B    & 5.769  &    	 \\
  104.75192 &    -55.99593 &    0.5363 & 3    & Chandra & A    & 4.509  & 42.02	 \\
  104.75401 &    -55.77038 &    0.368  & 2    & IRAC    & C    & 11.31  &    	 \\
  104.76315 &    -55.91658 &    0.0    & 2    & Chandra & A    & 4.362  &    	 \\
  104.76352 &    -55.77678 &    0.274  & 2    & IRAC    & C    & 11.05  &    	 \\
  104.76526 &    -55.93914 &    1.0288 & 2    & Chandra & A    & 4.018  & 42.56	 \\
  104.7668  &    -55.98671 &    0.9753 & 2    & Chandra & A    & 4.615  & 42.45	 \\
  104.77147 &    -55.96897 &    0.7058 & 1    & Chandra & A    & 4.349  & 42.05	 \\
  104.77202 &    -56.05343 &    0.5794 & 1    & Chandra & B    & 7.539  & 42.36	 \\
  104.77642 &    -55.79979 &    0.495  & 3    & IRAC    & C    & 9.958  &    	 \\
  104.77877 &    -55.78995 &    0.136  & 4    & Chandra & C    & 10.53  & 40.94	 \\
 \hline
  \end{tabular}
  \end{small}
\end{center}

$^{a}$ redshift quality flag, ranging from 1 for very insecure
     redshifts to 4 corresponding to secure ones. 1 in front of the
     flags, indicates objects classified as QSO. The flag 19 indicates
     spectra with a single isolated emission line, which are 100\%
     reliable; $^{b}$ instrument used to select the source; $^c$
     region where the source lands (A, B, C in Fig. \ref{image}, the
     symbol * indicates the red elliptical region in Fig. \ref{image},
     which has been excluded by the X-ray analysis); $^d$ distance
     from the cluster centre in arcmin; $^e$ log(2-10 keV observed
     luminosity, which has been evaluated using an absorbed power$-$law spectrum with
     $\Gamma=$1.9.
\label{sounozbullet}
\end{table*} 

\begin{table*}
  \ContinuedFloat
  \caption{New spectroscopic redshift of sources in the Bullet field}
     \begin{center}
  \begin{small}
    \begin{tabular}{lccccccc}      
      \hline
     R.A. & Dec. & z  &     fl$^{a}$ & Instr.$^{b}$  &  Reg.$^c$ & Dis.$^d$ & Log(Lx$_H$)$^e$ \\
 \hline
 \hline
  104.78366 &    -55.9283  &    0.578  & 1    & IRAC    & A    & 4.759  &    	 \\
  104.7873  &    -55.7961  &    0.136  & 4    & IRAC    & C    & 10.32  &    	 \\
  104.80312 &    -55.93729 &    0.35   & 3    & Chandra & A    & 5.295  & 41.74	 \\
  104.81268 &    -55.94435 &    0.0    & 3    & Chandra & B    & 5.575  &    	 \\
  104.81515 &    -55.98841 &    0.5798 & 1    & Chandra & B    & 6.121  & 42.73	 \\
  104.81929 &    -56.04149 &    1.3459 & 12   & Chandra & B    & 8.011  & 43.85	 \\
  104.8256  &    -55.96219 &    1.0792 & 13   & Chandra & B    & 6.052  & 44.0 	 \\
  104.8405  &    -56.05618 &    1.6134 & 12   & Chandra & C    & 9.134  & 44.14	 \\
  104.84364 &    -55.91464 &    0.349  & 1    & IRAC    & B    & 6.928  &    	 \\
  104.86249 &    -55.79964 &    0.36   & 2    & IRAC    & C    & 11.54  &    	 \\
  104.87967 &    -55.96294 &    0.219  & 3    & WFI     & B    & 7.862  &    	 \\
  104.88178 &    -55.80551 &    0.12   & 3    & IRAC    & C    & 11.7   &    	 \\
  104.88826 &    -55.78985 &    0.848  & 13   & Chandra & C    & 12.55  & 44.62	 \\
  104.88977 &    -55.98721 &    2.711  & 14   & Chandra & C    & 8.468  & 43.62	 \\
  104.89196 &    -55.82559 &    0.326  & 2    & WFI     & C    & 11.09  &    	 \\
  104.89279 &    -55.85004 &    0.566  & 2    & Chandra & C    & 10.19  & 43.36	 \\
  104.90293 &    -55.84408 &    0.0    & 1    & Chandra & C    & 10.68  &    	 \\
  104.9183  &    -56.00699 &    0.219  & 2    & WFI     & C    & 9.748  &    	 \\
  104.92801 &    -55.95046 &    0.348  & 1    & WFI     & C    & 9.445  &    	 \\
  104.93135 &    -55.82881 &    0.0    & 2    & WFI     & C    & 12.0   &    	 \\
  104.93182 &    -55.9974  &    2.143  & 13   & Chandra & C    & 9.993  & 44.2 	 \\
  104.93897 &    -55.77216 &    0.176  & 3    & WFI     & C    & 14.48  &    	 \\
  104.95363 &    -55.8173  &    1.246  & 13   & Chandra & C    & 13.01  & 44.12	 \\
  104.9603  &    -55.92678 &    0.578  & 2    & IRAC    & C    & 10.62  &    	 \\
  104.96163 &    -55.85552 &    0.245  & 2    & WFI     & C    & 11.99  &    	 \\
  104.96437 &    -55.97168 &    0.613  & 13   & Chandra & C    & 10.75  & 43.91	 \\
  104.98557 &    -55.97558 &    0.433  & 2    & WFI     & C    & 11.48  &    	 \\
  104.98769 &    -55.93163 &    0.222  & 1    & WFI     & C    & 11.5   &    	 \\
  105.04212 &    -55.94101 &    0.91   & 13   & Chandra & C    & 13.29  & 44.01  \\

  \hline
  \end{tabular}
  \end{small}
\end{center}

$^{a}$ redshift quality flag, ranging from 1 for very insecure
     redshifts to 4 corresponding to secure ones. 1 in front of the
     flags, indicates objects classified as QSO. The flag 19 indicates
     spectra with a single isolated emission line, which are 100\%
     reliable; $^{b}$ instrument used to select the source; $^c$
     region where the source lands (A, B, C in Fig. \ref{image}, the
     symbol * indicates the red elliptical region in Fig. \ref{image},
     which has been excluded by the X-ray analysis); $^d$ distance
     from the cluster centre in arcmin; $^e$ log(2-10 keV observed
     luminosity, which has been evaluated using an absorbed power$-$law spectrum with
     $\Gamma=$1.9.
\label{sounozbullet}
\end{table*}

\end{appendix}
\end{document}